% The following line automatically loads the mn macros if you are not
% using a format file.
\ifx\mnmacrosloaded\undefined 
\input mn\fi
\input psfig.sty

% defining some of my own macros

\def\eg{{e.g.~}}
\def\ie{{i.e.~}}
\def\etal{et~al.~}
\def\eV{e\kern-.15em V}                 % define eV macro
\def\keV{ke\kern-.15em V}                % define keV macro
\def\nh{$N_{\rm H}$}

\def\Ha{H$\alpha$}
\def\Hb{H$\beta$}
\def\oiii{[O{\sc iii}]}
\def\ciii{C{\sc iii}]}
\def\mgii{Mg{\sc ii}}
\def\oiiifull{[O{\sc iii}]$\lambda$5007}
\def\ciiifull{C{\sc iii}]$\lambda$1909}
\def\aliii{Al{\sc iii}$\lambda$1857}
\def\siiii{Si{\sc iii}]$\lambda$1892}
\def\civ{C{\sc iv}$\lambda$1549}
\def\mgiifull{Mg{\sc ii}$\lambda$2798}
\def\feii{Fe{\sc ii}}
\def\lha{$L_{{\rm H}\alpha}$}
\def\lhb{$L_{{\rm H}\beta}$}
\def\loiii{$L_{[{\rm O}{\sc iii}]}$}
\def\lciii{$L_{{\rm C}{\sc iii}]}$}
\def\lmgii{$L_{{\rm Mg}{\sc ii}}$}

\def\hafwhm{H$\alpha$~{\sc FWHM}}

\def\hbfwhm{H$\beta$~{\sc FWHM}}

\def\oiiifwhm{[O{\sc iii}]~{\sc FWHM}}
\def\fwhmciii{C{\sc iii}]~{\sc FWHM}}
\def\ciiifwhm{C{\sc iii}]~{\sc FWHM}}
\def\fwhmmgii{Mg{\sc ii}~{\sc FWHM}}
\def\mgiifwhm{Mg{\sc ii}~{\sc FWHM}}

\def\haew{H$\alpha$~{\sc EW}}

\def\hbew{H$\beta$~{\sc EW}}

\def\oiiiew{[O{\sc iii}]~{\sc EW}}

\def\ciiiew{C{\sc iii}]~{\sc EW}}

\def\mgiiew{Mg{\sc ii}~{\sc EW}}

\def\logne{log({\sl Ne})}

\def\aox{$\alpha_{\rm ox}$}
\def\aos{$\alpha_{\rm os}$}
\def\ax{$\alpha_{\rm x}$}
\def\aopt{$\alpha_{\rm opt}$}
\def\lhard{$L_{\rm 2keV}$}
\def\lsoft{$L_{\rm 0.2keV}$}

\def\lopt{$L_{\rm 5000}$}
\def\luv{$L_{2500}$}

\def\pcorr{$p_{\rm corr}$}
\def\nhgal{$N_{\rm HGal}$}
\def\kms{km~s$^{-1}$}

% If your system has the AMS fonts version 2.0 installed, MN.tex can be
% made to use them by uncommenting the line: %\AMStwofontstrue
%
% By doing this, you will be able to obtain upright Greek characters.
% e.g. \umu, \upi etc.  See the section on "Upright Greek characters" in
% this guide for further information.

\newif\ifAMStwofonts
%\AMStwofontstrue

\ifCUPmtplainloaded \else
  \NewTextAlphabet{textbfit} {cmbxti10} {}
  \NewTextAlphabet{textbfss} {cmssbx10} {}
  \NewMathAlphabet{mathbfit} {cmbxti10} {} % for math mode
  \NewMathAlphabet{mathbfss} {cmssbx10} {} %  "   "    "
  \ifAMStwofonts
    \NewSymbolFont{upmath} {eurm10}
    \NewSymbolFont{AMSa} {msam10}
    \NewMathSymbol{\upi}     {0}{upmath}{19}
    \NewMathSymbol{\umu}     {0}{upmath}{16}
    \NewMathSymbol{\upartial}{0}{upmath}{40}
    \NewMathSymbol{\leqslant}{3}{AMSa}{36}
    \NewMathSymbol{\geqslant}{3}{AMSa}{3E}

    \let\geq=\geqslant 
  \else
    \def\umu{\mu}
    \def\upi{\pi}
    \def\upartial{\partial}
  \fi
\fi

% Marginal adjustments using \pageoffset maybe required when printing
% proofs on a Laserprinter (this is usually not needed).
% Syntax: \pageoffset{ +/- hor. offset}{ +/- vert. offset}
% e.g.    \pageoffset{-3pc}{-4pc}

\pageoffset{-2.5pc}{0pc}

\loadboldmathnames

%\Referee   %  uncomment this for referee mode (double spaced)
%\Autonumber  %  auto-number sections, subsections and subsubsections

% \pagerange, \pubyear and \volume are defined at the Journals office and
% not by an author.

% \onecolumn        % enable one column mode
% \letters          % for `letters' articles
\pagerange{000--000}    % `letters' articles should use \pagerange{Ln--Ln}
\pubyear{0000}
\volume{000}
% \microfiche{}     % for articles with microfiche
% \authorcomment{}  % author comment for footline

\begintopmatter  %  start the two spanning material

\title{Optical and X-ray properties of the RIXOS AGN: II - Emission lines}
\author{E. M. Puchnarewicz,$^1$ K.~O.~Mason,$^1$ F.~J.~Carrera,$^{1,2}$  W. N.
Brandt,$^{3,4}$ F.~Cabrera-Guera,$^{5}$ R.~Carballo,$^{6,7}$ G.~Hasinger,$^8$
R.~G.~M$^c$Mahon,$^3$  J.~P.~D.~Mittaz,$^1$ M. J. Page,$^1$
I.~Perez-Fournon,$^{5}$ and A.~Schwope$^8$}

\affiliation{$^1$ Mullard Space Science  Laboratory, University College London,
Holmbury St. Mary, Dorking, Surrey RH5 6NT, UK.}
\affiliation{$^2$ Present address: Instituto de Fisica de Cantabria CSIC-UC,
Facultad de Ciencias, Avda. de los Castros s/n 39005, Santander, Spain.}
\affiliation{$^3$ Institute of Astronomy, Madingley Road, Cambridge CB3 0HA,
UK.}
\affiliation{$^4$ Present address: Harvard-Smithsonian Center for Astrophysics,
60 Garden Street, Cambridge, MA 02138, USA}
\affiliation{$^5$ Instituto de Astrofisica de Canarias Via Lactea, s/n 38200
La Laguna, Tenerife, Spain}
\affiliation{$^6$ Instituto de Fisica de Cantabria, Avda. de los 
Castros s/n 39005, Santander, Spain.}
\affiliation{$^7$ Dept. Fisica Moderna, Universidad de Cantabria, Avda. de los
Castros s/n 39005, Santander, Spain.}
\affiliation{$^8$ Astrophysikalisches Institut Potsdam, An der Sternwarte 16,
Potsdam, Germany.}

\shortauthor{E. M. Puchnarewicz et al.}
\shorttitle{Properties of RIXOS AGN: II - emission lines}

% \acceptedline is to be defined at the Journals office and not
% by an author.

%\acceptedline{Accepted 1988 December 15. Received 1988 December 14;
%  in original form 1988 October 11}

\abstract {We present the optical and UV emission line properties of 160 X-ray
selected AGN taken from the RIXOS survey (including \Ha, \Hb, \oiiifull,
\mgiifull\ and \ciiifull). This sample is believed to  contain a mixture of
absorbed and unabsorbed objects, with column densities up to
$\sim4\times10^{21}$ cm$^{-2}$.  Although the distribution of the \oiiifull\ EW
for the RIXOS AGN is  typical of optically-selected samples,  the Balmer line
EWs are relatively low. This is consistent with the presence of a dust absorber
between the broad and narrow line regions (\eg a molecular torus), and
intrinsically weak optical line emission. We find Baldwin effects in \ciiifull\
and \mgiifull, and a positive response of the \mgii\ line to its ionizing
continuum.  

There is a strong correlation between the EW and FWHM of \mgii, which may be
similar to that seen in other samples for \Hb. We demonstrate that this is
consistent with models which suggest two line-emitting zones, a `very broad
line region' (VBLR) and an `intermediate line region' (ILR). The correlation
between EW and FWHM in \mgii\ may be a physical characteristic of the ILR: \eg
the radiative acceleration of the line-emitting clouds; or it may reflect a
geometric dependence, \eg anisotropies in the line or continuum emission, or a
smaller covering factor of the ILR at large distances.

We found no correlation between the \hbfwhm\ and the slope of the X-ray
spectrum, despite reports of correlations in other samples.  However, this may
be due to the effects of dust absorption which suppresses the broad \Hb\
component, masking any relationship.  The \Ha\ FWHM does tend  to be narrow
when \ax\ is soft, and broadens as \ax\ hardens, although the formal
probability for this  correlation is low (91 per cent). If the distribution of
\ax\ in the RIXOS sample reflects the level of intrinsic absorption in these
AGN, the data suggest a possible link between the velocity of the Balmer
line-emitting region and the amount of absorbing material beyond

}


\maketitle  %  finish the two spanning material

\section{Introduction}

The investigation and interpretation of the broad emission line properties of
Seyferts and quasars is a complex task. Photoionization codes such as {\sc
cloudy} (Ferland 1993) and {\sc ion} (Netzer 1993) have been successful at
reproducing the observed ratios of many emission lines (\eg Ferland \& Netzer
1979; Kwan \& Krolik 1981; Rees, Netzer \& Ferland 1989), which supports
photoionization as the most likely physical means of line production in AGN.
However, finding an appropriate structure and geometry for the broad line
region (BLR) has been less straightforward.
                             
Many different models have been proposed for the BLR structure, often invoking
separate regions for the high- and low-ionization lines (HILs and LILs
respectively). For example, models based on the intensive NGC 5548 monitoring
campaign (Clavel \etal 1991; Peterson \etal 1991) suggest that the HILs (\eg
Ly$\alpha$, \ciii, C{\sc IV}, He{\sc I}, He{\sc II})  are emitted from an
inner, roughly spherical high-ionization zone  while the LILs (which include
most of the Balmer lines and \mgii) are emitted from  an outer,
flattened low-ionization zone (Krolik \etal 1991; O'Brien, Goad \& Gondhalekar
1994). Alternatively, it has been proposed that the HILs are emitted from an
outflow aligned with the radio axis, while the LILs come from a flattened
region lying in a plane perpendicular to the outflow (\eg Wills \etal 1993). In
the Collin-Souffrin \etal (1988) model, the LILs are formed in the outer parts
of an accretion disc while the HILs are produced in clouds of cooling gas
behind shocks in a hypersonic flow of interstellar matter. The winds or
envelopes of bloated stars have also been discussed as an origin of the BLR
(\eg Edwards 1980, Alexander \& Netzer 1994; Baldwin \etal 1996).

The viability of these models can be tested using the observed emission line
parameters of AGN, \eg the full width at half maximum (FWHM) of the line, its
luminosity and equivalent width (EW). The proximity of the BLR to the ionizing
continuum source suggests that much of the line emission is directed from the
inner face of a BLR cloud towards the central source  (Ferland \etal 1992); for
certain geometries (\eg disc-shaped regions) this introduces a degree of
anisotropy into the model which  may be reflected in the fluxes and EWs of the
lines. The FWHM, assuming that they are dominated by velocity broadening,
measure the line-of-sight component of the velocity of the line-emitting gas.
If the BLR is flattened so that the dominant motion of the clouds lies in a
plane, the observed FWHM will also be a function of  the viewing angle (\ie the
angle between our line of sight and the axis of the BLR). 

We have used a sample of 160 X-ray selected AGN taken from the RIXOS survey
(Mason \etal, in preparation) to test models of the BLR geometry in Seyfert 1s
and quasars. The optical/UV and X-ray continuum properties of the RIXOS AGN are
presented in Puchnarewicz \etal (1996a; hereafter Paper I). In this paper we
present the emission line parameters of \ciiifull, \mgiifull, \Ha, \Hb\ and
\oiiifull\ (see Section 2) and discuss sample properties and correlations for
emission lines and continua (Sections 3 and 4).  No redshift restriction has
been placed on the sample, therefore the lines measured for each AGN depend on
that AGN's redshift. This effectively results in five smaller `subsamples', one
for each emission line, which are compared with other line and continuum
parameters.  The implications of the results for the structure of the BLR and
the geometry of the central regions in AGN, taking into account the effects of
absorption (which are expected to be significant; see Paper I), are discussed
in Section 5.  

\section{Data reduction}

\subsection{The RIXOS Survey}

The RIXOS survey is compiled from serendipitous sources detected in pointed
observations made with the {\sl ROSAT} Position Sensitive Proportional Counter
(PSPC; Pfeffermann \etal\ 1986). To be included in the survey, an observation
must have an exposure time of at least 8~ksec and it must have been taken at a
Galactic latitude greater than 28$^\circ$. Only sources within 17~arcmin of the
centre of the field and with a flux greater than 3$\times10^{-14}$ erg
cm$^{-2}$ sec$^{-1}$ in the {\sl ROSAT} `hard' band (0.4-2.0~\keV) are used;
optical identification of survey sources  is 94 per cent complete to this flux
level over 15 deg$^2$.   

In total, 160 broad-line AGN (\ie including Seyfert 1s to 1.9s and quasars) 
from the RIXOS survey are included in this analysis, covering a range in
redshift from 0.03 to 2.92. Objects were included in this sample if they
exhibited an unambiguous detection of redshifted line emission, and if the
permitted line width was resolved and broad (\ie with a FWHM$\ga$1000~\kms). 
At redshifts of around 1.0, the only strong line which falls within the
observed range is \mgii, thus if only a single broad emission line was observed
and the expected positions of \Hb\ and \ciii\ were not covered by the spectrum,
the line was assumed to be \mgii. At $z\sim$0.3-0.4 the only permitted line
observed is \Hb. If the \Hb\ line is narrow then this object will have been
classified as a narrow emission line galaxy (NELG). However, it is also
possible that these objects may be Seyfert 1.9s or similar where the broad
component of \Hb\ is suppressed or obscured. There are four RIXOS objects, 
classified as NELGs, which may be Seyfert 1.9s; F226\_74, F254\_6,  F272\_10
and F273\_23; these are not included here but will be presented with other
RIXOS NELGs in Romero-Colmenero \etal (in preparation). Four AGN from the
complete sample whose X-ray spectra were poorly determined have been excluded
from this paper; emission line data for a further six were not available,
either because their redshift was too high to measure even \ciii, or because
they were catalogue identifications for which the relevant data could not be
found from other references. The full RIXOS source list including X-ray and
optical positions, fluxes etc., will be presented in a future paper (Mason
\etal, in preparation).

\begintable*{1}
\caption{{\bf Table 1.} Optical and UV emission line parameters.}
\halign{#\hfil                                                   % FID
       &\hskip0.1truecm#\hfil                                    % Snum
        &\hskip0.1truecm\hfil#\hskip0.2truecm                    % z
        &\hskip0.2truecm#\hfil\hskip0.05truecm
       &\hskip0.05truecm#\hfil\hskip0.05truecm
       &\hskip0.05truecm#\hfil\hskip0.2truecm
        &\hskip0.2truecm#\hfil\hskip0.05truecm
       &\hskip0.05truecm#\hfil\hskip0.05truecm
       &\hskip0.05truecm#\hfil\hskip0.2truecm
        &\hskip0.2truecm#\hfil\hskip0.05truecm
       &\hskip0.05truecm#\hfil\hskip0.05truecm
       &\hskip0.05truecm#\hfil\hskip0.2truecm
        &\hskip0.2truecm#\hfil\hskip0.05truecm
       &\hskip0.05truecm#\hfil\hskip0.05truecm
       &\hskip0.05truecm#\hfil\hskip0.2truecm
        &\hskip0.2truecm#\hfil\hskip0.05truecm
       &\hskip0.05truecm#\hfil\hskip0.05truecm
       &\hskip0.05truecm#\hfil\hskip0.2truecm
\cr
\noalign{\bigskip}
    FID   &
    SNo   &
    \hfil{\sl z} \hfil   &
    \multispan3{\hfil \Ha$\lambda$6562 \hfil}  &
    \multispan3{\hfil \Hb$\lambda$4861 \hfil}  &
    \multispan3{\hfil \oiiifull \hfil}  &
    \multispan3{\hfil \mgiifull \hfil}  &
    \multispan3{\hfil \ciiifull \hfil}  
 \cr
    & & 
    & \hfil{\sl lum} \hfil & \hfil{\sl ew} \hfil & \hfil{\sl fwhm} \hfil 
    & \hfil{\sl lum} \hfil & \hfil{\sl ew} \hfil & \hfil{\sl fwhm} \hfil 
    & \hfil{\sl lum} \hfil & \hfil{\sl ew} \hfil & \hfil{\sl fwhm} \hfil 
    & \hfil{\sl lum} \hfil & \hfil{\sl ew} \hfil & \hfil{\sl fwhm} \hfil 
    & \hfil{\sl lum} \hfil & \hfil{\sl ew} \hfil & \hfil{\sl fwhm} \hfil 
\cr
\noalign{\medskip}
 110 &   1 & 0.365 &  ---  &  ---  &  ---  & 42.1$^b$ & 50$^b$ & 2900$^b$ & 41.9 & 30 & 700 & 41.9 & 20 & 1900 &  ---  &  ---  &  ---  \cr
     &     &       &       &       &       & 42.0$^n$ & 20$^n$ &  900$^n$ \cr
 110 &   8 & 0.938 &  ---  &  ---  &  ---  &  ---  &  ---  &  ---  &  ---  &  ---  &  ---  & 42.9 & 80 & 8000 &  ---  &  ---  &  ---  \cr
% 110 &  34 & 1.235 &  ---  &  ---  &  ---  &  ---  &  ---  &  ---  &  ---  &  ---  &  ---  & 42.8 & 10 & 4400 & 41.2$^l$ & 30$^l$ & 4000$^f$ \cr
 110 &  35 & 0.582 &  ---  &  ---  &  ---  & 42.0$^u$ & 50$^u$ & 1500$^u$ & 42.2 & 90 & 400 & 42.1$^b$ & 30$^b$ & 1900$^b$ &  ---  &  ---  &  ---  \cr
     &     &       &       &       &       &       &       &       &       &       &       & 42.4$^v$ & 70$^v$ & 8500$^v$ \cr
 110 &  50 & 1.335 &  ---  &  ---  &  ---  &  ---  &  ---  &  ---  &  ---  &  ---  &  ---  & 42.8 & 20 & 4100 & 43.0 & 20 & 4700  \cr
 122 &   1 & 1.134 &  ---  &  ---  &  ---  &  ---  &  ---  &  ---  &  ---  &  ---  &  ---  & 43.0$^b$ & 10$^b$ & 1100$^b$ & 43.2 & 10 & 7100 \cr
     &     &       &       &       &       &       &       &       &       &       &       & 43.8$^v$ & 40$^v$ & 9200$^v$ \cr
 122 &  13 & 0.358 &  ---  &  ---  &  ---  & 41.4 & 10 & 3600$^u$ & 42.0 & 50 & 800 &  ---  &  ---  &  ---  &  ---  &  ---  &  ---  \cr
 122 &  14 & 0.380 &  ---  &  ---  &  ---  & 42.3$^b$ & 60$^b$ & 3100$^b$ & 41.8 & 20 & 500 & 42.3 & 20 & 2500 &  ---  &  ---  &  ---  \cr
     &     &       &       &       &       & 41.8$^n$ & 10$^n$ & 1200$^n$ \cr
 122 &  21 & 0.376 &  ---  &  ---  &  ---  & 42.1 & 30 & 1600$^u$ & 42.2 & 50 & 600 & 42.0$^l$ & 60$^l$ & 2000$^f$ &  ---  &  ---  &  ---  \cr
 123 &   1 & 0.281 &  ---  &  ---  &  ---  & 42.3 & 70 & 3000 & 41.9 & 30 & 400 &  ---  &  ---  &  ---  &  ---  &  ---  &  ---  \cr
 123 &  27 & 0.351 &  ---  &  ---  &  ---  & 41.5$^b$ & 30$^b$ & 2400$^{b,u}$ & 41.6 & 40 & 400 &  ---  &  ---  &  ---  &  ---  &  ---  &  ---  \cr
     &     &       &       &       &       & 40.8$^n$ & 10$^n$ & 400$^n$ & --- & --- & --- &  ---  &  ---  &  ---  &  ---  &  ---  &  ---  \cr
 123 &  28 & 0.212 & 41.8  & 40    & 2200  & 41.1  & 10    & 5100$^u$ & 41.1 & 10 & 500 &  ---  &  ---  &  ---  &  ---  &  ---  &  ---  \cr
 123 &  41 & 1.821 &  ---  &  ---  &  ---  &  ---  &  ---  &  ---  &  ---  &  ---  &  ---  & 43.4 & 10 & 2600 & 43.8 & 20 & 5200 \cr
 123 &  42 & 0.476 &  ---  &  ---  &  ---  & 41.3$^b$ & 20$^b$ & 2200$^{b,u}$ & 41.4 & 20 & 400 & 41.8 & 30 & 3000 &  ---  &  ---  &  ---  \cr
     &     &       &       &       &       & 40.7$^n$ & 10$^n$ & 300$^n$ \cr
 123 &  46 & 1.288 &  ---  &  ---  &  ---  &  ---  &  ---  &  ---  &  ---  &  ---  &  ---  & 43.4 & 40 & 5700 & 41.9$^l$ & 70$^l$ & 6000$^f$ \cr
 123 &  66 & 0.494 &  ---  &  ---  &  ---  & 42.2 & 50 & 3800$^u$ & 42.4 & 80 & 500 & 42.3 & 40 & 8000 &  ---  &  ---  &  ---  \cr
 123 &  85 & 0.652 &  ---  &  ---  &  ---  & 42.3 & 20 & 1100$^u$ & 41.7 & 10 & 300 & 42.4 & 10 & 1700 &  ---  &  ---  &  ---  \cr
 125 &  14 & 1.833 &  ---  &  ---  &  ---  &  ---  &  ---  &  ---  &  ---  &  ---  &  ---  & 43.8 & 30 & 4200 & 43.6 & 20 & 6400 \cr
 125 &  17 & 0.449 &  ---  &  ---  &  ---  & 42.1$^b$ & 60$^b$ & 2500$^b$ & 41.8 & 30 & 400 & 42.1 & 20 & 2500 &  ---  &  ---  &  ---  \cr
     &     &       &       &       &       & 41.3$^n$ & 10$^n$ & 400$^n$ \cr
 126 &   1 & 0.028 & 41.0$^b$ & 60$^b$ & 4800$^b$ & 40.3 & 10 & 4800$^u$ & 40.3 & 10 & 400 &  ---  &  ---  &  ---  &  ---  &  ---  &  ---  \cr
     &     &       & 40.4$^n$ & 10$^n$ & 800$^n$ \cr
% 126 &  27 & 3.305 &  ---  &  ---  &  ---  &  ---  &  ---  &  ---  &  ---  &  ---  &  ---  &  ---  &  ---  &  ---  &  ---  &  ---  &  ---  \cr
% 133 &  17 & 2.390 &  ---  &  ---  &  ---  &  ---  &  ---  &  ---  &  ---  &  ---  &  ---  &  ---  &  ---  &  ---  &  ---  &  ---  &  ---  \cr
 133 &  22 & 1.788 &  ---  &  ---  &  ---  &  ---  &  ---  &  ---  &  ---  &  ---  &  ---  & 44.3 & 20 & 4300 & 44.5 & 20 & 5800 \cr
 205 &   1 & 0.717 &  ---  &  ---  &  ---  &  ---  &  ---  &  ---  &  ---  &  ---  &  ---  & 42.5 & 30 & 2400 &  ---  &  ---  &  ---  \cr
 205 &   1 & 1.334 &  ---  &  ---  &  ---  &  ---  &  ---  &  ---  &  ---  &  ---  &  ---  & 43.0 & 20 & 3500 & --- &  ---  &  ---  \cr
 205 &  22 & 0.445 &  ---  &  ---  &  ---  & 42.6 & 80 & 5400 & 41.6 & 10 & 800 & 42.4 & 20 & 4000 &  ---  &  ---  &  ---  \cr
 205 &  23 & 0.618 &  ---  &  ---  &  ---  & 41.9$^u$ & 30$^u$ & 1100$^u$ & 42.0 & 30 & 700 & 41.9$^l$ & 40$^l$ & 2000$^f$ &  ---  &  ---  &  ---  \cr
 205 &  34 & 0.755 &  ---  &  ---  &  ---  &  ---  &  ---  &  ---  &  ---  &  ---  &  ---  & 42.3 & 50 & 5700 &  ---  &  ---  &  ---  \cr
 206 &   6 & 0.692 &  ---  &  ---  &  ---  & 42.3 & 30 & 2400$^u$ &  ---  &  ---  &  ---  & 42.7 & 40 & 3500 &  ---  &  ---  &  ---  \cr
 206 &   9 & 0.801 &  ---  &  ---  &  ---  &  ---  &  ---  &  ---  &  ---  &  ---  &  ---  & 42.4 & 30 & 2700 &  ---  &  ---  &  ---  \cr
 206 & 507 & 0.484 &  ---  &  ---  &  ---  & 41.1 & 10 & 2400$^u$ & 41.4 & 20 & 400 & 41.9 & 200 & 3800 &  ---  &  ---  &  ---  \cr
 206 & 522 & 0.740 &  ---  &  ---  &  ---  &  ---  &  ---  &  ---  &  ---  &  ---  &  ---  & 42.3 & 10 & 2300 &  ---  &  ---  &  ---  \cr
 208 &   2 & 0.387 &  ---  &  ---  &  ---  & 41.9 & 20 & 3900$^u$ & 41.8 & 20 & 600 & 42.2 & 30 & 2700 &  ---  &  ---  &  ---  \cr
 208 &  18 & 0.470 & 42.7 & 110 & 2100 & 42.0$^b$ & 20$^b$ & 3100$^b$ & 41.9 & 20 & 1500 & 42.6$^b$ & 30$^b$ & 2600$^b$ &  ---  &  ---  &  ---  \cr
     &     &       &       &    &      & 41.7$^n$ & 10$^n$ & 1100$^n$ &      &    &      & 43.0$^v$ & 70$^v$ & 16700$^v$ \cr
 208 &  55 & 1.718 &  ---  &  ---  &  ---  &  ---  &  ---  &  ---  &  ---  &  ---  &  ---  & 44.0 & 60 & 9800 & 43.7 & 30 & 7700 \cr
 211 &  30 & 1.419 &  ---  &  ---  &  ---  &  ---  &  ---  &  ---  &  ---  &  ---  &  ---  & 43.3 & 10 & 2000 & 44.0 & 20 & 8400 \cr
 211 &  35 & 0.465 &  ---  &  ---  &  ---  & 42.0$^l$ & 60$^l$ & 6000$^f$ & 41.2 & 10 & 500 & 42.4 & 90 & 5600 &  ---  &  ---  &  ---  \cr
 211 &  42 & 0.232 & 42.6$^b$ & 110$^b$ & 2200$^b$ & 42.0 & 30 & 2400 & 41.9 & 20 & 1300 &  ---  &  ---  &  ---  &  ---  &  ---  &  ---  \cr
     &     &       & 41.6$^n$ & 10$^n$ & 400$^n$ \cr
 212 &   6 & 1.000 &  ---  &  ---  &  ---  &  ---  &  ---  &  ---  &  ---  &  ---  &  ---  & 43.2 & 30 & 3700 &  ---  &  ---  &  ---  \cr
 212 &  16 & 0.842 &  ---  &  ---  &  ---  &  ---  &  ---  &  ---  &  ---  &  ---  &  ---  & 42.8 & 50 & 9100 &  ---  &  ---  &  ---  \cr
 212 &  25 & 0.802 &  ---  &  ---  &  ---  &  ---  &  ---  &  ---  &  ---  &  ---  &  ---  & 42.9 & 40 & 6700 &  ---  &  ---  &  ---  \cr
 212 &  32 & 0.927 &  ---  &  ---  &  ---  &  ---  &  ---  &  ---  &  ---  &  ---  &  ---  & 43.1 & 20 & 6200 &  ---  &  ---  &  ---  \cr
 212 &  32 & 0.924 &  ---  &  ---  &  ---  &  ---  &  ---  &  ---  &  ---  &  ---  &  ---  & 43.3 & 30 & 5100 &  ---  &  ---  &  ---  \cr
 213 &   7 & 0.542 &  ---  &  ---  &  ---  & 41.4 & 10 & 3900$^u$ & 41.9 & 40 & 600 & 42.3 & 50 & 3300 &  ---  &  ---  &  ---  \cr
 213 &  11 & 1.546 &  ---  &  ---  &  ---  &  ---  &  ---  &  ---  &  ---  &  ---  &  ---  & 43.2 & 60 & 7400 & 43.2 & 30 & 9600 \cr
 213 &  17 & 0.438 &  ---  &  ---  &  ---  & 41.8$^b$ & 30$^b$ & 4500$^b$ & 41.7 & 30 & 900 & 42.5 & 60 & 4100 &  ---  &  ---  &  ---  \cr
     &     &       &       &       &       & 40.7$^n$ & 10$^n$ & 300$^n$ \cr
 213 &  19 & 0.467 &  ---  &  ---  &  ---  & 42.1 & 40 & 3000 & 41.4 & 10 & 600 & 42.0 & 20 & 2000 &  ---  &  ---  &  ---  \cr
 213 &  20 & 0.664 &  ---  &  ---  &  ---  & 43.1 & 60 & 4700 & 42.3 & 10 & 400 & 42.9 & 20 & 2500 &  ---  &  ---  &  ---  \cr
 215 &   1 & 2.249 &  ---  &  ---  &  ---  &  ---  &  ---  &  ---  &  ---  &  ---  &  ---  & 44.2 & 10 & 2900 & 44.4 & 10 & 5000 \cr
 215 &  19 & 0.584 &  ---  &  ---  &  ---  & 42.7 & 40 & 2300$^u$ & 43.2 & 110 & 100 & 42.4 & 10 & 1600 &  ---  &  ---  &  ---  \cr
}                                                                                                                               
\endtable

\begintable*{2}
\caption{{\bf Table 1 } (continued): Optical and UV emission line parameters.}
\halign{#\hfil                                                   % FID
       &\hskip0.1truecm#\hfil                                    % Snum
        &\hskip0.1truecm#\hfil\hskip0.2truecm                    % z
        &\hskip0.2truecm#\hfil\hskip0.05truecm
       &\hskip0.05truecm#\hfil\hskip0.05truecm
       &\hskip0.05truecm#\hfil\hskip0.2truecm
        &\hskip0.2truecm#\hfil\hskip0.05truecm
       &\hskip0.05truecm#\hfil\hskip0.05truecm
       &\hskip0.05truecm#\hfil\hskip0.2truecm
        &\hskip0.2truecm#\hfil\hskip0.05truecm
       &\hskip0.05truecm#\hfil\hskip0.05truecm
       &\hskip0.05truecm#\hfil\hskip0.2truecm
        &\hskip0.2truecm#\hfil\hskip0.05truecm
       &\hskip0.05truecm#\hfil\hskip0.05truecm
       &\hskip0.05truecm#\hfil\hskip0.2truecm
        &\hskip0.2truecm#\hfil\hskip0.05truecm
       &\hskip0.05truecm#\hfil\hskip0.05truecm
       &\hskip0.05truecm#\hfil\hskip0.2truecm
\cr
\noalign{\bigskip}
    FID   &
    SNo   &
    \hfil{\sl z} \hfil   &
    \multispan3{\hfil \Ha$\lambda$6562 \hfil}  &
    \multispan3{\hfil \Hb$\lambda$4861 \hfil}  &
    \multispan3{\hfil \oiiifull \hfil}  &
    \multispan3{\hfil \mgiifull \hfil}  &
    \multispan3{\hfil \ciiifull \hfil}  
 \cr
    & & 
    & \hfil{\sl lum} \hfil & \hfil{\sl ew} \hfil & \hfil{\sl fwhm} \hfil 
    & \hfil{\sl lum} \hfil & \hfil{\sl ew} \hfil & \hfil{\sl fwhm} \hfil 
    & \hfil{\sl lum} \hfil & \hfil{\sl ew} \hfil & \hfil{\sl fwhm} \hfil 
    & \hfil{\sl lum} \hfil & \hfil{\sl ew} \hfil & \hfil{\sl fwhm} \hfil 
    & \hfil{\sl lum} \hfil & \hfil{\sl ew} \hfil & \hfil{\sl fwhm} \hfil 
\cr
\noalign{\medskip}
 215 &  32 & 0.612 &  ---  &  ---  &  ---  & 42.8$^u$ & 90$^u$ & 3700$^u$ & 41.6 & 10 & 600 & 42.5 & 20 & 2700 &  ---  &  ---  &  ---  \cr
 216 &   7 & 0.804 &  ---  &  ---  &  ---  &  ---  &  ---  &  ---  &  ---  &  ---  &  ---  & 43.2 & 50 & 5500 &  ---  &  ---  &  ---  \cr
 216 &  30 & 0.941 &  ---  &  ---  &  ---  &  ---  &  ---  &  ---  &  ---  &  ---  &  ---  & 43.0$^u$ & 210$^u$ & 7400$^u$ &  ---  &  ---  &  ---  \cr
 216 &  33 & 0.792 &  ---  &  ---  &  ---  & 42.8$^l$ & 90$^l$ & 5000$^f$ & 42.4$^l$ & 30$^l$ & 1000$^f$ & 42.5$^b$ & 10$^b$ & 3200$^b$ &  ---  &  ---  &  ---  \cr
     &     &       &       &       &       &          &        &          &          &        &          & 42.6$^v$ & 20$^v$ & 6500$^v$ &  ---  &  ---  &  ---  \cr
 217 &   3 & 0.989 &  ---  &  ---  &  ---  &  ---  &  ---  &  ---  &  ---  &  ---  &  ---  & 43.3 & 40 & 4600 &  ---  &  ---  &  ---  \cr
 217 &  21 & 0.561 &  ---  &  ---  &  ---  & 42.2$^l$ & 20$^l$ & 5000$^f$ & 41.4 & 10 & 700 & 42.0 & 30 & 4900 &  ---  &  ---  &  ---  \cr
 217 &  34 & 1.199 &  ---  &  ---  &  ---  &  ---  &  ---  &  ---  &  ---  &  ---  &  ---  & 43.2 & 20 & 3900 & 43.7$^u$ & 420$^u$ & 5500$^u$ \cr
 217 &  35 & 0.435 &  ---  &  ---  &  ---  & 41.4 & 10 & 3300 & 41.9 & 30 & 500 & 42.4 & 60 & 4600 &  ---  &  ---  &  ---  \cr
 217 &  59 & 0.587 &  ---  &  ---  &  ---  & 42.3$^l$ & 100$^l$ & 9000$^f$ & 41.5$^l$ & 20$^l$ & 1000$^f$ & 42.4 & 90 & 9300 &  ---  &  ---  &  ---  \cr
 218 &   1 & 0.545 &  ---  &  ---  &  ---  & 43.0$^b$ & 60$^b$ & 7400$^b$ & 42.6 & 30 & 500 & 43.2 & 40 & 5000 &  ---  &  ---  &  ---  \cr
     &     &       &       &       &       & 41.6$^n$ & 10$^n$ & 400$^n$ \cr
 218 &   9 & 0.703 &  ---  &  ---  &  ---  &  ---  &  ---  &  ---  &  ---  &  ---  &  ---  & 42.6 & 30 & 4100 &  ---  &  ---  &  ---  \cr
% 218 &  13 & 1.450 &  ---  &  ---  &  ---  &  ---  &  ---  &  ---  &  ---  &  ---  &  ---  &  ---  &  ---  &  ---  &  ---  &  ---  &  ---  \cr
 218 &  14 & 0.224 & 41.7$^b$ & 100$^b$ & 2700$^b$ & 40.8 & 10 & 1700 & 41.2 & 30 & 400 & --- & --- & --- & --- & --- & ---  \cr
     &     &       & 41.0$^n$ & 20$^n$  & 600$^n$  \cr
 218 &  21 & 0.79$^u$ &  ---  &  ---  &  ---  &  ---  &  ---  &  ---  &  ---  &  ---  &  ---  & 41.9$^u$ & 20$^u$ & 4700$^u$ &  ---  &  ---  &  ---  \cr
 218 &  27 & 0.629 &  ---  &  ---  &  ---  & 43.0 & 50 & 2400 & 42.4 & 10 & 900 & 43.1 & 30 & 3300 &  ---  &  ---  &  ---  \cr
 219 &  15 & 1.186 &  ---  &  ---  &  ---  &  ---  &  ---  &  ---  &  ---  &  ---  &  ---  & 43.0 & 20 & 2600 & 43.1 & 30 & 4900 \cr
 219 &  45 & 1.272 &  ---  &  ---  &  ---  &  ---  &  ---  &  ---  &  ---  &  ---  &  ---  & 43.4 & 70 & 6600 & 41.9$^l$ & 770$^l$ & 6000$^f$ \cr
 219 &  48 & 1.373 &  ---  &  ---  &  ---  &  ---  &  ---  &  ---  &  ---  &  ---  &  ---  & 43.0 & 30 & 9000 & 43.0 & 20 & 6600 \cr
 220 &  13 & 0.968 &  ---  &  ---  &  ---  &  ---  &  ---  &  ---  &  ---  &  ---  &  ---  & 42.6 & 10 & 3600 &  ---  &  ---  &  ---  \cr
 220 &  18 & 0.442 &  ---  &  ---  &  ---  & 41.4$^u$ & 20$^u$ & 1800$^u$ & 41.9 & 50 & 600 & 42.4 & 90 & 10100 &  ---  &  ---  &  ---  \cr
 220 &  23 & 0.193 & 42.5$^u$ & 70$^u$ & 2000$^u$ & 41.5$^u$ & 10$^u$ & 1200$^u$ & 41.5 & 10 & 500 &  ---  &  ---  &  ---  &  ---  &  ---  &  ---  \cr
 221 &   2 & 0.900 &  ---  &  ---  &  ---  & 42.5$^u$ & 20$^u$ & 1900$^u$ & 42.8 & 40 & 400 & 42.9 & 20 & 2300 &  ---  &  ---  &  ---  \cr
 221 &   7 & 0.292 &  ---  &  ---  &  ---  & 42.2 & 30 & 2300 & 42.1 & 20 & 800 &  ---  &  ---  &  ---  &  ---  &  ---  &  ---  \cr
 221 &  16 & 0.184 & 41.8 & 80 & 1300 & 41.2 & 20 & 1700 & 41.0 & 10 & 400 &  ---  &  ---  &  ---  &  ---  &  ---  &  ---  \cr
 221 &  35$^{r1}$ & 0.451 & 44.1 & 460 & 2800 & 43.5 & 130 & 4600 & 43.4 & 110 & 800 & 43.5 & 50 & 2400 &  ---  &  ---  &  ---  \cr
 223 &  17 & 0.288 &  ---  &  ---  &  ---  & 42.1$^b$ & 30$^b$ & 4000$^b$ & 41.7 & 10 & 500 &  ---  &  ---  &  ---  &  ---  &  ---  &  ---  \cr
     &     &       &       &       &       & 41.6$^n$ & 10$^n$ & 900$^n$ \cr
 224 & 201 & 1.547 &  ---  &  ---  &  ---  &  ---  &  ---  &  ---  &  ---  &  ---  &  ---  & 43.7 & 50 & 8100 & 43.7 & 30 & 7400 \cr
 225 &   1 & 0.487 &  ---  &  ---  &  ---  & 42.7 & 50 & 2900$^u$ & 42.5 & 70 & 1000 & 43.0$^l$ & 120$^l$ & 3000$^f$ &  ---  &  ---  &  ---  \cr
 226 &  41 & 1.316 &  ---  &  ---  &  ---  &  ---  &  ---  &  ---  &  ---  &  ---  &  ---  & 43.1 & 10 & 6400 & 43.8 & 30 & 5200 \cr
 226 & 114 & 1.021 &  ---  &  ---  &  ---  &  ---  &  ---  &  ---  &  ---  &  ---  &  ---  & 42.5 & 70 & 2300 &  ---  &  ---  &  ---  \cr
 227 &  19 & 1.859 &  ---  &  ---  &  ---  &  ---  &  ---  &  ---  &  ---  &  ---  &  ---  & 43.4 & 10 & 1700 & 43.9 & 20 & 6800 \cr
 227 &  37 & 1.417 &  ---  &  ---  &  ---  &  ---  &  ---  &  ---  &  ---  &  ---  &  ---  & 44.3 & 60 & 6700 & 44.2 & 50 & 9000 \cr
 227 & 513 & 0.958 &  ---  &  ---  &  ---  &  ---  &  ---  &  ---  &  ---  &  ---  &  ---  & 43.2 & 30 & 3600 &  ---  &  ---  &  ---  \cr
 228 &   1 & 1.722 &  ---  &  ---  &  ---  &  ---  &  ---  &  ---  &  ---  &  ---  &  ---  & 44.4 & 40 & 11300 & 43.9 & 10 & 2400 \cr
 232 &  16 & 0.227 & 42.0$^b$ & 100$^b$ & 2400$^b$ & 41.5 & 30 & 2100 & 41.2 & 10 & 400 &  ---  &  ---  &  ---  &  ---  &  ---  &  ---  \cr
     &     &       & 40.7$^n$ & 10$^n$ & 200$^n$ \cr
 232 & 301 & 0.384 & 42.9 & 120 & 2600 & 42.3 & 40 & 3300 & 42.3 & 40 & 900 &  ---  &  ---  &  ---  &  ---  &  ---  &  ---  \cr
 234 &   1 & 1.651 &  ---  &  ---  &  ---  &  ---  &  ---  &  ---  &  ---  &  ---  &  ---  & 44.2 & 30 & 6000 & 43.3 & 0 & 11800 \cr
 234 &  33 & 1.026 &  ---  &  ---  &  ---  &  ---  &  ---  &  ---  &  ---  &  ---  &  ---  & 43.0 & 20 & 7600 & 43.0 & 10 & 6300 \cr
 236 &   5 & 0.473 & 42.6 & 60 & 1300 & 42.2 & 40 & 2400$^u$ & 41.7 & 10 & 500 & 42.4 & 30 & 3100 &  ---  &  ---  &  ---  \cr
 236 &  21 & 1.128 &  ---  &  ---  &  ---  &  ---  &  ---  &  ---  &  ---  &  ---  &  ---  & 43.1 & 40 & 2800 & 43.5 & 130 & 8300 \cr
 236 &  22 & 0.048 & 41.5 & 70 & 3400 & 40.8 & 20 & 2700$^u$ & 40.5 & 10 & 600 &  ---  &  ---  &  ---  &  ---  &  ---  &  ---  \cr
 238 &  11 & 0.325 &  ---  &  ---  &  ---  & 42.1 & 80 & 6300$^u$ & 41.3 & 20 & 500 &  ---  &  ---  &  ---  &  ---  &  ---  &  ---  \cr
 240 &  15 & 1.277 &  ---  &  ---  &  ---  &  ---  &  ---  &  ---  &  ---  &  ---  &  ---  & 42.7 & 10 & 2100 & 42.8 & 10 & 6300 \cr
 240 &  82 & 0.519 &  ---  &  ---  &  ---  & 42.1 & 30 & 7400 &  ---  &  ---  &  ---  & 42.6 & 30 & 8600 &  ---  &  ---  &  ---  \cr
 245 &   4 & 0.711 &  ---  &  ---  &  ---  &  ---  &  ---  &  ---  &  ---  &  ---  &  ---  & 42.7 & 100 & 9300 &  ---  &  ---  &  ---  \cr
 248 &   2 & 0.274 & 42.0 & 180 & 2300 & 41.6$^l$ & 160$^l$ & 2000$^f$ & 41.3 & 60 & 800 &  ---  &  ---  &  ---  &  ---  &  ---  &  ---  \cr
 248 &  51$^{r2}$ & 0.242 & 42.7 & 170 & 3000 & 42.4 & 80 & 3200 & 41.8 & 20 & 500 &  ---  &  ---  &  ---  &  ---  &  ---  &  ---  \cr
 252 &   1 & 0.219 &  ---  &  ---  &  ---  & 42.3$^l$ & 90$^l$ & 2000$^f$ & 42.0 & 30 & 700 &  ---  &  ---  &  ---  &  ---  &  ---  &  ---  \cr
 252 &   9 & 0.673 &  ---  &  ---  &  ---  & 42.1 & 30 & 1900$^u$ &  ---  &  ---  &  ---  & 42.8 & 40 & 5800 &  ---  &  ---  &  ---  \cr
 252 &  31* & 1.413 &  ---  &  ---  &  ---  &  ---  &  ---  &  ---  &  ---  &  ---  &  ---  & 44.1$^u$ & 10$^u$ & 2800$^u$ & 44.1 & 10 & 5900 \cr
 252 &  31* & 1.415 &  ---  &  ---  &  ---  &  ---  &  ---  &  ---  &  ---  &  ---  &  ---  & 44.1$^u$ & 10$^u$ & 2800$^u$ & 44.1 & 10 & 5900 \cr
 252 &  34 & 0.679 &  ---  &  ---  &  ---  &  ---  &  ---  &  ---  &  ---  &  ---  &  ---  & 42.3 & 50 & 5500 &  ---  &  ---  &  ---  \cr
 252 &  36 & 1.038 &  ---  &  ---  &  ---  &  ---  &  ---  &  ---  &  ---  &  ---  &  ---  & 43.2 & 90 & 6000 & 42.9 & 40 & 4500 \cr
 252 &  38 & 0.216 & 42.3$^b$ & 230$^b$ & 1600$^b$ & 41.6$^b$ & 40$^b$ & 1300$^{b,u}$ & 41.9 & 80 & 400 &  ---  &  ---  &  ---  &  ---  &  ---  &  ---  \cr
     &     &       & 41.7$^n$ & 50$^n$ & 300$^n$ & 40.9$^n$ & 10$^n$ & 300$^n$ & 41.9 & 80 & 400 &  ---  &  ---  &  ---  &  ---  &  ---  &  ---  \cr
 252 &  46$^{r3}$ & 2.091 &  ---  &  ---  &  ---  &  ---  &  ---  &  ---  &  ---  &  ---  &  ---  &  ---  &  60  & 15300  &  ---  & 80 & 6300  \cr
}
\endtable

\begintable*{3} 
\caption{{\bf Table 1 } (continued): Optical and UV emission line parameters.} 
\halign{#\hfil                                                   % FID
       &\hskip0.1truecm#\hfil                                    % Snum
        &\hskip0.1truecm#\hfil\hskip0.2truecm                    % z
        &\hskip0.2truecm#\hfil\hskip0.05truecm
       &\hskip0.05truecm#\hfil\hskip0.05truecm
       &\hskip0.05truecm#\hfil\hskip0.2truecm
        &\hskip0.2truecm#\hfil\hskip0.05truecm
       &\hskip0.05truecm#\hfil\hskip0.05truecm
       &\hskip0.05truecm#\hfil\hskip0.2truecm
        &\hskip0.2truecm#\hfil\hskip0.05truecm
       &\hskip0.05truecm#\hfil\hskip0.05truecm
       &\hskip0.05truecm#\hfil\hskip0.2truecm
        &\hskip0.2truecm#\hfil\hskip0.05truecm
       &\hskip0.05truecm#\hfil\hskip0.05truecm
       &\hskip0.05truecm#\hfil\hskip0.2truecm
        &\hskip0.2truecm#\hfil\hskip0.05truecm
       &\hskip0.05truecm#\hfil\hskip0.05truecm
       &\hskip0.05truecm#\hfil\hskip0.2truecm
\cr
\noalign{\bigskip}
    FID   &
    SNo   &
    \hfil{\sl z} \hfil   &
    \multispan3{\hfil \Ha$\lambda$6562 \hfil}  &
    \multispan3{\hfil \Hb$\lambda$4861 \hfil}  &
    \multispan3{\hfil \oiiifull \hfil}  &
    \multispan3{\hfil \mgiifull \hfil}  &
    \multispan3{\hfil \ciiifull \hfil}  
 \cr
    & & 
    & \hfil{\sl lum} \hfil & \hfil{\sl ew} \hfil & \hfil{\sl fwhm} \hfil 
    & \hfil{\sl lum} \hfil & \hfil{\sl ew} \hfil & \hfil{\sl fwhm} \hfil 
    & \hfil{\sl lum} \hfil & \hfil{\sl ew} \hfil & \hfil{\sl fwhm} \hfil 
    & \hfil{\sl lum} \hfil & \hfil{\sl ew} \hfil & \hfil{\sl fwhm} \hfil 
    & \hfil{\sl lum} \hfil & \hfil{\sl ew} \hfil & \hfil{\sl fwhm} \hfil 
\cr
\noalign{\medskip}
 253 &   5 & 1.204 &  ---  &  ---  &  ---  &  ---  &  ---  &  ---  &  ---  &  ---  &  ---  & 43.1 & 40 & 6300 & 43.2 & 30 & 11600 \cr
 254 &  10 & 0.935 &  ---  &  ---  &  ---  & 44.3 & 40 & 2600 & 44.2 & 30 & 1100 & 44.0 & 10 & 2700 & 44.2 & 10 & 3400 \cr
 254 &  11 & 1.166 &  ---  &  ---  &  ---  &  ---  &  ---  &  ---  &  ---  &  ---  &  ---  & 44.1 & 20 & 4100 & 44.2 & 20 & 9700 \cr
 254 &  41 & 0.486 &  ---  &  ---  &  ---  & 42.4 & 50 & 6900 & 41.6 & 10 & 700 & 42.7 & 40 & 8000 &  ---  &  ---  &  ---  \cr
 255 &   7 & 0.260 & 41.8 & 70 & 2100 & 41.0$^l$ & 20$^l$ & 4000$^f$ & 40.9 & 10 & 500 &  ---  &  ---  &  ---  &  ---  &  ---  &  ---  \cr
 255 &  13 & 0.581 &  ---  &  ---  &  ---  & 42.4$^l$ & 120$^l$ & 4000$^f$ & 42.0$^l$ & 50$^l$ & 1000$^f$ & 42.5 & 40 & 3900 &  ---  &  ---  &  ---  \cr
 255 &  19 & 0.862 &  ---  &  ---  &  ---  &  ---  &  ---  &  ---  &  ---  &  ---  &  ---  & 42.9 & 30 & 5000 &  ---  &  ---  &  ---  \cr
 255 &  23 & 0.759 &  ---  &  ---  &  ---  &  ---  &  ---  &  ---  &  ---  &  ---  &  ---  & 42.8$^u$ & 80$^u$ & 14100$^u$ &  ---  &  ---  &  ---  \cr
 257 &   1 & 1.029 &  ---  &  ---  &  ---  &  ---  &  ---  &  ---  &  ---  &  ---  &  ---  & 43.7$^u$ & 10$^u$ & 14100$^u$ & 43.6$^u$ & 10$^u$ & 8800$^u$ \cr
 257 &  14 & 1.096 &  ---  &  ---  &  ---  &  ---  &  ---  &  ---  &  ---  &  ---  &  ---  & 42.3 & 40 & 2300 & 42.6$^u$ & 850$^u$ & 3400$^u$ \cr
 257 &  20 & 1.302 &  ---  &  ---  &  ---  &  ---  &  ---  &  ---  &  ---  &  ---  &  ---  & 43.5 & 20 & 5700 & 43.6$^u$ & 20$^u$ & 10900$^u$ \cr
 257 &  37 & 0.329 &  ---  &  ---  &  ---  & 40.7$^u$ & 20$^u$ & 300$^u$ & 40.9 & 40 & 400 &  ---  &  ---  &  ---  &  ---  &  ---  &  ---  \cr
 257 &  38 & 1.260 &  ---  &  ---  &  ---  &  ---  &  ---  &  ---  &  ---  &  ---  &  ---  & 43.8 & 20 & 4400 & 43.9 & 10 & 10300 \cr
 258 &   1 & 0.698 &  ---  &  ---  &  ---  & 41.9 & 10 & 600$^u$ &  ---  &  ---  &  ---  & 42.3$^u$ & 40$^u$ & 3300$^u$ &  ---  &  ---  &  ---  \cr
 258 &   5 & 0.811 &  ---  &  ---  &  ---  &  ---  &  ---  &  ---  &  ---  &  ---  &  ---  & 42.6 & 20 & 2200 &  ---  &  ---  &  ---  \cr
 258 &  30 & 0.847 &  ---  &  ---  &  ---  &  ---  &  ---  &  ---  &  ---  &  ---  &  ---  & 42.5 & 30 & 5000 &  ---  &  ---  &  ---  \cr
 258 &  32 & 1.615 &  ---  &  ---  &  ---  &  ---  &  ---  &  ---  &  ---  &  ---  &  ---  & 43.3$^u$ & 20$^u$ & 5500$^u$ & 42.7$^b$ & 10$^b$ & 1200$^b$ \cr
     &     &       &       &       &       &       &       &       &       &       &       &          &        &          & 43.2$^v$ & 10$^v$ & 10200$^v$ \cr
 259 &   5 & 0.984 &  ---  &  ---  &  ---  &  ---  &  ---  &  ---  &  ---  &  ---  &  ---  & 43.1 & 40 & 8400 & 42.8 & 10 & 5300 \cr
 259 &   7 & 0.408 &  ---  &  ---  &  ---  & 41.9 & 40 & 1900 & 41.5 & 20 & 600 & 41.9 & 20 & 3200 &  ---  &  ---  &  ---  \cr
 259 &  11 & 0.995 &  ---  &  ---  &  ---  &  ---  &  ---  &  ---  &  ---  &  ---  &  ---  & 42.9 & 50 & 5800 & 42.2$^u$ & 10$^u$ & 5500$^u$ \cr
% 259 &  30 & 1.940 &  ---  &  ---  &  ---  &  ---  &  ---  &  ---  &  ---  &  ---  &  ---  &  ---  &  ---  &  ---  &  ---  &  ---  &  ---  \cr
 260 &   8 & 1.822 &  ---  &  ---  &  ---  &  ---  &  ---  &  ---  &  ---  &  ---  &  ---  & 44.3 & 10 & 7400 & 44.5 & 20 & 4600 \cr
 260 &  44 & 1.502 &  ---  &  ---  &  ---  &  ---  &  ---  &  ---  &  ---  &  ---  &  ---  & 43.6 & 30 & 3500 & 43.6 & 20 & 4100 \cr
 262 &   1 & 0.882 &  ---  &  ---  &  ---  &  ---  &  ---  &  ---  &  ---  &  ---  &  ---  & 43.0 & 50 & 4800 &  ---  &  ---  &  ---  \cr
% 262 &   2 & 1.202 &  ---  &  ---  &  ---  &  ---  &  ---  &  ---  &  ---  &  ---  &  ---  &  ---  &  ---  &  ---  &  ---  &  ---  &  ---  \cr
% 262 &  10 & 0.336 &  ---  &  ---  &  ---  &  ---  &  ---  &  ---  &  ---  &  ---  &  ---  &  ---  &  ---  &  ---  &  ---  &  ---  &  ---  \cr
 262 &  12 & 0.923 &  ---  &  ---  &  ---  & 43.5$^l$ & 540$^l$ & 5000$^f$ & 43.0$^l$ & 180$^l$ & 1000$^f$ & 43.0$^b$ & 20$^b$ & 3100$^b$ & 43.0 & 30 & 2100 \cr
     &     &       &       &       &       &          &         &          &          &         &          & 42.9$^v$ & 20$^v$ & 8700$^v$ \cr
 262 &  34 & 0.311 & 42.6$^b$ & 250$^b$ & 6300$^b$ & 42.3 & 60 & 4300 & 41.6 & 10 & 700 & 42.4 & 40 & 3900 &  ---  &  ---  &  ---  \cr
     &     &       & 42.6$^n$ & 120$^n$ & 1660$^n$ \cr
 265 &   1 & 2.336 &  ---  &  ---  &  ---  &  ---  &  ---  &  ---  &  ---  &  ---  &  ---  &  ---  &  ---  &  ---  & 42.2$^l$ & 10$^l$ & 9000$^f$ \cr
 265 &  17 & 0.448 &  ---  &  ---  &  ---  & 41.9 & 40 & 3400 & 41.3 & 10 & 400 & 41.8 & 10 & 1900 &  ---  &  ---  &  ---  \cr
 268 &  11 & 1.189 &  ---  &  ---  &  ---  &  ---  &  ---  &  ---  &  ---  &  ---  &  ---  & 42.7 & 50 & 7900 & 42.7 & 60 & 7600 \cr
 268 &  24 & 0.252 & 42.5$^b$ & 80$^b$ & 2100$^b$ & 42.2 & 30 & 3800$^u$ & 41.5 & 10 & 800 &  ---  &  ---  &  ---  &  ---  &  ---  &  ---  \cr
     &     &       & 42.1$^n$ & 30$^n$ & 800$^n$ \cr
 271 &   2 & 0.444 & 43.0 & 240 & 2900 & 41.9 & 10 & 1400$^u$ & 41.5 & 10 & 800 & 42.4 & 40 & 3100 &  ---  &  ---  &  ---  \cr
 271 &   7 & 1.035 &  ---  &  ---  &  ---  &  ---  &  ---  &  ---  &  ---  &  ---  &  ---  & 43.2 & 20 & 4300 & 43.3 & 20 & 7000 \cr
 272 &   8 & 1.817 &  ---  &  ---  &  ---  &  ---  &  ---  &  ---  &  ---  &  ---  &  ---  & 43.3 & 30 & 3500 & 43.1 & 10 & 4600 \cr
 272 &  18 & 0.604 &  ---  &  ---  &  ---  & 42.5$^b$ & 30$^b$ & 4400$^b$ & 42.4 & 20 & 600 & 43.1 & 50 & 6100 &  ---  &  ---  &  ---  \cr
     &     &       &       &       &       & 41.6$^n$ & 10$^n$ & 400$^n$ \cr
 272 &  23 & 0.096 & 41.9 & 130 & 2600 & 41.3 & 50 & 5900$^u$ & 41.3 & 50 & 1400 &  ---  &  ---  &  ---  &  ---  &  ---  &  ---  \cr
 272 &  28 & 0.444 &  ---  &  ---  &  ---  & 40.9$^u$ & 10$^u$ & 600$^u$ & 42.0 & 30 & 600 & 42.4 & 100 & 6500 &  ---  &  ---  &  ---  \cr
 273 &   4 & 1.045 &  ---  &  ---  &  ---  &  ---  &  ---  &  ---  &  ---  &  ---  &  ---  & 43.6 & 20 & 2600 & 43.6 & 10 & 4100 \cr
 273 &   6 & 0.269 & 43.2 & 230 & 9200 & 42.5 & 40 & 11000 & 41.6 & 10 & 1000 &  ---  &  ---  &  ---  &  ---  &  ---  &  ---  \cr
 273 &  18 & 0.361 &  ---  &  ---  &  ---  & 41.7 & 40 & 2600 & 41.9 & 70 & 500 & 41.7 & 20 & 4500 &  ---  &  ---  &  ---  \cr
 273 &  22 & 1.075 &  ---  &  ---  &  ---  &  ---  &  ---  &  ---  &  ---  &  ---  &  ---  & 43.9 & 10 & 3700 & 43.9 & 10 & 7700 \cr
%273 &  23 & 0.433 &  ---  &  ---  &  ---  & 41.9$^l$ & 30$^l$ & 3000$^f$ & 41.4 & 10 & 600 & 42.0 & 40 & 2700 &  ---  &  ---  &  ---  \cr
 274 &   8 & 0.156 & 42.3 & 50 & 1200 & 41.9 & 20 & 2200 & 41.9 & 20 & 900 &  ---  &  ---  &  ---  &  ---  &  ---  &  ---  \cr
 278 &   9 & 0.948 &  ---  &  ---  &  ---  & 43.7 & 50 & 2600 & 42.9 & 10 & 800 & 43.6 & 20 & 3800 & 43.7 & 10 & 5600 \cr
 278 &  10 & 0.091 & 42.9$^b$ & 260$^b$ & 7100$^b$ & 41.6 & 10 & 800$^u$ & 42.6 & 120 & 800 &  ---  &  ---  &  ---  &  ---  &  ---  &  ---  \cr
     &     &       & 42.5$^n$ & 60$^n$ & 1300$^n$ \cr
 281 &  11 & 2.919 &  ---  &  ---  &  ---  &  ---  &  ---  &  ---  &  ---  &  ---  &  ---  &  ---  &  ---  &  ---  & 43.5 & 10 & 4300 \cr
 281 &  21 & 0.347 & 42.7 & 170 & 3200 & 42.4 & 90 & 9100$^u$ & 41.6 & 20 & 700 & 42.3 & 200 & 5300 &  ---  &  ---  &  ---  \cr
 283 &   6 & 1.219 &  ---  &  ---  &  ---  &  ---  &  ---  &  ---  &  ---  &  ---  &  ---  & 44.1 & 30 & 5200 & 43.9 & 20 & 4200 \cr
 283 &  11 & 0.272 & 43.1$^b$ & 180$^b$  & 3500$^b$ & 42.5$^b$ & 30$^b$ & 4900$^b$ & 42.0 & 10 & 1200 &  ---  &  ---  &  ---  &  ---  &  ---  &  ---  \cr
     &     &       & 42.6$^n$ & 60$^n$ & 1700$^n$ & 42.0$^n$ & 10$^n$ & 1600$^n$ \cr
 283 &  14 & 0.284 & 42.5$^b$ & 110$^b$ & 3900$^b$ & 42.3 & 40 & 3300 & 41.8 & 20 & 1600 &  ---  &  ---  &  ---  &  ---  &  ---  &  ---  \cr
     &     &       & 42.4$^n$ & 70$^n$ & 1700$^n$ \cr
 283 &  21 & 0.719 &  ---  &  ---  &  ---  & 43.3$^l$ & 170$^l$ & 6000$^f$ & 42.7 & 50 & 1400 & 43.0 & 40 & 6300 &  ---  &  ---  &  ---  \cr
 286 &   2 & 1.498 &  ---  &  ---  &  ---  &  ---  &  ---  &  ---  &  ---  &  ---  &  ---  & 44.4 & 30 & 9800 & 44.0 & 10 & 11900 \cr
 290 &  21 & 2.573 &  ---  &  ---  &  ---  &  ---  &  ---  &  ---  &  ---  &  ---  &  ---  &  ---  &  ---  &  ---  & 44.5 & 10 & 4800 \cr
}
\endtable

\begintable*{4} 
\caption{{\bf Table 1 } (continued): Optical and UV emission line parameters.} 
\halign{#\hfil                                                   % FID
       &\hskip0.1truecm#\hfil                                    % Snum
        &\hskip0.1truecm#\hfil\hskip0.2truecm                    % z
        &\hskip0.2truecm#\hfil\hskip0.05truecm
       &\hskip0.05truecm#\hfil\hskip0.05truecm
       &\hskip0.05truecm#\hfil\hskip0.2truecm
        &\hskip0.2truecm#\hfil\hskip0.05truecm
       &\hskip0.05truecm#\hfil\hskip0.05truecm
       &\hskip0.05truecm#\hfil\hskip0.2truecm
        &\hskip0.2truecm#\hfil\hskip0.05truecm
       &\hskip0.05truecm#\hfil\hskip0.05truecm
       &\hskip0.05truecm#\hfil\hskip0.2truecm
        &\hskip0.2truecm#\hfil\hskip0.05truecm
       &\hskip0.05truecm#\hfil\hskip0.05truecm
       &\hskip0.05truecm#\hfil\hskip0.2truecm
        &\hskip0.2truecm#\hfil\hskip0.05truecm
       &\hskip0.05truecm#\hfil\hskip0.05truecm
       &\hskip0.05truecm#\hfil\hskip0.2truecm
\cr
\noalign{\bigskip}
    FID   &
    SNo   &
    \hfil{\sl z} \hfil   &
    \multispan3{\hfil \Ha$\lambda$6562 \hfil}  &
    \multispan3{\hfil \Hb$\lambda$4861 \hfil}  &
    \multispan3{\hfil \oiiifull \hfil}  &
    \multispan3{\hfil \mgiifull \hfil}  &
    \multispan3{\hfil \ciiifull \hfil}  
 \cr
    & & 
    & \hfil{\sl lum} \hfil & \hfil{\sl ew} \hfil & \hfil{\sl fwhm} \hfil 
    & \hfil{\sl lum} \hfil & \hfil{\sl ew} \hfil & \hfil{\sl fwhm} \hfil 
    & \hfil{\sl lum} \hfil & \hfil{\sl ew} \hfil & \hfil{\sl fwhm} \hfil 
    & \hfil{\sl lum} \hfil & \hfil{\sl ew} \hfil & \hfil{\sl fwhm} \hfil 
    & \hfil{\sl lum} \hfil & \hfil{\sl ew} \hfil & \hfil{\sl fwhm} \hfil 
\cr
\noalign{\medskip}
 293 &   1 & 0.824 &  ---  &  ---  &  ---  &  ---  &  ---  &  ---  &  ---  &  ---  &  ---  & 43.1 & 50 & 3200 &  ---  &  ---  &  ---  \cr
 293 &   6 & 0.081 & 42.2 & 70 & 1900 & 41.8$^l$ & 20$^l$ & 2000$^f$ & 41.4 & 10 & 1000 &  ---  &  ---  &  ---  &  ---  &  ---  &  ---  \cr
 293 &  10 & 0.758 &  ---  &  ---  &  ---  &  ---  &  ---  &  ---  &  ---  &  ---  &  ---  & 42.7$^b$ & 20$^b$ & 3000$^b$ &  ---  &  ---  &  ---  \cr
     &     &       &       &       &       &       &       &       &       &       &       & 43.1$^v$ & 50$^v$ & 14200$^v$ \cr
 293 &  12 & 0.917 &  ---  &  ---  &  ---  & 42.6$^u$ & 20$^u$ & 1800$^u$ & 42.8$^l$ & 60$^l$ & 1000$^f$ & 42.7 & 20 & 2500 & 42.8 & 10 & 3900 \cr
 293 &  13 & 0.189 & 41.2$^u$ & 80$^u$ & 1700$^u$ & 41.1$^l$ & 70$^l$ & 2000$^f$ & 40.3 & 10 & 200 &  ---  &  ---  &  ---  &  ---  &  ---  &  ---  \cr
 294 &   1 & 0.713 &  ---  &  ---  &  ---  &  ---  &  ---  &  ---  &  ---  &  ---  &  ---  & 42.8 & 50 & 8200 &  ---  &  ---  &  ---  \cr
 302 &  14 & 0.809 &  ---  &  ---  &  ---  & 43.1$^l$ & 120$^l$ & 2000$^f$ & 42.8$^l$ & 60$^l$ & 1000$^f$ & 42.9 & 20 & 2200 &  ---  &  ---  &  ---  \cr
 302 &  18 & 0.924 &  ---  &  ---  &  ---  &  ---  &  ---  &  ---  &  ---  &  ---  &  ---  & 43.7 & 40 & 3300 &  ---  &  ---  &  ---  \cr
 305 &  11 & 0.251 & 42.0 & 90 & 3500 & 41.7 & 30 & 5500$^u$ & 41.3 & 10 & 700 & 42.4$^u$ & 170$^u$ & 5900$^u$ &  ---  &  ---  &  ---  \cr
 305 &  18 & 0.387 & 42.3$^b$ & 110$^b$ & 3800$^b$ & 41.9 & 30 & 2900$^u$ & 41.7 & 20 & 600 & 42.1 & 30 & 3100 &  ---  &  ---  &  ---  \cr
     &     &       & 41.8$^n$ & 30$^n$ & 500$^n$ \cr
 305 &  34 & 0.854 &  ---  &  ---  &  ---  & 42.4$^u$ & 30$^u$ & 1400$^u$ & 42.4 & 50 & 500 & 42.6 & 30 & 2100 &  ---  &  ---  &  ---  \cr
\noalign{\bigskip}
\multispan6{Sample statistics\hfil}\cr
       &
       &
    \hfil{\sl z} \hfil   &
    \multispan3{\hfil \Ha$\lambda$6562 \hfil}  &
    \multispan3{\hfil \Hb$\lambda$4861 \hfil}  &
    \multispan3{\hfil \oiiifull \hfil}  &
    \multispan3{\hfil \mgiifull \hfil}  &
    \multispan3{\hfil \ciiifull \hfil}  
 \cr
    & & 
    & \hfil{\sl lum} \hfil & \hfil{\sl ew} \hfil & \hfil{\sl fwhm} \hfil 
    & \hfil{\sl lum} \hfil & \hfil{\sl ew} \hfil & \hfil{\sl fwhm} \hfil 
    & \hfil{\sl lum} \hfil & \hfil{\sl ew} \hfil & \hfil{\sl fwhm} \hfil 
    & \hfil{\sl lum} \hfil & \hfil{\sl ew} \hfil & \hfil{\sl fwhm} \hfil 
    & \hfil{\sl lum} \hfil & \hfil{\sl ew} \hfil & \hfil{\sl fwhm} \hfil 
\cr
\noalign{\medskip}
\multispan2{\sl median\hfil} & 0.719 & 42.3 & 110 & 2600 & 42.1 & 34 & 3300 & 41.8 & 20 & --- & 42.9 & 29 & 3900 & 43.6 & 16 & 5800\cr
\multispan2{\sl mean\hfil}   & 0.82  & 42.9 & 140 & 3100 & 42.8 & 38 & 3900 & 42.5 & 29 & --- & 43.4 & 37 & 4600 & 43.9 & 22 & 6300\cr
\multispan2{$\sigma$ \hfil}  & 0.53  & \hfil0.6 & \hfil90 & 1800 & \hfil0.7 & 23 & 2000 & \hfil0.8 & 26 & --- & \hfil0.5 & 30 & 2300 &  \hfil0.4 & 21 & 2600\cr
\multispan2{\sl error\hfil}  & 0.04  & \hfil0.2 & \hfil20 & \hfil400 & \hfil0.2
& \hfil3 & \hfil400 & \hfil0.2 & \hfil3 & --- & \hfil0.1 &
\hfil3\hskip0.3truecm &
\hfil200\hskip0.3truecm &
\hfil0.1 & \hfil3 & \hfil400 \cr
}
\tabletext{FID: RIXOS field number, SNo: RIXOS source number, {\sl z}: AGN
redshift, {\sl lum}: Logarithm of the line luminosity in ergs s$^{-1}$,  {\sl
ew}: Rest-frame equivalent width in \AA, {\sl fwhm}: FWHM in km s$^{-1}$;
$^b$~broad component only; $^n$~narrow component only; $^v$~very broad
component only; $^u$~parameter uncertain; $^l$~line luminosities and EWs are
upper limits;  $^f$~FWHM fixed (\ie for calculating upper limits); $^{r1}$~data
from Puchnarewicz \etal (1992); $^{r2}$~data from Stephens (1989); $^{r3}$~data
from Brotherton \etal (1994); *~F252\_31 is the double quasar E0957+561; the
optical spectrum of only one component (from Puchnarewicz \etal 1992) was
available and this was assumed to be the same for the other component as well.
The median, mean, $\sigma$ (\ie standard deviation on the mean) and error on
the mean have been calculated from the sample shown, excluding uncertain
parameters and neglecting upper limits; source 6 in field 273 was also excluded
because of its unusual nature (see Section 4.5.1). Where two components have
been fitted to permitted line profiles, only the broad components are used for
these statistics.} 
\endtable

\begintable*{5}
\caption{{\bf Table 2}: Optical/UV and X-ray continuum parameters.}
\halign{#\hfil
       &\quad#\hfil
       &\hskip0.2truecm\hfil#\hfil\hskip0.2truecm
       &\hskip0.2truecm\hfil#\hfil\hskip0.2truecm
       &\hskip0.2truecm\hfil#\hfil\hskip0.2truecm
       &\hskip0.2truecm#\hfil\hskip0.2truecm
       &\hskip0.2truecm\hfil#\hfil\hskip0.2truecm
       &\hskip0.2truecm\hfil#\hfil\hskip0.2truecm
       &\hskip0.2truecm#
       &\hskip0.2truecm\hfil#\hskip0.2truecm
       &\hskip0.2truecm\hfil#\hfil\hskip0.2truecm
       &\hskip0.2truecm\hfil#\hfil
\cr
\noalign{\bigskip}
    FID   &
    S No  &
    Ins   &
     z    &
    \nhgal\ &
    \hfil $L_{\rm 2500}$ &
    $L_{\rm 0.2keV}$ &
    $L_{\rm 2keV}$ &
    \hskip0.05truecm$\alpha_{\rm opt}$ &
    $\alpha_{\rm x}$\hfil &
    $\alpha_{\rm os}$ &
    $\alpha_{\rm ox}$ 
 \cr
\noalign{\smallskip}
      (1)\hfil 
    & (2)\hfil 
    & (3)
    & (4)
    & (5)
    & \hfil (6)
    & (7)
    & (8)
    & \hskip0.15truecm(9)\hfil
    & (10)\hfil
    & (11)
    & (12) \cr
\noalign{\medskip}
 122 &  13 & ISIS & 0.358 & 4.1 & 28.2$^{i}$ & 26.1$^{+0.5}_{-0.6}$ & 26.1$^{+0.2}_{-0.2}$ & \hskip0.15truecm3.1 & 0.00$^{+0.28}_{-0.28}$ & 1.3$^{+0.6}_{-0.5}$ & 0.8$^{+0.2}_{-0.2}$ \cr
 122 &  21$^{(43)}$ & ISIS & 0.376 & 4.1 & 28.5$^{i}$ & 26.4$^{+0.5}_{-0.6}$ & 26.2$^{+0.2}_{-0.2}$ & \hskip0.15truecm2.3 & 0.23$^{+0.26}_{-0.28}$ & 1.3$^{+0.6}_{-0.5}$ & 0.9$^{+0.2}_{-0.2}$ \cr
 123 &  28 & ISIS & 0.212 & 1.2 & 28.6$^{i}$ & 26.2$^{+0.4}_{-0.4}$ & 25.0$^{+0.2}_{-0.2}$ & \hskip0.15truecmnp & 1.16$^{+0.13}_{-0.14}$ & 1.5$^{+0.4}_{-0.4}$ & 1.3$^{+0.2}_{-0.2}$ \cr
 125 &  14 & ISIS & 1.833 & 5.0 & 30.2 & 28.6$^{+0.9}_{-1.7}$ & 27.7$^{+0.2}_{-0.3}$ & \hskip0.15truecm1.6 & 0.96$^{+0.40}_{-0.51}$ & 1.0$^{+1.2}_{-0.8}$ & 1.0$^{+0.2}_{-0.2}$ \cr
 125 &  17$^{(36)}$ & ISIS & 0.449 & 5.0 & 28.9 & 27.7$^{+0.4}_{-0.5}$ & 26.3$^{+0.1}_{-0.1}$ & \hskip0.15truecm0.6 & 1.48$^{+0.16}_{-0.20}$ & 0.7$^{+0.5}_{-0.4}$ & 1.0$^{+0.2}_{-0.2}$ \cr
 126 &   1 & ISIS & 0.028 & 2.0 & 27.5 & 24.0$^{+0.5}_{-0.7}$ & 23.9$^{+0.4}_{-0.1}$ & \hskip0.15truecm2.3$^{p}$ & 0.14$^{+0.33}_{-0.35}$ & 2.1$^{+0.6}_{-0.5}$ & 1.4$^{+0.2}_{-0.3}$ \cr
% 133 &  17 & DK & 2.390 & 1.2 & 30.2 & --- & --- & \hskip0.15truecm0.2$^{r}$ &  ---\hfil & --- & --- \cr
 205 &   1 & ISIS & 0.717 & 4.3 & 29.2 & 28.8$^{+0.7}_{-1.0}$ & 26.4$^{+0.2}_{-0.2}$ & \hskip0.15truecm0.6 & 1.43$^{+0.34}_{-0.32}$ & 0.2$^{+0.8}_{-0.6}$ & 1.0$^{+0.2}_{-0.2}$ \cr
 205 &   1 & ISIS & 1.334 & 4.3 & 29.6 & 28.6$^{+0.8}_{-1.1}$ & 27.2$^{+0.2}_{-0.2}$ & \hskip0.15truecm1.2 & 1.43$^{+0.34}_{-0.32}$ & 0.6$^{+0.9}_{-0.7}$ & 0.9$^{+0.2}_{-0.2}$ \cr
 205 &  22 & ISIS & 0.445 & 4.3 & 29.3 & 27.8$^{+0.4}_{-0.4}$ & 26.2$^{+0.2}_{-0.2}$ & \hskip0.15truecm0.4 & 1.65$^{+0.16}_{-0.17}$ & 0.9$^{+0.5}_{-0.4}$ & 1.2$^{+0.2}_{-0.2}$ \cr
 205 &  23 & ISIS & 0.618 & 4.3 & 28.3 & 27.6$^{+0.6}_{-0.8}$ & 26.5$^{+0.2}_{-0.2}$ & \hskip0.15truecmnp & 1.14$^{+0.27}_{-0.29}$ & 0.4$^{+0.7}_{-0.6}$ & 0.7$^{+0.2}_{-0.2}$ \cr
 205 &  34 & ISIS & 0.755 & 4.3 & 28.7$^{i}$ & 27.5$^{+1.1}_{-1.8}$ & 26.3$^{+0.3}_{-0.3}$ & \hskip0.15truecm0.9 & 1.20$^{+0.51}_{-0.62}$ & 0.7$^{+1.3}_{-0.8}$ & 0.9$^{+0.2}_{-0.2}$ \cr
 206 & 507 & ISIS & 0.484 & 3.7 & 27.7 & 26.8$^{+0.7}_{-1.0}$ & 25.9$^{+0.3}_{-0.3}$ & \hskip0.15truecm3.9 & 0.98$^{+0.34}_{-0.40}$ & 0.5$^{+0.8}_{-0.6}$ & 0.7$^{+0.2}_{-0.2}$ \cr
 206 & 522 & ISIS & 0.740 & 3.7 & 29.3 & 27.7$^{+0.6}_{-0.7}$ & 26.4$^{+0.2}_{-0.2}$ & --0.0 & 1.37$^{+0.23}_{-0.27}$ & 0.9$^{+0.6}_{-0.6}$ & 1.1$^{+0.2}_{-0.2}$ \cr
 208 &  18 & IDS & 0.470 & 0.7 & 29.3$^{i}$ & 27.6$^{+0.3}_{-0.3}$ & 26.1$^{+0.1}_{-0.2}$ & \hskip0.15truecm0.6 & 1.49$^{+0.11}_{-0.09}$ & 1.0$^{+0.4}_{-0.4}$ & 1.2$^{+0.2}_{-0.2}$ \cr
 211 &  35$^{(18)}$ & ISIS & 0.465 & 4.0 & 28.7$^{i}$ & 27.0$^{+0.7}_{-1.0}$ & 25.9$^{+0.3}_{-0.3}$ & \hskip0.15truecm1.8 & 1.12$^{+0.33}_{-0.40}$ & 1.0$^{+0.8}_{-0.6}$ & 1.1$^{+0.2}_{-0.2}$ \cr
 212 &  32* & ISIS & 0.927 & 1.2 & 30.0 & 28.1$^{+0.5}_{-0.5}$ & 26.7$^{+0.2}_{-0.2}$ & --0.3 & 1.46$^{+0.12}_{-0.15}$ & 1.2$^{+0.5}_{-0.5}$ & 1.3$^{+0.2}_{-0.2}$ \cr
 212 &  32* & ISIS & 0.924 & 1.2 & 30.0 & 28.1$^{+0.5}_{-0.5}$ & 26.7$^{+0.2}_{-0.2}$ & \hskip0.15truecm0.4 & 1.46$^{+0.12}_{-0.15}$ & 1.1$^{+0.5}_{-0.5}$ & 1.3$^{+0.2}_{-0.2}$ \cr
 213 &   7 & ISIS & 0.542 & 4.4 & 28.7 & 27.0$^{+0.9}_{-1.3}$ & 26.3$^{+0.3}_{-0.1}$ & \hskip0.15truecm1.0 & 0.71$^{+0.48}_{-0.57}$ & 1.1$^{+1.0}_{-0.8}$ & 0.9$^{+0.2}_{-0.2}$ \cr
 213 &  11 & ISIS & 1.546 & 4.4 & 29.4 & 26.1$^u$ & 26.9$^u$ & --0.1 & --0.49$^{+1.0}_{-1.6}$ & 2.0$^u$ & 0.9$^u$ \cr
 213 &  20 & ISIS & 0.664 & 4.4 & 29.8 & 27.6$^{+0.8}_{-1.2}$ & 26.6$^{+0.3}_{-0.3}$ & \hskip0.15truecm0.5 & 1.02$^{+0.37}_{-0.47}$ & 1.4$^{+1.0}_{-0.7}$ & 1.2$^{+0.2}_{-0.2}$ \cr
 216 &  30$^{(50)}$ & ISIS & 0.941 & 3.5 & 28.8$^{i}$ & 27.8$^{+0.6}_{-0.8}$ & 26.8$^{+0.1}_{-0.2}$ & \hskip0.15truecmnp & 0.99$^{+0.26}_{-0.30}$ & 0.6$^{+0.7}_{-0.6}$ & 0.8$^{+0.2}_{-0.2}$ \cr
 216 &  33 & IDS & 0.792 & 3.5 & 29.3 & 27.5$^{+0.8}_{-1.1}$ & 26.5$^{+0.2}_{-0.2}$ & \hskip0.15truecm0.4 & 1.04$^{+0.35}_{-0.39}$ & 1.1$^{+0.9}_{-0.7}$ & 1.1$^{+0.2}_{-0.2}$ \cr
 218 &   9 & ISIS & 0.703 & 3.0 & 29.3 & 27.7$^{+0.9}_{-1.1}$ & 26.2$^{+0.3}_{-0.4}$ & \hskip0.15truecm0.2 & 1.51$^{+0.36}_{-0.36}$ & 0.9$^{+0.9}_{-0.8}$ & 1.2$^{+0.3}_{-0.2}$ \cr
 218 &  13$^{(25)}$ & HB & 1.450 & 3.0 & 30.7 & 28.1$^{+1.3}_{-2.4}$ & 27.1$^{+0.3}_{-0.5}$ & \hskip0.15truecm--- & 1.01$^{+0.54}_{-0.67}$ & 1.6$^{+1.7}_{-1.0}$ & 1.4$^{+0.3}_{-0.2}$ \cr
 218 &  14$^{(25)}$ & ISIS & 0.224 & 3.0 & 27.7 & 26.7$^{+0.6}_{-0.7}$ & 25.5$^{+0.3}_{-0.3}$ & \hskip0.15truecm2.6 & 1.21$^{+0.53}_{-0.80}$ & 0.6$^{+0.6}_{-0.5}$ & 0.9$^{+0.2}_{-0.2}$ \cr
 218 &  21 & ISIS & 0.790 & 3.0 & 28.7 & 26.2$^{+1.4}_{-2.7}$ & 26.6$^{+0.3}_{-0.4}$ & \hskip0.15truecmnp & --0.32$^{+0.68}_{-0.91}$ & 1.6$^{+1.9}_{-1.1}$ & 0.8$^{+0.3}_{-0.2}$ \cr
 219 &  45$^{(28)}$ & ISIS & 1.272 & 1.3 & 29.4$^{i}$ & 27.9$^{+0.4}_{-0.4}$ & 27.4$^{+0.1}_{-0.1}$ & \hskip0.15truecm2.4 & 0.55$^{+0.13}_{-0.14}$ & 0.9$^{+0.4}_{-0.4}$ & 0.8$^{+0.2}_{-0.2}$ \cr
 220 &  13 & ISIS & 0.968 & 3.9 & 29.6 & 25.0$^u$ & 26.5$^u$ & \hskip0.15truecmnp & --0.9$^{+1.3}_{-1.8}$ & 2.8$^u$ & 1.2$^u$ \cr
 220 &  18 & ISIS & 0.442 & 3.9 & 28.8$^{i}$ & 26.7$^{+1.1}_{-2.3}$ & 25.9$^{+0.6}_{-0.4}$ & \hskip0.15truecm0.6 & 0.78$^{+0.57}_{-0.88}$ & 1.3$^{+1.6}_{-0.9}$ & 1.1$^{+0.3}_{-0.3}$ \cr
 221 &   2 & HB & 0.900 & 2.9 & 29.6 & 27.6$^{+0.8}_{-1.2}$ & 26.6$^{+0.2}_{-0.2}$ & \hskip0.15truecm0.4 & 0.94$^{+0.36}_{-0.39}$ & 1.3$^{+0.9}_{-0.7}$ & 1.1$^{+0.2}_{-0.2}$ \cr
 221 &  16 & ISIS & 0.184 & 2.9 & 28.3 & 26.6$^{+0.4}_{-0.5}$ & 25.0$^{+0.2}_{-0.3}$ & \hskip0.15truecm1.4 & 1.67$^{+0.21}_{-0.13}$ & 1.0$^{+0.5}_{-0.5}$ & 1.3$^{+0.2}_{-0.2}$ \cr
 224 & 201$^{(25)}$ & ISIS & 1.547 & 1.0 & 30.0$^{i}$ & 28.6$^{+0.4}_{-0.4}$ & 27.3$^{+0.1}_{-0.2}$ & \hskip0.15truecm1.8 & 1.27$^{+0.14}_{-0.11}$ & 0.8$^{+0.5}_{-0.4}$ & 1.0$^{+0.2}_{-0.2}$ \cr
 236 &   5 & FOS & 0.473 & 2.6 & 29.2 & 27.1$^{+0.9}_{-1.0}$ & 25.7$^{+0.4}_{-0.1}$ & \hskip0.15truecm0.7 & 1.36$^{+0.54}_{-0.42}$ & 1.2$^{+0.8}_{-0.8}$ & 1.3$^{+0.2}_{-0.3}$ \cr
 236 &  21 & ISIS & 1.128 & 2.6 & 29.5 & 28.2$^{+1.1}_{-1.5}$ & 26.7$^{+0.3}_{-0.4}$ & \hskip0.15truecm1.1 & 1.47$^{+0.43}_{-0.46}$ & 0.8$^{+1.1}_{-0.9}$ & 1.1$^{+0.3}_{-0.2}$ \cr
 236 &  22 & FOS & 0.048 & 2.6 & 28.1 & 25.1$^{+0.7}_{-0.9}$ & 23.7$^{+0.5}_{-0.5}$ & \hskip0.15truecmnp & 1.34$^{+0.31}_{-0.36}$ & 1.8$^{+0.7}_{-0.6}$ & 1.7$^{+0.3}_{-0.3}$ \cr
 238 &  11 & ISIS & 0.325 & 4.1 & 28.3 & 26.0$^{+1.0}_{-1.8}$ & 25.7$^{+0.5}_{-0.4}$ & \hskip0.15truecm1.9 & 0.29$^{+0.58}_{-0.67}$ & 1.4$^{+1.3}_{-0.8}$ & 1.0$^{+0.3}_{-0.3}$ \cr
 245 &   4$^{(36)}$ & ISIS & 0.711 & 8.8 & 28.8 & 27.5$^{+1.1}_{-1.4}$ & 26.2$^{+0.3}_{-0.1}$ & \hskip0.15truecm1.1 & 1.33$^{+0.55}_{-0.62}$ & 0.7$^{+1.1}_{-0.9}$ & 1.0$^{+0.1}_{-0.2}$ \cr
 248 &  51 & HB & 0.242 & 1.5 & 29.8 & 27.2$^{+0.2}_{-0.2}$ & 25.7$^{+0.1}_{-0.1}$ & \hskip0.15truecm--- & 1.46$^{+0.07}_{-0.05}$ & 1.6$^{+0.3}_{-0.3}$ & 1.6$^{+0.2}_{-0.2}$ \cr
 252 &   1$^{(43)}$ & ISIS & 0.219 & 0.8 & 27.7$^{i}$ & 26.1$^{+0.4}_{-0.5}$ & 25.4$^{+0.2}_{-0.3}$ & \hskip0.15truecm3.7 & 0.71$^{+0.17}_{-0.19}$ & 1.0$^{+0.5}_{-0.5}$ & 0.9$^{+0.2}_{-0.2}$ \cr
 252 &  31** & HB & 1.413 & 0.8 & 31.3 & 29.2$^{+0.1}_{-0.1}$ & 28.1$^{+0.1}_{-0.1}$ & \hskip0.15truecm0.6 & 1.15$^{+0.03}_{-0.04}$ & 1.3$^{+0.3}_{-0.3}$ & 1.2$^{+0.1}_{-0.1}$ \cr
 252 &  31** & HB & 1.415 & 0.8 & 31.3 & 29.2$^{+0.1}_{-0.1}$ & 28.1$^{+0.1}_{-0.1}$ & \hskip0.15truecm0.6 & 1.15$^{+0.03}_{-0.04}$ & 1.3$^{+0.3}_{-0.3}$ & 1.2$^{+0.1}_{-0.1}$ \cr
 252 &  46 & HB & 2.091 & 0.8 & 30.9 & 29.3$^{+0.5}_{-0.5}$ & 27.8$^{+0.2}_{-0.2}$ & \hskip0.15truecm--- & 1.47$^{+0.17}_{-0.10}$ & 1.0$^{+0.5}_{-0.5}$ & 1.2$^{+0.2}_{-0.2}$ \cr
 255 &   7 & ISIS & 0.260 & 5.1 & 27.8 & 26.5$^{+1.1}_{-2.6}$ & 25.2$^{+0.8}_{-0.6}$ & \hskip0.15truecm2.6 & 1.22$^{+0.66}_{-0.99}$ & 0.8$^{+1.8}_{-0.9}$ & 1.0$^{+0.3}_{-0.4}$ \cr
 255 &  13 & ISIS & 0.581 & 5.1 & 29.1 & 27.6$^{+1.0}_{-1.6}$ & 25.9$^{+0.4}_{-0.4}$ & \hskip0.15truecm0.1 & 1.63$^{+0.49}_{-0.53}$ & 0.9$^{+1.2}_{-0.8}$ & 1.2$^{+0.3}_{-0.3}$ \cr
 255 &  23 & ISIS & 0.759 & 5.1 & 29.0 & 27.4$^{+1.3}_{-2.4}$ & 26.4$^{+0.4}_{-0.4}$ & --0.2 & 1.01$^{+0.61}_{-0.83}$ & 1.0$^{+1.7}_{-1.0}$ & 1.0$^{+0.3}_{-0.3}$ \cr
 257 &  37 & ISIS & 0.329 & 2.2 & 27.9 & 26.4$^{+0.7}_{-1.0}$ & 25.4$^{+0.4}_{-0.4}$ & \hskip0.15truecm1.0$^{w}$ & 1.05$^{+0.31}_{-0.36}$ & 0.9$^{+0.8}_{-0.6}$ & 1.0$^{+0.3}_{-0.3}$ \cr
 258 &   5 & ISIS & 0.811 & 3.4 & 29.2 & 28.1$^{+0.7}_{-0.8}$ & 26.4$^{+0.2}_{-0.2}$ & \hskip0.15truecm1.2 & 1.64$^{+0.24}_{-0.26}$ & 0.7$^{+0.7}_{-0.6}$ & 1.1$^{+0.2}_{-0.2}$ \cr
 258 &  30 & ISIS & 0.847 & 3.4 & 29.1 & 24.6$^u$ & 26.4$^u$ & \hskip0.15truecmnp & --1.1$^{+1.7}_{-1.4}$ & 2.7$^u$ & 1.0$^u$ \cr
 259 &  30 & HB & 1.940 & 2.0 & 29.7 & 27.7$^{+0.9}_{-1.4}$ & 27.6$^{+0.2}_{-0.3}$ & \hskip0.15truecm--- & 0.15$^{+0.35}_{-0.42}$ & 1.2$^{+1.1}_{-0.7}$ & 0.8$^{+0.2}_{-0.2}$ \cr
}
\endtable

\eject

\begintable*{6}
\caption{{\bf Table 2} (continued): Optical/UV and X-ray continuum parameters.}
\halign{#\hfil
       &\quad#\hfil
       &\hskip0.2truecm\hfil#\hfil\hskip0.2truecm
       &\hskip0.2truecm\hfil#\hfil\hskip0.2truecm
       &\hskip0.2truecm\hfil#\hfil\hskip0.2truecm
       &\hskip0.2truecm#\hfil\hskip0.2truecm
       &\hskip0.2truecm\hfil#\hfil\hskip0.2truecm
       &\hskip0.2truecm\hfil#\hfil\hskip0.2truecm
       &\hskip0.2truecm#
       &\hskip0.2truecm\hfil#\hskip0.2truecm
       &\hskip0.2truecm\hfil#\hfil\hskip0.2truecm
       &\hskip0.2truecm\hfil#\hfil
\cr
\noalign{\bigskip}
    FID   &
    S No  &
    Ins   &
     z    &
    \nhgal\ &
    \hfil $L_{\rm 2500}$ &
    $L_{\rm 0.2keV}$ &
    $L_{\rm 2keV}$ &
    \hskip0.05truecm$\alpha_{\rm opt}$ &
    $\alpha_{\rm x}$\hfil &
    $\alpha_{\rm os}$ &
    $\alpha_{\rm ox}$ 
 \cr
\noalign{\smallskip}
      (1)\hfil 
    & (2)\hfil 
    & (3)
    & (4)
    & (5)
    & \hfil (6)
    & (7)
    & (8)
    & \hskip0.15truecm(9)\hfil
    & (10)\hfil
    & (11)
    & (12) \cr
\noalign{\medskip}
 262 &   1 & ISIS & 0.882 & 3.4 & 29.4 & 28.0$^{+0.7}_{-0.8}$ & 26.5$^{+0.2}_{-0.2}$ & \hskip0.15truecm0.5 & 1.52$^{+0.25}_{-0.28}$ & 0.8$^{+0.7}_{-0.6}$ & 1.1$^{+0.2}_{-0.2}$ \cr
 262 &   2 & HB & 1.202 & 3.4 & 30.7 & 28.2$^{+0.6}_{-0.8}$ & 27.0$^{+0.2}_{-0.2}$ & \hskip0.15truecm--- & 1.16$^{+0.25}_{-0.24}$ & 1.5$^{+0.7}_{-0.6}$ & 1.4$^{+0.2}_{-0.2}$ \cr
 262 &  10 & HB & 0.336 & 3.4 & 29.8 & 27.5$^{+0.2}_{-0.2}$ & 26.3$^{+0.1}_{-0.1}$ & \hskip0.15truecm--- & 1.24$^{+0.10}_{-0.08}$ & 1.4$^{+0.3}_{-0.3}$ & 1.4$^{+0.2}_{-0.2}$ \cr
 262 &  12 & IDS & 0.923 & 3.4 & 29.5 & 28.0$^{+0.6}_{-0.6}$ & 26.8$^{+0.1}_{-0.2}$ & \hskip0.15truecm0.9$^{w}$ & 1.28$^{+0.22}_{-0.21}$ & 0.9$^{+0.6}_{-0.5}$ & 1.1$^{+0.2}_{-0.2}$ \cr
 262 &  34 & IDS & 0.311 & 3.4 & 29.0 & 27.2$^{+0.3}_{-0.4}$ & 25.8$^{+0.2}_{-0.2}$ & \hskip0.15truecm0.8$^{p}$ & 1.44$^{+0.17}_{-0.11}$ & 1.1$^{+0.4}_{-0.4}$ & 1.2$^{+0.2}_{-0.2}$ \cr
 272 &   8 & ISIS & 1.817 & 4.7 & 29.8$^{i}$ & 29.1$^{+1.2}_{-1.8}$ & 27.5$^{+0.3}_{-0.5}$ & \hskip0.15truecm1.0 & 1.54$^{+0.43}_{-0.51}$ & 0.4$^{+1.3}_{-0.9}$ & 0.9$^{+0.3}_{-0.2}$ \cr
 272 &  18 & ISIS & 0.604 & 4.7 & 29.6 & 27.7$^{+0.7}_{-1.0}$ & 26.7$^{+0.2}_{-0.2}$ & \hskip0.15truecm0.4 & 1.08$^{+0.33}_{-0.39}$ & 1.1$^{+0.8}_{-0.6}$ & 1.1$^{+0.2}_{-0.2}$ \cr
 272 &  28 & ISIS & 0.444 & 4.7 & 28.4 & 27.5$^{+0.7}_{-0.8}$ & 26.0$^{+0.3}_{-0.3}$ & \hskip0.15truecmnp & 1.46$^{+0.30}_{-0.31}$ & 0.5$^{+0.7}_{-0.6}$ & 0.9$^{+0.2}_{-0.2}$ \cr
 273 &  22 & FOS & 1.075 & 2.8 & 30.8 & 28.9$^{+1.2}_{-0.7}$ & 26.7$^{+0.3}_{-0.1}$ & \hskip0.15truecm0.7 & 2.15$^{+0.31}_{-0.35}$ & 1.2$^{+0.6}_{-0.9}$ & 1.6$^{+0.1}_{-0.2}$ \cr
%273 &  23 & ISIS & 0.433 & 2.8 & 28.6 & 26.9$^{+0.9}_{-2.1}$ & 25.6$^{+0.6}_{-0.5}$ & \hskip0.15truecm2.3 & 1.24$^{+0.59}_{-0.67}$ & 1.0$^{+1.5}_{-0.7}$ & 1.2$^{+0.3}_{-0.3}$ \cr
 281 &  11 & ISIS & 2.919 & 5.8 & 30.1 & 29.1$^{+1.8}_{-3.1}$ & 28.0$^{+0.6}_{-1.0}$ & \hskip0.15truecm1.8 & 1.15$^{+0.66}_{-0.79}$ & 0.6$^{+2.1}_{-1.3}$ & 0.8$^{+0.5}_{-0.3}$ \cr
 281 &  21 & IDS & 0.347 & 5.8 & 28.5 & 26.6$^{+0.9}_{-1.2}$ & 26.0$^{+0.4}_{-0.2}$ & \hskip0.15truecm2.4 & 0.68$^{+0.49}_{-0.57}$ & 1.2$^{+0.9}_{-0.8}$ & 1.0$^{+0.2}_{-0.3}$ \cr
 283 &  11 & FOS & 0.272 & 10. & 29.7 & 24.9$^{+1.0}_{-2.3}$ & 25.7$^{+0.7}_{-0.5}$ & \hskip0.15truecm0.5 & --0.67$^{+0.60}_{-0.77}$ & 2.9$^{+1.6}_{-0.8}$ & 1.6$^{+0.3}_{-0.4}$ \cr
 283 &  21 & IDS & 0.719 & 10. & 29.6 & 26.1$^{+1.0}_{-2.1}$ & 26.5$^{+0.4}_{-0.1}$ & \hskip0.15truecm0.8 & --0.35$^{+0.56}_{-0.69}$ & 2.1$^{+1.5}_{-0.8}$ & 1.2$^{+0.1}_{-0.3}$ \cr
 293 &   1 & ISIS & 0.824 & 4.6 & 29.5 & 27.5$^{+1.1}_{-2.0}$ & 26.7$^{+0.2}_{-0.2}$ & --0.8 & 0.83$^{+0.52}_{-0.71}$ & 1.2$^{+1.4}_{-0.9}$ & 1.1$^{+0.2}_{-0.2}$ \cr
 293 &   6$^{(43)}$ & FOS & 0.081 & 4.6 & 29.4 & 26.0$^{+0.4}_{-0.5}$ & 24.8$^{+0.2}_{-0.2}$ & \hskip0.15truecm2.9 & 1.26$^{+0.21}_{-0.22}$ & 2.1$^{+0.5}_{-0.4}$ & 1.8$^{+0.2}_{-0.2}$ \cr
 293 &  12 & IDS+ISIS & 0.917 & 4.6 & 29.6 & 27.3$^{+1.5}_{-3.0}$ & 26.7$^{+0.3}_{-0.1}$ & \hskip0.15truecm0.4 & 0.68$^{+0.85}_{-0.97}$ & 1.4$^{+2.0}_{-1.1}$ & 1.1$^{+0.1}_{-0.2}$ \cr
 294 &   1 & ISIS & 0.713 & 4.3 & 29.2 & 27.8$^{+0.7}_{-0.8}$ & 26.5$^{+0.2}_{-0.2}$ & \hskip0.15truecmnp & 1.29$^{+0.29}_{-0.30}$ & 0.9$^{+0.7}_{-0.6}$ & 1.0$^{+0.2}_{-0.2}$ \cr
 302 &  14 & IDS & 0.809 & 1.2 & 29.7 & 27.7$^{+0.5}_{-0.5}$ & 26.6$^{+0.2}_{-0.2}$ & \hskip0.15truecm0.6 & 1.15$^{+0.16}_{-0.17}$ & 1.2$^{+0.5}_{-0.5}$ & 1.2$^{+0.2}_{-0.2}$ \cr
 302 &  18$^{(36)}$ & ISIS & 0.924 & 1.2 & 30.1 & 27.7$^{+0.4}_{-0.5}$ & 27.0$^{+0.1}_{-0.1}$ & \hskip0.15truecm0.3 & 0.75$^{+0.14}_{-0.17}$ & 1.5$^{+0.5}_{-0.5}$ & 1.2$^{+0.2}_{-0.2}$ \cr
 305 &  11 & IDS & 0.251 & 2.3 & 28.8 & 27.1$^{+0.3}_{-0.3}$ & 25.7$^{+0.2}_{-0.2}$ & \hskip0.15truecm0.9 & 1.44$^{+0.13}_{-0.10}$ & 1.0$^{+0.4}_{-0.4}$ & 1.2$^{+0.2}_{-0.2}$ \cr
 305 &  18$^{(39)}$ & IDS & 0.387 & 2.3 & 28.8 & 26.4$^{+0.6}_{-0.8}$ & 26.1$^{+0.2}_{-0.2}$ & \hskip0.15truecm1.1 & 0.28$^{+0.27}_{-0.31}$ & 1.5$^{+0.7}_{-0.5}$ & 1.0$^{+0.2}_{-0.2}$ \cr
 305 &  34$^{(46)}$ & IDS & 0.854 & 2.3 & 29.8 & 27.8$^{+0.6}_{-0.7}$ & 26.7$^{+0.2}_{-0.2}$ & --0.3 & 1.06$^{+0.21}_{-0.23}$ & 1.3$^{+0.6}_{-0.6}$ & 1.2$^{+0.2}_{-0.2}$ \cr
\noalign{\bigskip}
\multispan6{Sample statistics (includes data from Paper I)\hfil}\cr
    &
    &
    &
     z    &
    \nhgal\ &
    \hfil $L_{\rm 2500}$ &
    $L_{\rm 0.2keV}$ &
    $L_{\rm 2keV}$ &
    \hskip0.15truecm$\alpha_{\rm opt}$ &
    $\alpha_{\rm x}$\hfil &
    $\alpha_{\rm os}$ &
    $\alpha_{\rm ox}$ 
 \cr
\noalign{\medskip}
\multispan3{\sl median\hfil} & 0.719 & 2.3 & 29.3 & 27.6 & 26.5 & \hfil0.8 & 1.08\hfil & 1.1 & 1.12 \cr
\multispan3{\sl mean\hfil}   & 0.82  & 2.8 & 30.2 & 28.2 & 27.0 & \hfil0.9 & 1.07\hfil & 1.4 & 1.16 \cr
\multispan3{$\sigma$\hfil}    & 0.53  & 1.9 & \hfil0.6 & \hfil0.5 & \hfil0.5 & \hfil0.9 & 0.51\hfil & 0.6 & 0.18 \cr
\multispan3{\sl error\hfil}  & 0.04  & 0.2 & \hfil0.1 & \hfil0.1 & \hfil0.1 & \hfil0.1 & 0.04\hfil & 0.1 & 0.02 \cr
}

\tabletext{{\sl (1)} RIXOS field number (see Mason et al, in preparation); {\sl (2)} RIXOS
source number (Mason et al, in preparation) - the number in brackets is the  radius of the
extraction circle used for the X-ray data (in arcseconds) where it is less than
54 arcsec (see Paper I); {\sl (3)} the instrument with which the spectrum was
taken - see Section 2.2; HB - Hewitt \& Burbridge (1989; spectrum not taken);
MSS - Gioia \etal (1984; spectrum not taken) {\sl (4)} redshift;  {\sl (5)}
Galactic column density (10$^{20}$ cm$^{-2}$) - errors are $\sim$10 per cent
(see also Paper I); {\sl (6)} log of the monochromatic UV luminosity at
2500~\AA\ (erg s$^{-1}$ Hz$^{-1}$) - error is estimated to be $\sim$50 percent
(Section 2.2.1);   {\sl (7)} log of the monochromatic X-ray luminosity at
0.2~\keV\ (erg s$^{-1}$ Hz$^{-1}$) - errors are calculated from the 90\% errors
on the fits (Paper I);   {\sl (8)} log of the monochromatic X-ray luminosity at
2.0~\keV\ (erg s$^{-1}$ Hz$^{-1}$)  - errors are calculated from the 90\%
errors on  the fits (Paper I);   {\sl (9)} energy index of the best-fitting
power-law to the optical/UV continuum - error is estimated to be $\pm$0.5 (Paper
I);  {\sl (10)} energy index of the best-fitting power-law to the X-ray data -
errors are 90\% (Paper I); {\sl (11)} and {\sl (12)}  for definitions see
Section 2.3 - errors are calculated from the quoted errors on \luv, \lsoft\ and
\lhard; $^i$~- CCD image taken; $^w$~- optical/UV continuum weak;
$^p$~optical/UV power-law slope poor fit; $^u$~X-ray power-law slope poor fit,
data uncertain; np - spectrum not taken at the parallactic angle; *~F212\_32 is
a double quasar; the optical data have been measured from spectra of the
individual components, but the components are not resolved in the PSPC image,
thus \ax\ is the slope for the total X-ray emission while the \lsoft\ and
\lhard\ for each component comprise one half of the total; **~F252\_31 is the
double quasar E0957+561; the optical spectrum of only one component (from
Puchnarewicz \etal 1992) was available and this was assumed to be the same for
both. The components are not resolved in the X-ray image, thus \ax\ is the
slope for the total X-ray emission while the \lsoft\ and \lhard\ for each
component comprise one half of the total. The median, mean, $\sigma$ (\ie
standard deviation on the  mean) and error on the mean have been calculated for
the full sample used (excluding uncertain parameters), \ie including those
whose continuum parameters are given in Paper I.} 
                                        
\endtable

\subsection{Optical}

The optical spectra were obtained over several observing runs with the Isaac
Newton (INT) and William Herschel Telescopes (WHT) at the Observatorio del
Roque de los Muchachos, La Palma. Three different instruments were used, the
Faint Object Spectrograph (FOS) and the Intermediate Dispersion Spectrograph
(IDS) on the INT and the Intermediate-Dispersion Spectrograph and Imaging
System (ISIS) on the WHT. The FOS spectra cover a range of 3500\AA\ to
10000\AA\ with a resolution of 15-20\AA\ FWHM in the red and 8-10\AA\ FWHM in
the blue, while the IDS spectra cover the same range with a resolution of
$\sim$10\AA\ FWHM. The ISIS spectra cover 3000\AA\ to 9000\AA\ with a
resolution of 3\AA\ FWHM in the red and 2\AA\ in the blue.  All spectra were
taken using a narrow slit (typically $\sim1^{\prime\prime}-1.5^{\prime\prime}$)
which was positioned at the parallactic angle except where indicated in Table
2. On average, the seeing was $\sim1^{\prime\prime}-1.5^{\prime\prime}$.

Photometrically-calibrated CCD images of 36 sources were also obtained at the
INT, Jacobus Kapteyn (JKT), and Nordic Optical (NOT) Telescopes and these were
used to check for the amount of light typically lost around the narrow slit.
For these AGN,  the average ratio of flux measured from the CCD images to flux
measured from the spectra was 1.2$\pm0.1$ (error on the mean). Objects where
there was a strong galactic contribution to the CCD images were not included in
the calculation of this ratio. Flux measurements from spectra where CCD images
are available were corrected for any difference between spectral and CCD flux
levels. Where CCD images are not available, the flux measured from the spectrum
was increased by the mean factor of 1.2.

\subsubsection{Line parameters}

Due to the redshift coverage of the sample and the limited observed wavelength
range, not all of the five emission lines discussed here (\ie \Ha, \Hb, \oiii,
\mgii\ and \ciii) could be measured for all AGN. Effectively, this produces
five subsamples, one for each line. There are 30 measurements of \Ha, 71
measurements of \Hb\ (plus 13 upper limits), 75 measurements of \oiiifull\
(plus 6 upper limits), 127 measurements of \mgii\ (plus 3 upper limits) and 49
measurements of \ciii\ (plus 4 upper limits). 

Line fluxes, positions, rest-frame equivalent widths (EWs) and FWHM were
measured from the spectrum by fitting one or more Gaussian profiles to each
line. Where the emission line of interest was  blended with other lines or
blends (such as \feii), additional Gaussians were used to model the
contaminants. Where the line of interest could not be well represented by a
single Gaussian profile, additional components were used. The continuum was
modelled as a second-order polynomial in the fits. The FWHM of the lines are
the FWHM of the best-fitting Gaussian model, while fluxes and EWs were measured
directly from the spectrum after the fitted continuum and Gaussian model fluxes
of any other contaminants had been subtracted. The FWHM have been deconvolved
from the instrumental width.

Line luminosities were calculated assuming a value of 50 km s$^{-1}$ Mpc$^{-1}$
for the Hubble constant ($H_0$) and 0 for the deceleration parameter ($q_0$;
these values are assumed throughout this paper). Luminosities of sources for
which CCD data were also available, were explicitly corrected for any light
lost around the slit; the remaining luminosities were increased by the mean
correction factor of 1.2 (see Section 2.2). Line EWs, FWHM and luminosities are
given in Table 1.  

In some cases, two Gaussian components were required to fit the profile of the
permitted lines and data for the separate fits are given in Table 1. For the
optical lines (\ie \Ha\ and \Hb), these have been assumed to represent emission
from the broad and narrow line regions and are indicated `{\sl b}' and `{\sl
n}' respectively in the table. For the UV lines (\mgii\ and \ciii), an
additional  `very broad' underlying component with a FWHM of $\sim$10000~\kms\
was sometimes observed; these are also given in Table 1 and indicated `{\sl v}'
(the lower velocity component for the UV lines is assumed to come from the
`normal' BLR and is tagged `{\sl b}').

A particular problem in the measurement of emission-line parameters is that of
line and continuum blending. Blends of \feii\ can be strong throughout the
optical and UV and especially around \Hb, \oiii\ and \mgii. Modelling of the
\feii\ such as that of Wills, Netzer \& Wills (1985), has been used by previous
authors to both measure the \feii\ blends themselves and to remove the \feii\
features before measuring residual line and continuum features. Although we
have not used such techniques here (as these require data with a very high
signal to noise ratio), obvious blends around the \Hb/\oiii\ and \mgii\ regions
have been removed in the fitting procedure  before measuring the lines. The
\ciii\ line is often blended with \aliii\ and \siiii; Steidel \& Sargent (1991)
found that \aliii\ may contribute up to 12 percent of the flux attributed to
\ciii, while the \siiii\ component is not significant. The quality of the RIXOS
spectra do not permit the reliable deblending of these features and we caution
the reader of possible contamination in the line parameters presented here.

Balmer continuum and optical \feii\ emission can also blend to produce a
``quasi''-continuum at around 3000~\AA\ (Wills \etal 1985) and this may affect
measurements of the optical/UV spectral slope, \aopt. This has been reduced by
removing all absorption and emission features and regions with a very low
signal-to-noise ratio before measuring \aopt. By also using the  broadest
wavelength range available ($\sim$5000~\AA\ in the observer-frame), we have
ensured that contamination in \aopt\ by the \feii/Balmer continuum has been
minimized, although it may still affect AGN with very low-$z$, or with
$z\sim$1.9, where the Balmer continuum region falls at the extremes of the
observed spectra. 

\paragraph{Errors and upper limits}

Errors on the fluxes, EWs and FWHM have been estimated by comparing parameters
from repeated observations of the same AGN (\ie to assess systematic errors)
and by repeating measurements on the same spectrum (to assess uncertainties
induced by the fitting techniques by, for example, the allowed extremes in
continuum level). Typical errors are $\pm$0.3 on the logarithm of the
luminosity, $\pm$0.3 on the logarithm of the EW and $\pm$0.15 on the logarithm
of the FWHM (except at FWHM$<$2000\kms\ which are less certain and have errors
of $\sim\pm$0.25 dex). Parameters with larger errors than these are indicated
by the superscript `$u$' in Table~1 (\ie the value quoted is uncertain).
Possible problems with the subtraction of contaminating \feii\ and/or the
Balmer continuum will add to the  uncertainties. Also, additional errors 
should be considered on the logarithm of the line luminosities when individual
lines are compared with the X-ray parameters as these have been taken at a
different epoch. The magnitude of such errors are difficult to quantify but,
for example, the X-ray spectra of many  AGN typically show flux variability of
$\sim\pm$50 per cent (Mushotzky, Done \& Pounds 1993; Nicholson, Mittaz \&
Mason 1997). While these errors are relatively high, the large number of AGN
should allow a meaningful sample analysis to be made nonetheless.

Upper limits on line fluxes and EWs were calculated from the parameters of a
Gaussian profile which was judged by eye to overestimate the emission from a
`true' line. The position of the Gaussian was fixed at the expected position of
the line; the FWHM for upper limits on permitted lines was fixed to the width
of the strongest permitted line detected in the same spectrum, while for \oiii\
it was fixed at 1000~\kms. Upper limits on EWs and luminosities are indicated
by `$l$' in Table 1; fixed FWHM are indicated by `$f$'.

\subsection{Continuum slopes and luminosities}

The optical and X-ray continuum parameters of many AGN used in this sample have
already been presented in Paper I, to which we refer the reader. Slopes and
luminosities for objects not covered by Paper I are listed here in Table 2, and
have been derived in an identical manner. All power-law indices or slopes,
$\alpha$, have been defined such that $F_\nu\propto\nu^{-\alpha}$.

We shall be describing much of the changes observed in the RIXOS optical and
X-ray spectra in terms of the gradients of the continuum slopes. Different
terminologies are commonly used in the literature for the slopes in the optical
and the X-ray  ranges, thus to avoid any confusion, we have adopted a single
convention for both, \ie using the terms ``soft'' and ``hard''. A ``soft''
slope falls towards high energies and has a relatively high energy index
$\alpha$ (recalling here that the negative sign is implicit in our definition
of $\alpha$); in the optical region, a soft slope corresponds to a ``red''
continuum. A ``hard'' slope {\sl rises} towards high energy and has a low or
negative $\alpha$; this corresponds to a ``blue'' slope in the optical. To
describe changes in slope, we will use the terms ``soften'' (\ie $\alpha$
increasing) and ``harden'' (where $\alpha$ is decreasing).

\subsubsection{Optical}

The logarithms of the UV and optical continuum luminosities in the quasar
rest-frame at 2500~\AA\ and 5000~\AA\ (\luv\ and \lopt\ respectively)  were
calculated from the best-fitting power-law model fits to the spectra.
Luminosities of sources for which CCD data were also available were explicitly
corrected for any light lost around the slit; the remaining luminosities were
increased by the mean correction factor of 1.2 (see Section 2.2).  While \lopt\
is only measured where the rest-frame flux at 5000~\AA\ falls within the
observed wavelength range, \luv\ has been extrapolated where necessary  from
the power-law fits. Taking into account possible variability, light losses
around the slit and errors on the power-law fits,  we estimate that
uncertainties on the logarithm of the optical and UV luminosities are typically 
$\sim\pm$0.3.

\subsubsection{X-ray}

The PSPC counts were divided into three bands [0.1 to 0.4~\keV\  (channels 8 to
41); 0.5 to 0.9~\keV\ (channels 52 to 90) and 0.9 to 2.0~\keV\  (channels 91 to
201)] and these were combined to produce `spectra' with three data points for
each source. The spectra were fitted with single power-law models using the
method described  in Mittaz \etal (1997), which finds the best-fit by
minimizing a Poissonian-based statistic. In the fits, the absorbing column
density was fixed at the Galactic column (\nhgal) measured from the 21~cm
survey of Stark \etal\ (1992). All instrumental effects, including vignetting,
dead-time corrections and particle contamination, were folded into the fitting
process. Mittaz \etal (1997) have demonstrated that this  model
provides a good fit to the data, reproducing the observed X-ray colours to
within 1$\sigma$ in more than 90 percent of cases, and within 3$\sigma$ for
{\sl all} AGN.

\beginfigure*{1} 
\psfig{figure=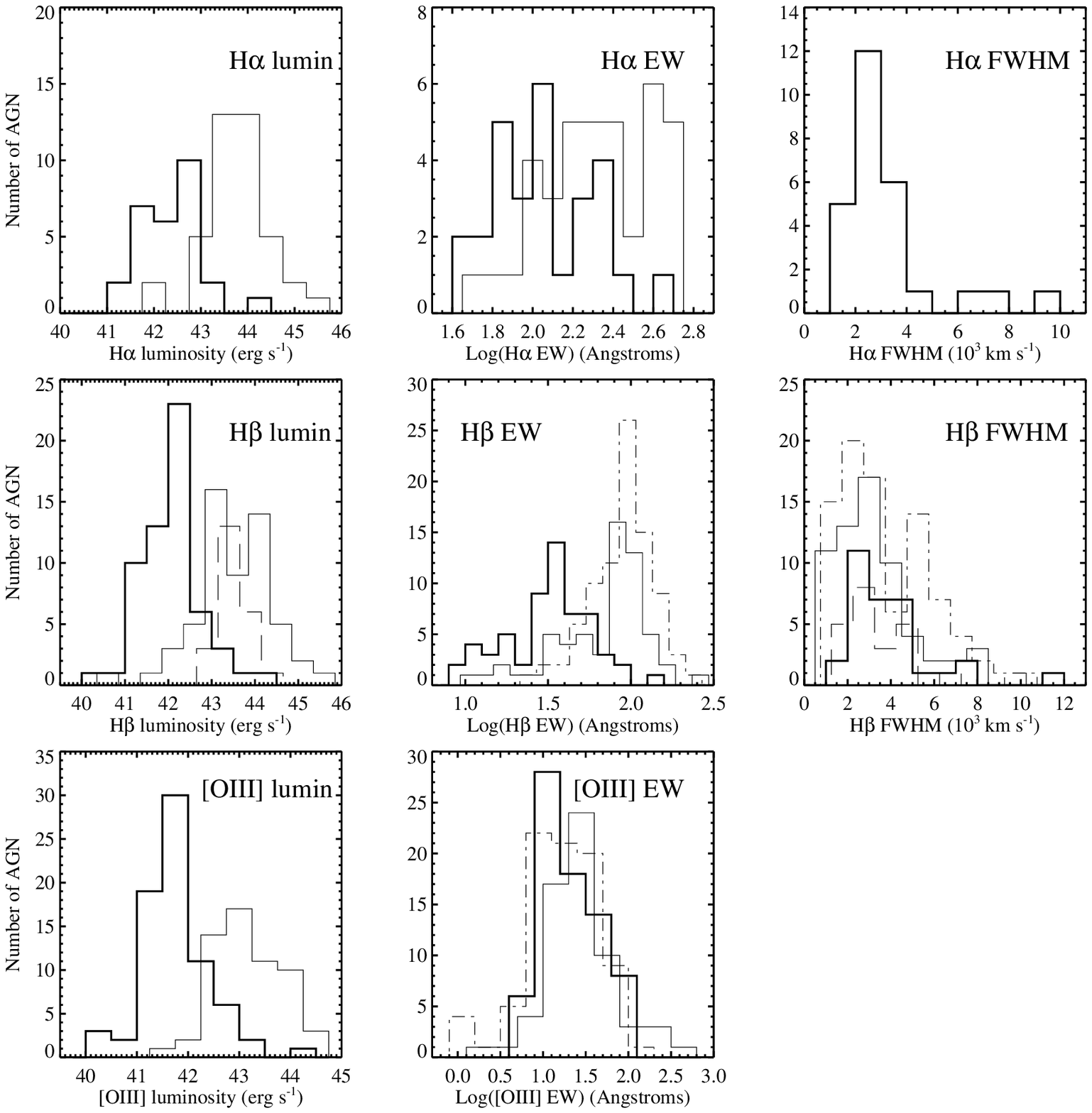,height=7in,width=7in,angle=0}
\caption{{\bf Figure 1.} Number distributions of the optical emission line
parameters for the RIXOS AGN, including luminosities, EWs and FWHM of \Ha, \Hb\
and \oiii. Other samples have also been plotted for comparison: the Stephens
(1989) data (\lha, \haew, \lhb, \hbew, \hbfwhm, \loiii\ and \oiiiew) are drawn
with a thin, solid line; the Boroson \& Green (1992) data (\hbew, \hbfwhm\ and
\oiiiew) are drawn with a dot-dashed line; and the Laor \etal (1997) data
(\lhb\ and \hbfwhm) are plotted with a long-dashed line.}
\endfigure

\beginfigure*{2} 
\psfig{figure=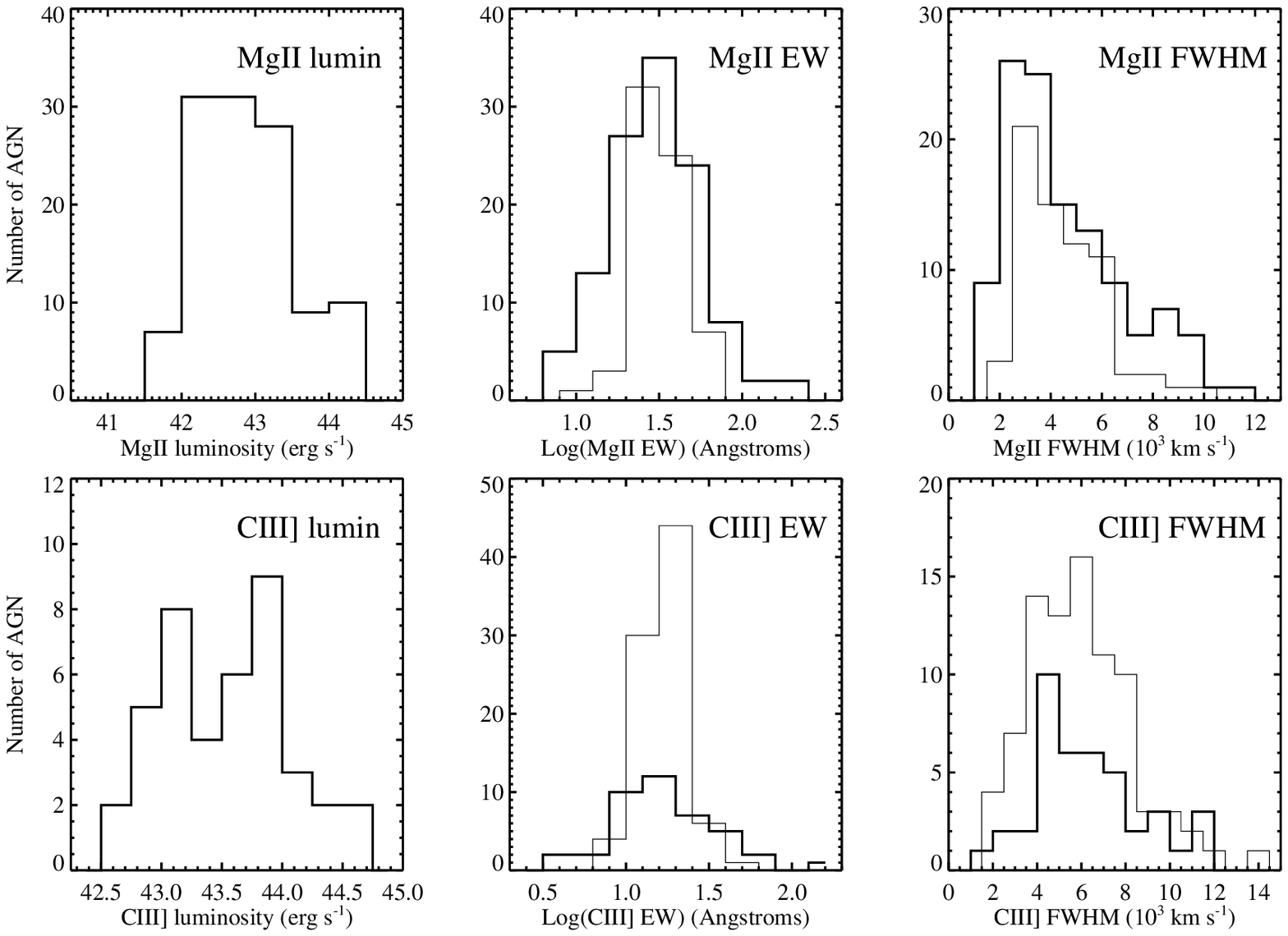,height=5in,width=7in,angle=0}
\caption{{\bf Figure 2.} Number distributions of the UV emission line
parameters for the RIXOS AGN (drawn as a thick histogram), including
luminosities, EWs and FWHM of \mgii\ and \ciii. The data from Steidel \&
Sargent (1991) and Brotherton \etal (1994) are also drawn for comparison as
thin lines.}
\endfigure

\beginfigure*{3} 
\psfig{figure=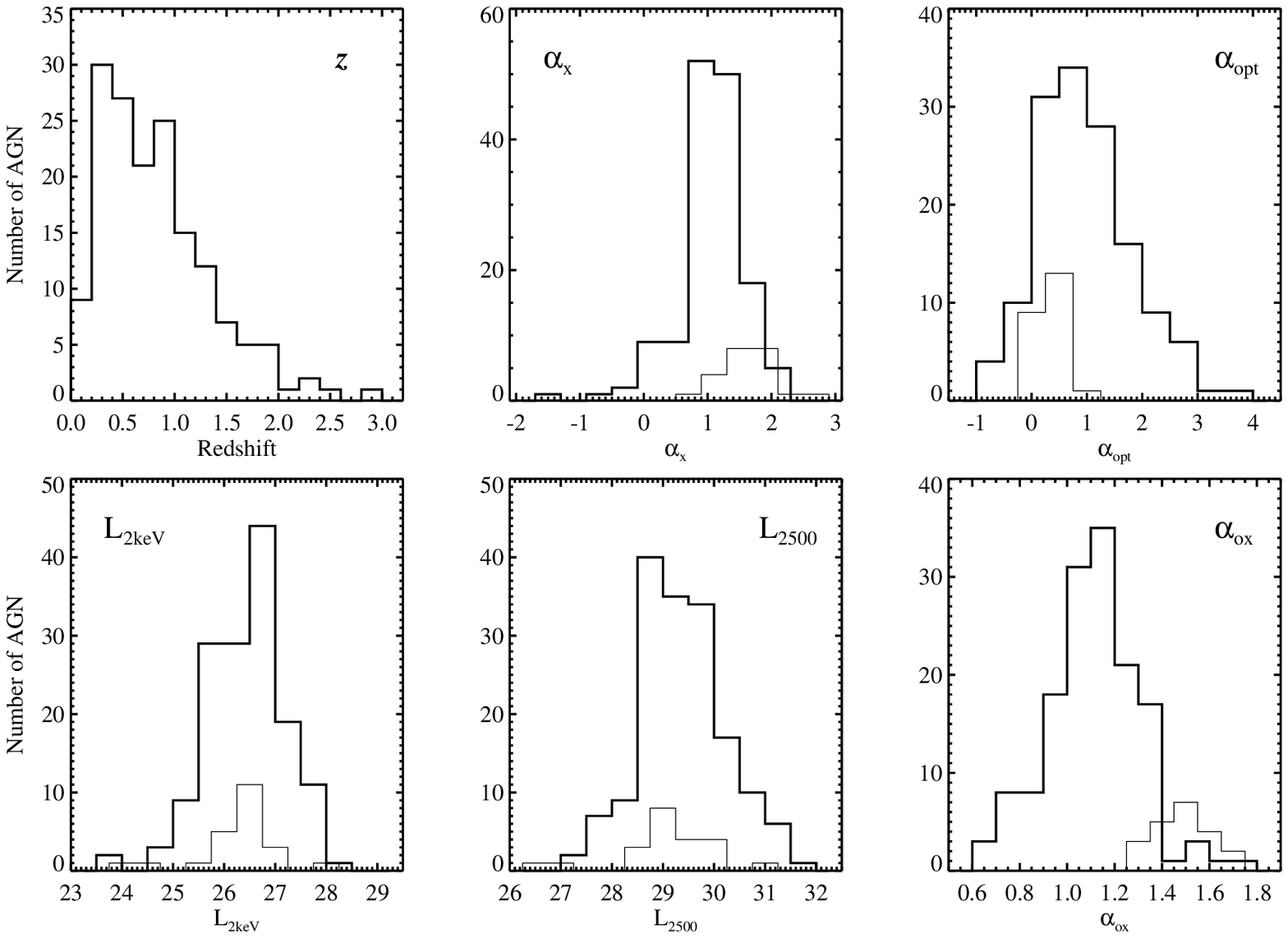,height=5in,width=7in,angle=0}
\caption{{\bf Figure 3.} Number distributions of optical/UV and X-ray
continuum parameters for the RIXOS AGN, including redshift, X-ray and
optical/UV power-law slopes (\ax\ and \aopt), X-ray and UV continuum
luminosities (\lhard\ and \luv) and \aox. Histograms for the RIXOS AGN are
drawn as thick histograms; also plotted as thin histograms  are the
distributions for the Laor \etal (1997) sample.}
\endfigure

The logarithms of the X-ray luminosities at 0.2~\keV\ (\lsoft) and 2~\keV\
(\lhard)  in ergs s$^{-1}$ Hz$^{-1}$ in the rest-frame of the quasar were
calculated using the best-fit power-law model for each individual source.  The
ratios of UV (\ie \luv) to X-ray luminosity at 2.0~\keV\ and  0.2~\keV\
are parameterized using \aox\ (Tananbaum \etal 1979)  and \aos\ (Puchnarewicz
\etal 1992) respectively. X-ray luminosities, power-law indices, \aox, \aos\
and their 90\% errors are given in Table 2 and Paper I. 

\section{Properties of the RIXOS AGN sample}

Number distributions of all optical and UV emission line parameters for the
RIXOS AGN (except the \oiiifwhm\ which are unresolved) are plotted in Figs 1
and 2, while the distributions of redshift and of the optical/UV and X-ray
continuum parameters are plotted in Fig 3. The median, mean, $\sigma$ (\ie
standard deviation on the mean) and error on the mean for all continuum and
emission line parameters are given in Tables 1 and 2. Uncertain parameters (see
Tables 1 and 2) were excluded from the distributions shown and the sample
statistics;  source number 6 in field 273 was also excluded because of its
unusual nature (see section 4.5.1). Upper limits have not been taken into
account. For the permitted lines where two components were fitted to the
profile, only the `broad' components where included in these calculations.
Further details of the continuum properties for the RIXOS AGN may be found in
Paper I.

\subsection{Balmer lines}

The mean \Ha\ EW for the RIXOS AGN is 140$\pm$20~\AA\ (the error quoted is the
standard error on the mean) which is lower than that which we calculate for 
the Stephens (1989) sample of brighter X-ray selected AGN (\ie 233$\pm$23~\AA;
see Fig 1). The mean \Ha\ FWHM of the RIXOS AGN is 3100$\pm$400~\kms. 

The mean RIXOS \Hb\ EW is 38$\pm$3~\AA, which like \Ha, is lower than that
which we calculate for the Stephens (1989) sample (a mean of 77$\pm$7~\AA). It
is also lower than the mean for the optically-selected  Boroson \& Green (1992;
hereafter BG) sample; 96$\pm$4~\AA\ (see Fig 1). Goodrich (1989) also noted
that the \Hb\ EWs of X-ray selected AGN were lower than those of
optically-selected objects  and suggested that this was related to the X-ray
flux.  

The mean \Hb\ FWHM is 3900$\pm$400~\kms\ (excluding objects with poorly
determined broad components; these are indicated with the superscript $u$ in
Table 1). This is similar to the mean for the BG sample, which we calculate to
be  3800$\pm$200~\kms\ (see Fig 1), and suggests that although the \Hb\
emission is weak relative to the optical continuum, the velocity of the BLRs in
the RIXOS AGN may be comparable to those of optically-selected AGN (although
the \hbfwhm\ may be lowered by the effects of absorption in the RIXOS AGN; see
Section 4.5). The mean for the Stephens sample is 3200$\pm$200~\kms, which is
also similar to that of the RIXOS AGN. 

\subsection{\oiii}

The mean EW of the \oiii\ emission in the RIXOS AGN is 29$\pm$3~\AA; this is
similar to BG and slightly lower than Stephens which have mean \oiii\ EWs of
24$\pm$3~\AA\ and 44$\pm$8~\AA\ respectively. All three samples are compared in
Fig 1 and show very similar distributions. (The \oiii\ lines were unresolved in
most cases thus their FWHM distribution and mean properties have not been
given.)

The similarity of the RIXOS AGN \oiiiew\ distribution to the BG sample (which
are UV-excess selected and thus presumably relatively unobscured) is a
surprising result, if indeed the RIXOS sample suffers the effects of dust
absorption (Paper I). If the absorption occurs between the BLR and the NLR,
perhaps in a molecular torus, then while the \oiii\ luminosity would be
unaffected, the optical continuum would be absorbed, leading to generally
higher values of the \oiiiew\ in absorbed objects (see also Section 5.3).  The
presence of an absorbing medium between the BLR and NLR which modifies the
optical continuum, was also suggested by Baker (1997) for her sample of
radio-loud quasars. However, relatively high \oiii\ EWs are not seen in the
absorbed (\ie RIXOS) objects when compared with the unobscured (\ie BG
quasars).

We have estimated the effect that absorption would have on the \oiiiew\
distribution of the RIXOS AGN, using the X-ray continuum slope as a
`measurement' of the cold gas absorption and assuming a Galactic gas-to-dust
ratio [\ie where  an E(B-V) of 1 corresponds to \nh=6$\times10^{21}$ cm$^{-2}$:
Ryter, Cesarsky \& Audouze 1975; Gorenstein 1975].  Using the \ax\ and redshift
of each individual AGN, the cold  gas column density was interpolated from the
model curves shown in Figure 17 of Paper I. By assuming that all objects have
an \oiiiew\ of 25~\AA\ when unaffected by absorption (\ie similar to the BG
mean), then using the reddening curves of Cardelli, Clayton \& Mathis (1989),
we find that the mean \oiiiew\ of the RIXOS AGN would be $\sim$63~\AA;
significantly higher than that which we observe.

\begintable*{7} 
\caption{{\bf Table 3.} Emission line and continuum correlation probabilities [($r_S$), \pcorr]}
\halign{        #\hfil 
     &\quad\hfil(#) &\hfil#\quad
     &\quad\hfil(#) &\hfil#\quad
     &\quad\hfil(#) &\hfil#\quad
     &\quad\hfil(#) &\hfil#\quad
     &\quad\hfil(#) &\hfil#\quad
     &\quad\hfil(#) &\hfil#\quad
     &\quad\hfil(#) &\hfil#\quad
     &\quad\hfil(#) &\hfil#\quad \cr
          & \multispan2{\hfil\lopt\hfil} & \multispan2{\hfil\luv\hfil}  
          & \multispan2{\hfil\lsoft\hfil} & \multispan2{\hfil\lhard\hfil} 
          & \multispan2{\hfil\aopt\hfil} & \multispan2{\hfil\ax\hfil}   
          & \multispan2{\hfil\aos\hfil}   & \multispan2{\hfil\aox\hfil}    \cr
          & $r_S$& \% & $r_S$& \% & $r_S$& \% & $r_S$& \% 
          & $r_S$& \% & $r_S$& \% & $r_S$& \% & $r_S$& \% \cr
\noalign{\medskip} 
\lha\      &  0.9& $>$99 &  0.9& $>$99 &  0.4&    59 &  0.5&    88   & -0.5&  84 &  0.1&     1 &  0.2&     1 &  0.0&     1 \cr
\lhb\      &  0.9& $>$99 &  0.8& $>$99 &  0.4&    97 &  0.5& $>$99   & -0.4&  98 &  0.1&     1 &  0.1&     1 &  0.2&    36 \cr
\loiii\    &  0.7& $>$99 &  0.6& $>$99 &  0.2&    39 &  0.5& $>$99   & -0.4&  98 & -0.1&     1 &  0.2&     3 &  0.1&     1 \cr
\lmgii\    &  0.7& $>$99 &  0.7& $>$99 &  0.3&    97 &  0.5& $>$99   & -0.1&   1 & -0.1&     1 &  0.3& $>$99 &  0.3&    96 \cr
\lciii\    &  ---&   --- &  0.9& $>$99 &  0.3&    68 &  0.5&    98   &  0.2&   1 &  0.0&     1 &  0.4&    91 &  0.6& $>$99 \cr
\noalign{\medskip} 
\haew\     &  0.4&   62  &  0.4&    59 &  0.4&    54 &  0.7& $>$99   & -0.2&     1 & -0.1&     1 & -0.1&     1 & -0.5&    84 \cr
\hbew\     &  0.3&   47  &  0.4&    94 &  0.3&    91 &  0.3&    55   & -0.4&    88 &  0.1&     1 &  0.0&     1 &  0.1&     1 \cr
\oiiiew\   & -0.1&    1  & -0.1&     1 & -0.1&     1 &  0.2&     1   &  0.0&     1 & -0.2&    16 & -0.1&     1 & -0.3&    95 \cr
\mgiiew\   & -0.4&   95  & -0.5& $>$99 & -0.4& $>$99 & -0.2&    79   &  0.1&     1 & -0.2&    89 & -0.2&    56 & -0.5& $>$99 \cr
\ciiiew\   &  ---&  ---  & -0.5&    98 & -0.3&    34 & -0.2&     1   &  0.3&    52 & -0.1&     1 & -0.2&     1 & -0.4&    89 \cr
\noalign{\medskip}                                                
\hafwhm\   &  0.1&     1 &  0.1&     1 & -0.1&     1 &  0.4&    53   &  0.1&   1 & -0.5&    91 &  0.3&    8 & -0.2&     1 \cr
\hbfwhm\   &  0.3&    19 &  0.5&    94 &  0.2&     1 &  0.4&    67   & -0.6&  99 & -0.2&     1 &  0.1&     1 &  0.1&     1 \cr
%\oiiifwhm\ &  0.4& $>$99 &  0.3&    93 &  0.1&     1 &  0.1&     1   &  0.1&   1 &  0.1&     1 &  0.1&     1 &  0.3&    65 \cr
\mgiifwhm\ & -0.3&   74  &  0.0&     1 &  0.0&     1 &  0.1&     1   &  0.1&   1 & -0.2&    52 &  0.0&     1 & -0.2&    67 \cr
\ciiifwhm\ &  ---&   --- &  0.0&     1 &  0.3&    26 &  0.0&     1   & -0.1&   1 &  0.3&    38 & -0.1&     1 &  0.1&     1 \cr
} 
\tabletext{The two quantities given for each correlation are the Spearman
correlation coefficient ($r_S$, shown in brackets) and \pcorr. Correlation
coefficients between two different luminosities were calculated after having
corrected for redshift (see Section 4). Percentages are not given where there
were fewer than 7 data pairs. The \oiii\ lines are unresolved thus correlations
of the \oiiifwhm\ with other data have not been made.}
\endtable 

This cannot be due to a relatively strong optical continuum in the RIXOS AGN;
the mean absolute V-band magnitude of the BG sample is --24.2$\pm$0.2,  whereas
for the RIXOS objects with measured \oiii, the mean is --21.3$\pm$0.2, \ie an
order of magnitude fainter. Instead it may be that the \oiii\ line luminosity 
is itself relatively low in the RIXOS AGN. Since the \Hb\ and \Ha\ EWs also
appear to be weak relative to the BG sample (Section 3.1; EWs of lines emitted
from within the absorber will be unaffected if the continuum also comes from
within), this suggests  either that the ionizing flux on the Balmer line region
and the NLR is relatively low, or that the covering factors of these regions
are low.

\subsection{\mgii\ and \ciii}

The mean \mgii\ EWs and FWHM for the RIXOS AGN are 37$\pm$3~\AA\ and
4600$\pm$200~\kms\ respectively. The mean EW is similar to that which we
calculate for the Steidel \& Sargent (1991; hereafter SS91) sample of 
radio-loud and radio-quiet objects (34$\pm$2~\AA). The mean FWHM is similar to
that of Brotherton \etal (1994; hereafter B94) who found a mean of
4440$\pm$200~\kms\ (the B94 sample is drawn from SS91). Thus there appears to
be no overall differences in the EWs and FWHM of the \mgii\ line, compared with
the general AGN population (this is also borne out in Fig 2). 

The mean \ciii\ EW for the RIXOS AGN is 22$\pm$3\AA;  this is similar to those
of the Green (1996) sample (16.5$\pm$1.9~\AA) and the SS91 sample
(18.2$\pm$0.7\AA; see Fig 2).  The mean FWHM of the \ciii\ emission in the
RIXOS AGN is 6300$\pm$400~\kms\ which is similar to that of the radio-quiet
objects in the B94 sample (6880$\pm$340~\kms; the mean \ciii\ FWHM of the 
radio-loud AGN in the B94 sample was significantly lower, \ie
4990$\pm$310~\kms) and to the UV-bright sample of Green (1996) where the mean
\ciii\ FWHM was 6340$\pm$520~\kms. Like \mgii\ then, the \ciii\ strengths and
widths are similar to the AGN population as a whole.

\section{Correlations}

In this section, we search for relationships between the various line and
continuum parameters, using the Spearman rank-order correlation coefficient
which is a non-parametric method insensitive to outlying points. Results of the
correlation analysis between line parameters and continuum slopes and
luminosities are given in Table 3, while correlations between line EWs,
luminosities and FWHM are listed in Tables 4a and b. Correlations for the
continuum parameters only, are presented in Paper I; although we have used a
slightly different sample of objects to that in Paper I, the correlations
between the continuum parameters for the AGN in Paper I are consistent with the
sample used here. 

\begintable*{8}
\caption{{\bf Table 4a} Emission line correlation probabilities; luminosities
and EWs correlated with luminosities, EWs and FWHM [($r_S$), \pcorr].}
\halign{        #\hfil 
     &\quad\hfil(#) &\hfil#
     &\hfil(#) &\hfil#
     &\hfil(#) &\hfil#
     &\hfil(#) &\hfil#
     &\hfil(#) &\hfil#\quad
     &\quad\hfil(#) &\hfil#
     &\hfil(#) &\hfil#
     &\hfil(#) &\hfil#
     &\hfil(#) &\hfil#
     &\hfil(#) &\hfil# \cr
\noalign{\bigskip}
 & \multispan2{\hfil\lha\hfil}
 & \multispan2{\hfil\lhb\hfil}
 & \multispan2{\hfil\loiii\hfil}
 & \multispan2{\hfil\lmgii\hfil}
 & \multispan2{\hfil\lciii\hfil}
 & \multispan2{\hfil\haew\hfil}
 & \multispan2{\hfil\hbew\hfil}
 & \multispan2{\hfil\oiiiew\hfil}
 & \multispan2{\hfil\mgiiew\hfil}
 & \multispan2{\hfil\ciiiew\hfil}\cr
          & $r_S$& \% & $r_S$& \% & $r_S$& \% & $r_S$& \% 
          & $r_S$& \% & $r_S$& \% & $r_S$& \% & $r_S$& \% 
          & $r_S$& \% & $r_S$& \% \cr
\noalign{\medskip} 
\lhb    & 0.8& $>$99  \cr
\loiii\ & 0.8& $>$99 &  0.6& $>$99 \cr
\lmgii\ & 0.6&    57 &  0.8& $>$99 & 0.6&  $>$99 \cr
\lciii\ & ---&   --- &  ---&   --- & ---&    --- &  0.8& $>$99 \cr
\noalign{\medskip} 
\haew\      & 0.7& $>$99 &   0.5&    87 &   0.6&    98 &   0.1&  17 &   ---& --- \cr 
\hbew\      & 0.4&    67 &   0.7& $>$99 &   0.3&    61 &   0.3&  25 &   ---& --- & 0.5&  91 \cr
\oiiiew\    & 0.1&     1 &  -0.1&     1 &   0.5& $>$99 &   0.0&   1 &   ---& --- & 0.4&  84  &   0.1&   1 \cr
\mgiiew\    & 0.6&    62 &  -0.2&     1 &  -0.1&     1 &  -0.1&   1 &  -0.3& 69 & 0.6&  62  &  -0.2&   1 &  0.2&  1 \cr
\ciiiew\    & ---&   --- &   ---&   --- &   ---&   --- &  -0.3&  41 &  -0.1&  1 & ---& ---  &   ---& --- &  ---& --- & 0.5&  96 \cr
\noalign{\medskip} 
\hafwhm\    & 0.3&   7 &   0.2&   1 &   0.0&   1 &  -0.7&  75 &   ---& ---  & 0.5&  86  &   0.1&  1 &  -0.1&   1 & 0.2&    21 & ---& --- \cr
\hbfwhm\    & 0.9&  99 &   0.4&  86 &  0.1&   1 &   0.3&   9 &   ---& --- & 0.8&  99  &   0.3& 57 &  -0.2&   1 & 0.5&    79 & ---& --- \cr
\mgiifwhm\ & -0.4&  37 &   0.0&   1 &  -0.1&   1 &   0.3&  96 &   0.0&  1  & ---& ---  &   0.0&  1 &   0.0&   1 & 0.6& $>$99 & 0.1& 1 \cr
\ciiifwhm\  & ---& --- &   ---& --- &   ---& --- &   0.1&   1 &   0.2&  1 & ---& ---  &   ---& --- &  ---& --- & 0.2&    12 & 0.2& 1 \cr
%\oiiifwhm\  & 0.6&  98 &   0.3&  61 &   0.2&  57 &   0.3&  68 &   ---& --- 
}
\tabletext{Coefficients and probabilities between two different luminosities
have been calculated after correction for redshift (see Section 4).
Percentages are not given where there
were fewer than 7 data pairs.}
\endtable

\begintable{9}
\caption{{\bf Table 4b} Emission line correlation probabilities;\break FWHM
correlated with FWHM [($r_S$), \pcorr].}
\halign{        #\hfil\quad 
     &\quad\hfil(#) &\hfil#\quad
     &\quad\hfil(#) &\hfil#\quad
     &\quad\hfil(#) &\hfil#\quad \cr
\noalign{\bigskip}
 & \multispan2{\hfil\hafwhm\hfil}
 & \multispan2{\hfil\hbfwhm\hfil}
 & \multispan2{\hfil\mgiifwhm\hfil}\cr
          & $r_S$& \% & $r_S$& \% & $r_S$& \% \cr
\noalign{\medskip} 
\hbfwhm\   & 0.8& 99  \cr
\mgiifwhm\ & 0.8& 75  & 0.4& 72 \cr 
\ciiifwhm\ & ---& --- & ---& ---  & 0.2& 13 \cr
}
\tabletext{Percentages are not given where there
were fewer than 7 data pairs.}
\endtable

\beginfigure*{4} 
\psfig{figure=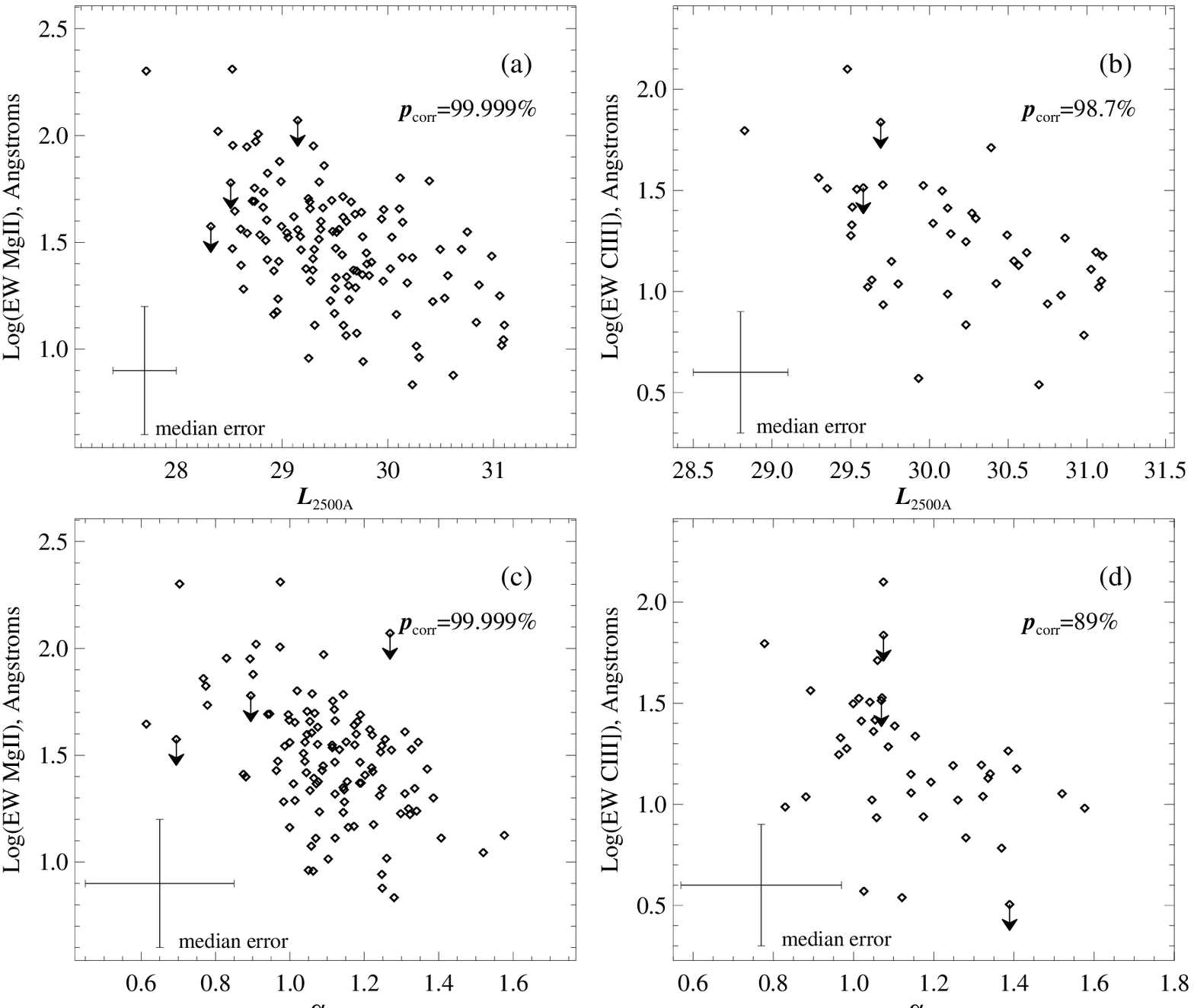,height=6in,width=7in,angle=0}
\caption{{\bf Figure 4.} An illustration of the Baldwin effect in \mgii\ and
\ciii. Arrows indicate upper limits on the line EWs. {\sl (a)} \mgii\ EW
plotted against \luv, and {\sl (b)} \ciii\ EW plotted against \luv. There are
also correlations with \aox; {\sl (c)} \mgii\ EW plotted against \aox, {\sl
(d)} \ciii\ EW plotted against \aox.}  
\endfigure

The `correlation probabilities', \pcorr, which are  given in the following
sections and in the tables, are expressed as percentages and are equal to 100\%
minus the percentage probability of being drawn from a random sample. Neither
uncertain line parameters, upper limits nor data for double quasars have been
included in the correlations.  In general the number of upper limits represents
only 7 per cent of the line parameters, thus their exclusion will have a
small effect on the correlation analysis. The emission line with the largest
fraction of upper limits (15 per cent) is \Hb; in this case the effect of
including upper limits or uncertain values are considered where appropriate.
Also excluded were {\sl (1)} data taken from the literature; {\sl (2)} weak
optical continua or optical continua which were poorly fitted by a simple
power-law model; and {\sl (3)} X-ray data for sources which lay less than 30
arcseconds away from a neighbouring X-ray source.  Where two components are
given for the emission lines in Table 1, only the broad components (\ie those
indicated {\sl b} in Table 1) have been used.

One problem in searching for correlations in flux-limited samples of this kind,
is that luminosities are dominated by the correction for the distance to the
quasar, so that the lines and  continua of low-redshift objects have low
luminosities while parameters of high-$z$ objects tend to have high
luminosities. This usually results in very strong correlations between
luminosities of different parameters (lines and continua)  and may not reflect
a real dependence.  In order to correct for this, when comparing parameters
which are both luminosities, each pair of parameters is divided by a factor
which is appropriate to the redshift of that object. The factor used is the
rest-frame luminosity calculated for a flux of 1$\times10^{-15}$ ergs cm$^{-2}$
s$^{-1}$ measured in the frame of the observer (arbitrarily chosen to match the
observed-flux typically measured in the optical emission lines).

When comparing optical/UV properties with X-ray parameters, it must be borne in
mind that these optical data are not simultaneous with the X-ray observations
and the optical data are not photometric, although efforts have been made to
correct for this (see Section 2.2). While these effects should average out over
the large sample used here, they will introduce an additional degree of scatter
into quantities such as \aox\ which are calculated from both optical and X-ray
spectra. Correlations between separate optical/UV and X-ray parameters may also
have an additional element of scatter for these reasons.

\subsection{Continuum luminosities}

All emission line luminosities are strongly correlated ($>$99 per cent) with
the optical/UV continuum luminosities, \lopt\ and \luv. The \Hb, \oiii\ and
\mgii\ luminosities are also strongly correlated with \lhard\ ($>$99 per cent);
there is a weaker (98 per cent) correlation for \ciii\ but for \Ha, the \pcorr\
with \lhard\ is only 88 per cent. Correlations between the \Hb\ luminosity and
the optical/UV and X-ray continuum luminosities have been reported in previous
samples (\eg Kriss \& Canizares 1982; Blumenthal, Keel \& Miller 1982; Stephens
1989, Laor \etal 1997). 

There are weak correlations between the \Hb\ and \mgii\ luminosities and the
soft X-rays at 0.2~\keV\ (\pcorr=98 per cent). However these may be induced;
for example, looking at the subsample for which \mgii\ has been measured, 
strong correlations are found in \lmgii\ versus \lhard, and in \lmgii\ versus
\luv. For the same subsample, \lsoft\ is also related to \lhard\ and \luv, thus
it is possible that the \lmgii\ versus \lsoft\ correlation is induced. The
correlation between \lhb\ and \lsoft\ may be similarly affected. In contrast,
for AGN with coverage of \Ha, the continuum luminosities \lhard\ and \lopt\
are not correlated with \lsoft; neither is there a correlation between \lha\
and \lsoft. 

Given that the intrinsic spectrum of AGN is expected to peak in the largely
unmeasureable EUV (which provides much of the ionizing photon flux), the
luminosity at 0.2~\keV\ should be most closely related to the strength of the
broad emission lines (except perhaps for \mgii\ which is produced by
0.6-0.8~\keV\ photons; Krolik \& Kallman 1988), yet all of the lines have the
weakest dependence on \lsoft\ (\ie compared to \lopt, \luv\ and \lhard). In
Paper~I we suggested that the RIXOS AGN sample has a large proportion of
moderately absorbed AGN with column densities of the order of
10$^{21}$~cm$^{-2}$. Columns of this size will greatly modify \lsoft\ but will
have a much lesser effect on the optical/UV lines and continuum. Since our
measurement of \lsoft\ has not been corrected for any intrinsic absorption, we
would not then expect it to be a reliable indicator of the strength of the
local ionizing flux and thus the observed line and 0.2~\keV\ luminosities would
tend not to be related.

While there is only a weak dependence of the \Ha\ luminosity on \lhard, there
is a strong correlation between the {\sl EW} of \Ha\ and \lhard, suggesting a
relationship between the strength of \Ha\ relative to the local optical
continuum, and the strength of the hard X-ray power-law. The \mgii\ and \ciii\
EWs are anti-correlated with the local UV continuum luminosity (\ie \luv); this
is the `Baldwin effect' (see Figure 4a-b and Section 4.3). There are also
anti-correlations between the \mgii\ and \ciii\ EWs and \aox\ (Fig 4c-d); these
may be induced by the strong \luv\ versus \aox\ correlation for those objects,
however since the measurements of \aox\ have additional scatter (see Section
4), a primary correlation between the \mgii\ and \ciii\ EWs and \aox\ is also
possible. 

\subsection{Line-line correlations}

All of the line luminosities are strongly correlated with each other (the
correlation between \lha\ and \lmgii\ is relatively weak but is calculated from
only 7 data pairs). In Table 4a, it can be seen that for the \Ha, \Hb\ and
\oiii\ lines, their EWs are strongly correlated with their own luminosities.
This is discussed further in Section 4.3 which addresses the evidence for
Baldwin effects in the RIXOS AGN emission lines. 

\subsubsection{FWHM-EW correlations}

Relationships between the FWHM and EW of AGN emission lines have been reported
previously; a correlation has been observed in \Hb\  (Osterbrock 1977; Gaskell
1985; Osterbrock \& Pogge 1985; Goodrich 1989: we also find a \pcorr=99.8 per
cent correlation in the radio-quiet quasars from the BG sample) while {\sl
anti-}correlations are seen in Ly$\alpha$ and \civ\ (Francis \etal 1992; Wills
\etal 1993). In the case of \Hb, Gaskell (1985) suggested that this is evidence
for radiative acceleration of the line-emitting clouds; the EWs are low because
the cloud density is high (\logne$>10$ cm$^{-3}$: this suppresses Balmer line
emission) and the cloud velocity is low because the denser clouds experience a
weaker radiative acceleration. Osterbrock \& Pogge (1985) suggested instead
that the  lower-velocity Balmer line regions have a smaller covering factor. 

\beginfigure{5} 
\psfig{figure=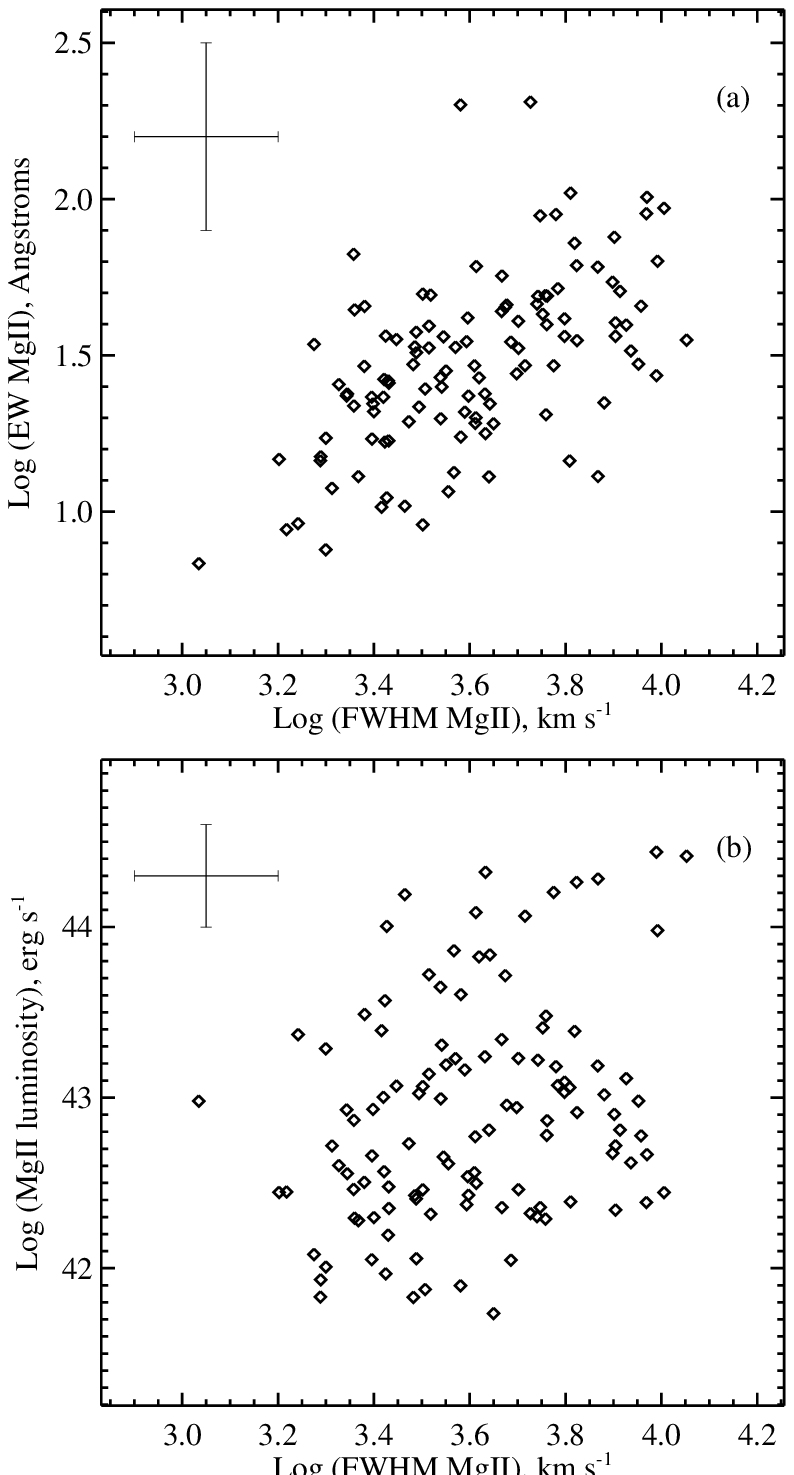,height=6.0in,width=3.3in,angle=0}
\caption{{\bf Figure 5.} The relationships between the \mgii\ parameters; {\sl
(a)} EW plotted as a function of FWHM; {\sl (b)} luminosity plotted against
FWHM. Typical error bars are also shown on the plots.}
\endfigure

For Ly$\alpha$ and \civ, Francis \etal (1992) and Wills \etal (1993)
demonstrated that the anti-correlation is due to the dominance of a changing
core component relative to invariant underlying wings. As the core component
grows, the EW increases while the FWHM falls. Two models have been proposed to
explain this; {\sl (1)} a two phase region of the BLR, \ie a ``very broad line
region'' which emits the broad base of the lines and an ``intermediate width
emission-line region'' from where the line core originates (Francis \etal 1992;
Wills \etal 1993); and {\sl (2)} a `bipolar' model for the BLR where HILs come
from a bipolar outflow and LILs from a disc-shaped region (Wills \etal 1993;
although this latter model is intended to apply to radio-loud objects which
have a well-defined axis).

For the RIXOS AGN, we find a strong {\sl correlation}  between the EW and the
FWHM of \mgii\  (see Fig 5a), \ie in the same sense as that seen in the
low-ionization Balmer lines. A weak (90 per cent) correlation in \mgii\ was
also seen by B94. There is no evidence for any relationship between the FWHM
and EW of \ciii\ in the RIXOS AGN; neither was any found by B94. We find a weak
(94 per cent) correlation between the FWHM and EW of \Ha, although none was
seen in \Hb. The lack of a correlation for \Hb\ may be due to a poor
determination of the intrinsic \Hb\ parameters, again possibly influenced by
the effects of absorption (see also Section 4.5). 

\subsubsection{Other line-line correlations}

For \mgii, in addition to the FWHM-EW relationship, there is also a weaker (96
per cent) correlation between the FWHM and the line luminosity, thus it is
conceivable that these dependences may have been  induced by the improper
subtraction of underlying \feii\ (see Section 2.2.1). However, there is {\sl
no} relationship between the luminosity (or the flux) of the \mgii\ line and
its EW; there is also a large degree of scatter in the \lmgii-FWHM correlation
(see Fig 5b). Neither does the luminosity-FWHM dependence seem to be related to
the Baldwin effect in \mgii; there is no correlation between the \mgiifwhm\ and
\luv. Rather it implies that when the line is broad, it is strong  with respect
to the local continuum and its luminosity tends to be high; firm conclusions
can only be drawn from higher signal-to-noise spectra however. 

B94 have reported  an anti-correlation between the \mgii/\ciii\ flux ratio and
\fwhmciii\ in their sample of 85 quasars, but none between the \mgii/\ciii\
flux ratio and \fwhmmgii. For the RIXOS AGN, we find the opposite, \ie a strong
{\sl correlation} between the \mgii/\ciii\ flux ratio  and \fwhmmgii\ but none
for \fwhmciii. Using the B94 data, we find that the correlation between the
\mgii/\ciii\ flux ratio and \fwhmciii\ is very strong in the radio-loud objects
but not seen in the radio-quiet AGN. Since we expect that the RIXOS AGN are
predominantly radio-quiet, this may account for the discrepancy between the two
samples for \mgii/\ciii\ versus \fwhmciii.  The correlation between this flux
ratio and the \mgii\ FWHM in the RIXOS AGN may be induced by the dependence of
\lmgii\ on \mgiifwhm.

\beginfigure{6} 
\psfig{figure=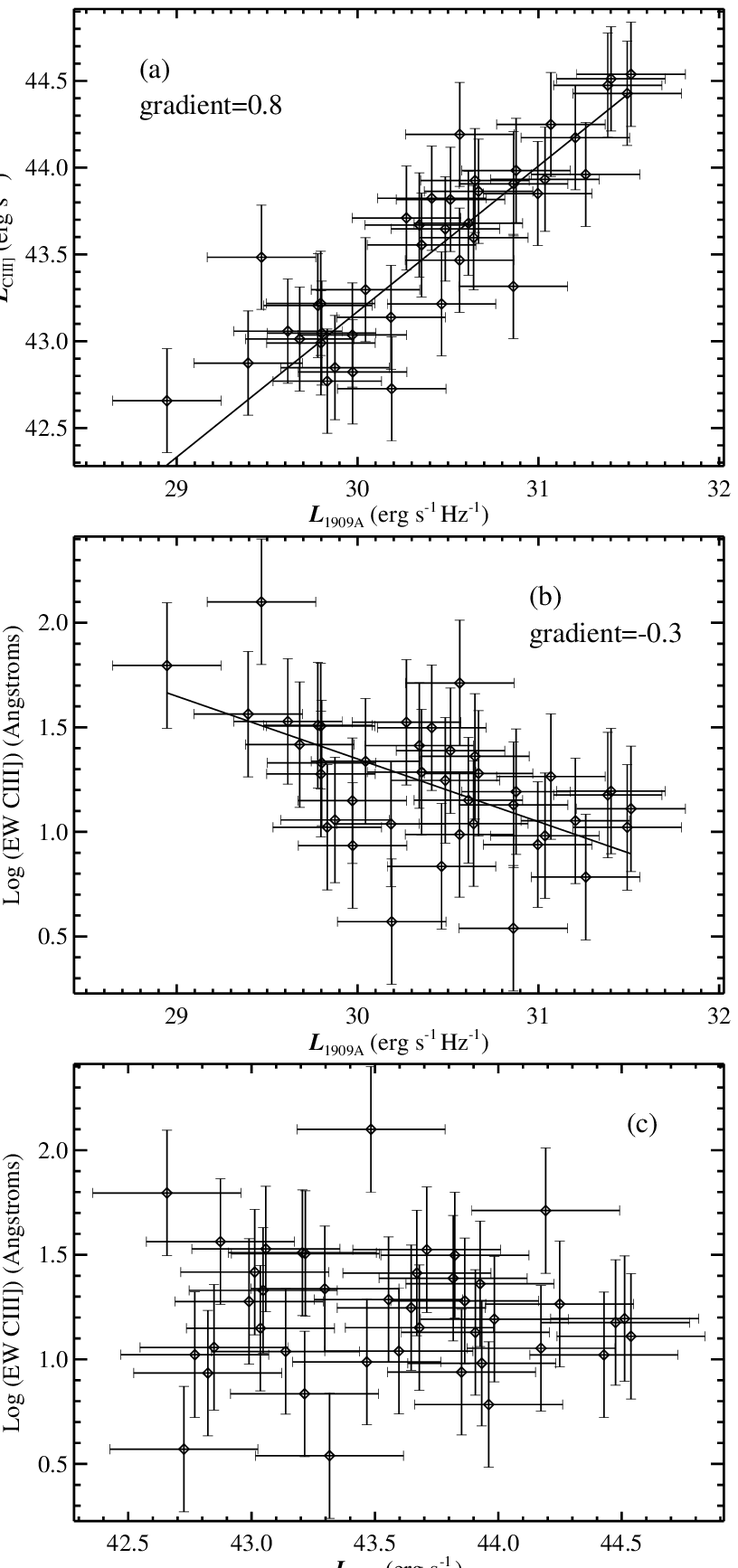,height=7.0in,width=3.3in,angle=0}
\caption{{\bf Figure 6.} A demonstration of the Baldwin effect in \ciii. {\sl
(a)} \ciii\ line luminosity is plotted as a function of the continuum luminosity
at the line centre (\ie at 1909~\AA; $L_{\rm 1909A}$). {\sl (b)} EW plotted
against the 1909~\AA\ continuum luminosity; this is the Baldwin effect. {\sl
(c)} The \ciii\ EW plotted against its own luminosity - there is no significant
correlation between these two parameters.}
\endfigure

\subsection{The Baldwin effect}

The `Baldwin effect' is a term describing the anti-correlation of a line's EW
with the continuum luminosity and was originally discovered in C{\sc
iv}$\lambda$1549 (Baldwin 1977). We find evidence of a Baldwin effect in \mgii\
and \ciii\ (with \pcorr$>$99.9 per cent and 99.3 per cent respectively; see
Figure 4a and b). SS91 also found Baldwin effects in \mgii\ and \ciii; in both
lines the effect was stronger in the radio-loud subsample (\ie when compared to
radio-quiet).  There is no evidence of a Baldwin effect in the optical lines of
the RIXOS AGN (\ie \Ha, \Hb\ and \oiii).

One might expect that the effects of absorption would reduce any Baldwin
effects in a sample, because while the UV continuum luminosity (\ie \luv)
would be reduced by any dust absorption, the EWs would remain unchanged. 
However, for a typical column density expected for the RIXOS AGN ($\sim10^{21}$
cm$^{-2}$; Paper I), the difference in \luv\ is only $\sim$0.4~dex (for a
Galactic dust-to-gas ratio), which is small compared to the scatter on the data
(see Fig 4a). Therefore, it would not be unreasonable to expect the Baldwin
effect to remain significant in a sample of moderately absorbed AGN such as
this.

The Baldwin effect arises because there is a tight correlation between the
luminosity of an emission line and the luminosity of its local continuum. We
can demonstrate this by fitting a linear slope to the line versus continuum
luminosity data (in log-log space) and using the gradient as a measure of the
relative rate of change in the line luminosity. For example in Fig 6a, \lciii\
is plotted as a function of the continuum at the position of \ciii\ (\ie at
1909~\AA). A straight line fitted to these data (taking into account the
errors) gives a gradient of 0.8$\pm$0.1, \ie the line luminosity increases more
slowly than the continuum, so a plot of EW (the ratio of the line to continuum
luminosities) versus continuum data should show an anti-correlation with a
gradient of $\sim$--0.2. Indeed the log(\ciiiew) versus continuum data are
strongly anti-correlated; a straight line fitted to the data has a gradient of
--0.3$\pm$0.1 (Fig 6b). Note in Fig 6c, the lack of a significant correlation
between the \ciiiew\ and the \ciii\ luminosity.

Repeating this for \Ha, we calculate a gradient of 1.3$\pm$0.3 for the \lha\
versus \Ha\ continuum data, \ie the \Ha\ luminosity increases {\sl faster}
than the local continuum. In this case we might expect a correlation between
the \Ha\ EW and the continuum (\ie an `inverse' Baldwin effect) but none is
actually seen. However, there is a  strong correlation between the \Ha\ EW and
the \Ha\ luminosity which suggests that the line luminosity (rather than the
continuum) has the strongest influence on the behaviour of the EW. A similar
situation occurs for \Hb\ and \oiii, \ie although inverse Baldwin effects might
be expected, the line EWs are correlated with the line luminosity rather than
the continuum. 

The Baldwin effect must be possible because the strength of the continuum local
to the line is closely related to that part of the continuum which produces the
line emission. The photons primarily responsible for producing \Ha\ and \Hb\
lie in the 13.6-54.4~\eV\ and 0.3-0.8~\keV\ ranges, for \ciii\ the Lyman
continuum (13.6-24.5~\eV) is most important, while \mgii\ requires photons with
energies between 0.6-0.8~\keV\ (Krolik \& Kallman 1988). Thus the detection of
Baldwin effects in \ciii\ and \mgii\ suggests that the UV continuum is strongly
related to the EUV and soft X-rays, while the lack of Baldwin effects in the
optical lines implies that the optical continuum is less so. This is to be
expected since the UV and soft X-ray continua are probably dominated by the
same component, \ie the big blue bump (BBB; Walter \& Fink 1993; Laor \etal
1994,1997; Fiore \etal 1995; Paper I), while emission from the host galaxy (and
perhaps an optical power-law component) becomes more significant at longer
wavelengths, so that the optical continuum at $\lambda\geq$5000~\AA\  is less
likely to be as strongly related to the strength of the BBB.

We can measure the direct response of one of the emission lines given in this
paper to its ionizing continuum;  \mgii\ is produced primarily by 0.6-0.8~\keV\
photons which fall within the range of the PSPC data. A straight-line fitted to
a plot of \lmgii\ versus the 0.7~\keV\ luminosity (calculated from the
best-fitting power-law models to the X-ray data) has a slope of 1.2$\pm$0.1 \ie
the line luminosity increases faster than that of its {\sl ionizing} continuum,
but more slowly than its {\sl local} continuum. This could be due to the
effects of absorption, removing more of the ionizing continuum flux than the
\mgii\ flux. However, for a Galactic dust-to-gas ratio  and assuming the
dereddening curves of Cardelli \etal (1989), we find that the dust removes {\sl
more} light at 2800~\AA\ [60 per cent for an E(B-V) of 0.17] than an
`equivalent' column of cold gas removes at 0.7~\keV\ (40 per cent for
\nh=10$^{21}$ cm$^{-2}$). It may be that the dust-to-gas ratio is lower than
the Galactic value in the RIXOS AGN, or this may indicate a non-linear
relationship between the strength of \mgii\ and its soft X-ray ionizing
continuum. Measurements of \lmgii\ and the 0.7~\keV\ luminosity in a sample of
unabsorbed AGN would provide clearer evidence of this.

\beginfigure{7} 
\psfig{figure=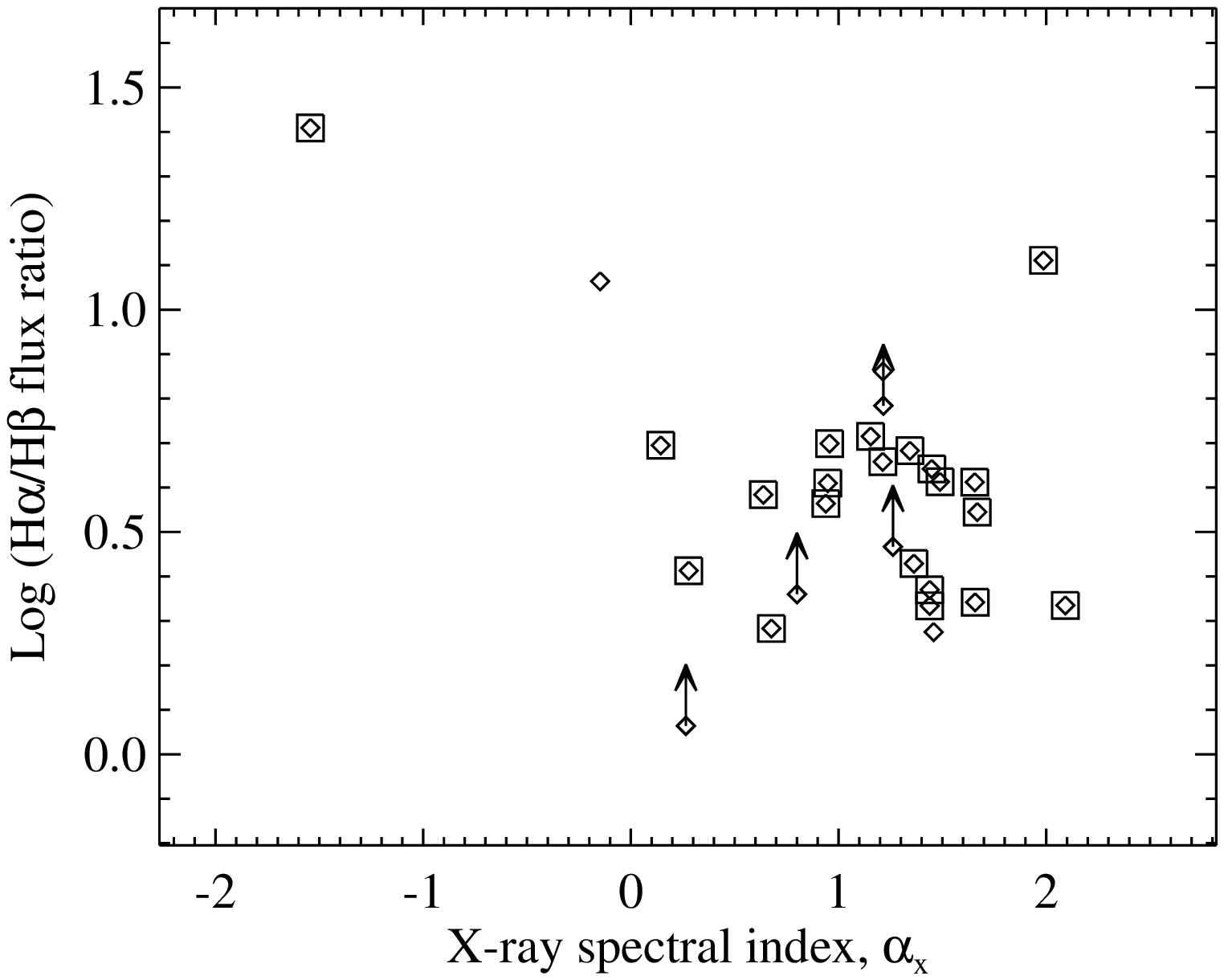,height=2.5in,width=3.2in,angle=0} 
\caption{{\bf Figure 7.} Balmer decrement (\ie \Ha/\Hb\ flux ratio) for the
RIXOS AGN plotted as a function of X-ray spectral index, \ax. All AGN,
including those with uncertain measurements of the \Hb\ line, are plotted as
diamonds. Boxes are drawn around the more secure measurements of \Hb, while the
arrows indicate those objects with upper limits on the \Hb\ flux.}
\endfigure

\subsection{Balmer decrement}

If absorption does have a major influence on properties of the sample, \ie
hardening \ax\ and softening \aopt, then one might also expect to see its
effects on other parameters, for example on the Balmer decrement (as measured
by the \Ha/\Hb\ flux ratio). If it can be assumed that the case B recombination
value for the Balmer decrement always applies, then the {\sl intrinsic} ratio
will be $\sim$3. Any dust extinction along the line of sight will
preferentially  absorb the optical spectrum at shorter wavelengths, thus the
Balmer decrement increases with the amount of dust.  

The Balmer decrements are plotted as a function of the X-ray spectral index,
\ax, in Fig 7. There is no formal correlation between these parameters  using
the Spearman test, although the mean Balmer decrement for the RIXOS AGN is
5$\pm$1 (1$\sigma$ standard deviation; neglecting uncertain values for \Hb),
suggesting some degree of dust absorption present in the sample. Using the
reddening curve of Cardelli \etal  (1989) and assuming a Galactic dust to gas
ratio, we find a typical \nh\ from the RIXOS AGN Balmer decrements of
$\sim0-2\times10^{21}$ cm$^{-2}$ (\ie for those AGN with both \Ha\ and \Hb\
measurements; 23 objects). This is consistent with the continuum modelling of
Paper I, where we found column densities up to $\sim3\times10^{21}$ cm$^{-2}$. 

\beginfigure*{8}  
\psfig{figure=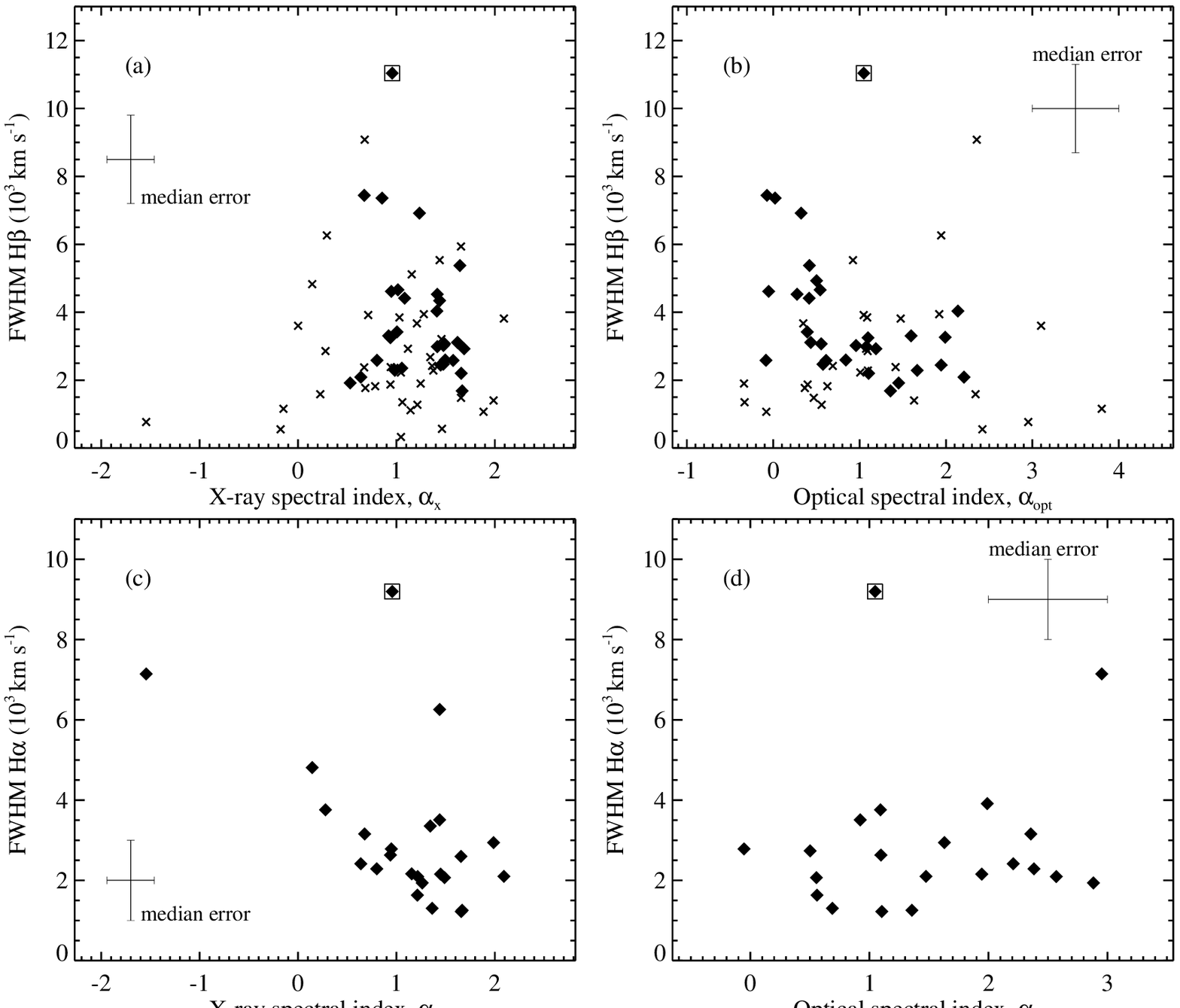,height=6in,width=7in,angle=0} 
\caption{{\bf Figure 8.} Balmer line width plotted as a function of the X-ray 
spectral index, \ax, and optical index, \aopt.  Good data are plotted as
diamonds while AGN with uncertain \Hb\ FWHM are drawn as crosses. A box is
drawn around source F273\_006 (RX~J1042+1212) which exhibits unusually broad,
double-peaked Balmer line emission. {\sl (a)} \Hb\ FWHM versus \ax, {\sl (b)}
\Hb\ FWHM versus \aopt, {\sl (c)} \Ha\ FWHM  versus \ax\ and {\sl (d)} \Ha\
FWHM versus \aopt. }  
\endfigure

\beginfigure*{9} 
\psfig{figure=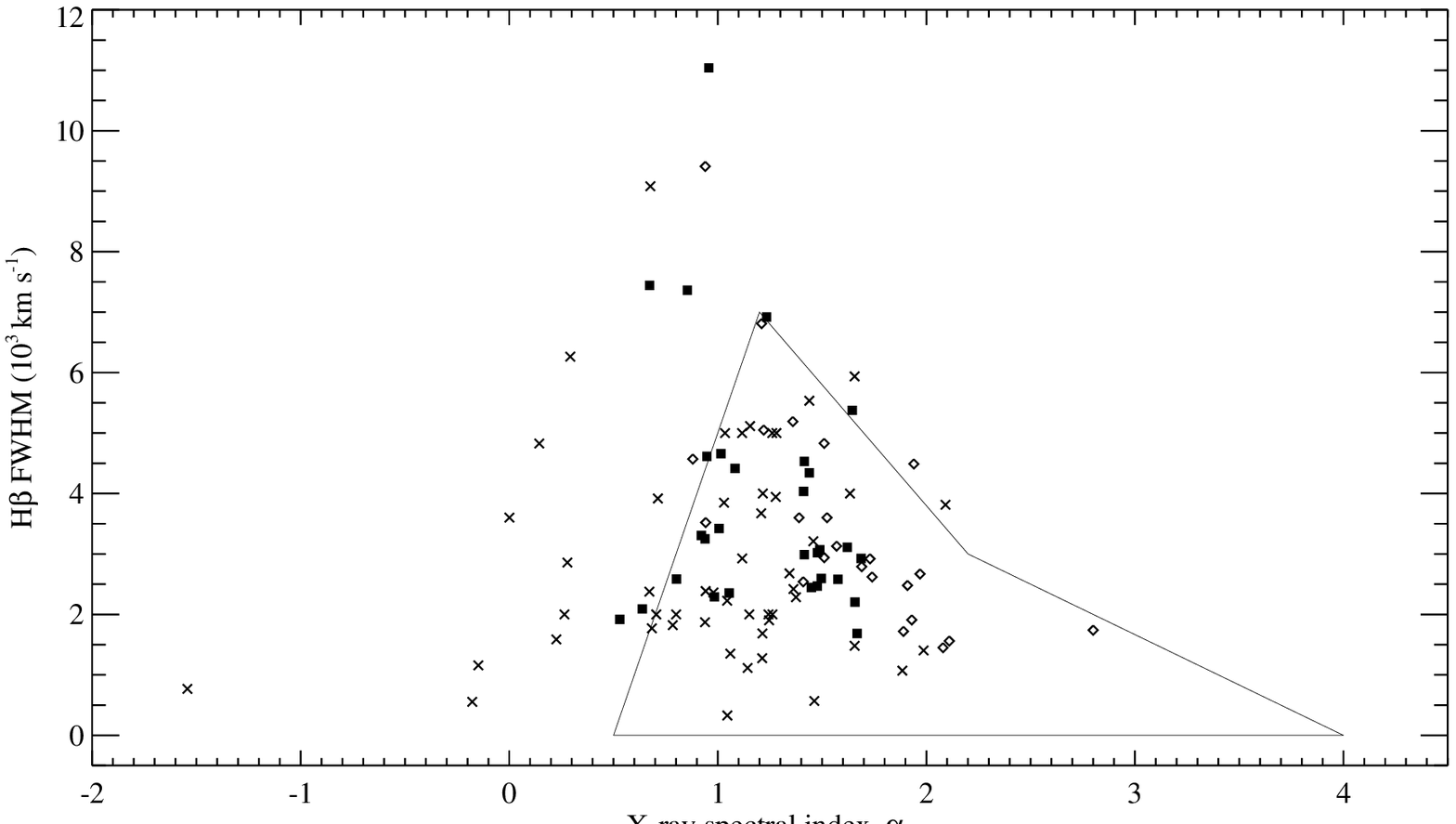,height=4in,width=7in,angle=0}  
\caption{{\bf Figure 9.} \Hb\ FWHM plotted as a function of the X-ray  spectral
index, \ax, for the RIXOS (solid squares), Laor \etal (1997;  diamonds) and BBF
(region enclosed by thin line). RIXOS AGN where the uncertainties on the \Hb\
FWHM are relatively high, are plotted as crosses.} 
\endfigure

\subsection{Continuum slopes}

\subsubsection{Balmer line FWHM}

The relationships between Balmer line FWHM and optical/UV and X-ray slopes  are
plotted in Fig 8. The object with the broadest FWHM which lies away from the
main distributions is source number 6 in field 273 (\ie F273\_006), a
radio-quiet AGN with double-peaked Balmer lines  (Puchnarewicz, Mason \&
Carrera 1996b). The incidence of double-peaked line emission is very rare 
in radio-quiet AGN and may indicate that the nature of this object is quite
different from the general population, \eg it may contain two BLRs each
orbiting its own supermassive black hole in a binary system, or the line
emission may be formed in opposing jets or radiation cones. 

\subsubsection{X-ray spectral slope and Balmer line FWHM}

Although strong relationships between the \Hb\ FWHM and the soft X-ray spectrum
have been observed in other samples of AGN, we have found no similar dependence
in the RIXOS AGN (Fig 8a). Puchnarewicz \etal\ (1992) found a tendency for
ultra-soft X-ray AGN to have narrow Balmer lines, while  Laor \etal (1994,1997)
reported a strong anti-correlation between \ax\ and the FWHM of \Hb\ for their
sample of PG quasars (although they reported none for \Ha). Boller, Brandt \&
Fink (1996; hereafter BBF) showed that narrow-line Seyfert 1s (NLS1s; Seyfert
galaxies whose permitted lines are broader than the forbidden lines, yet have
FWHM of less than 2000~\kms; Osterbrock \& Pogge 1985) tend to have soft X-ray
spectral slopes. They also demonstrated that there seem to be no AGN with both
broad \hbfwhm\ and soft 0.2-2~\keV\ spectra. 

We combine the Laor \etal (1997) and BBF results on \Hb\ with the RIXOS AGN in
Fig 9.  All of the measured FWHM for the RIXOS sample, including `uncertain'
values for the FWHM, are shown.  The RIXOS data extend to harder values of \ax\
but are largely consistent with the BBF sample at softer slopes, \ax$\ga$1. The
Laor \etal\ data are consistent with both samples. Together, these data seem to
confirm a lack of lines with broad FWHM at soft \ax, rather than a true
anti-correlation between the two quantities.  Note that none of the RIXOS AGN
which have an \ax\ harder than 0.5 (where presumably the absorption is
greatest) also had a well-defined \Hb\ line. These objects tend to have weak
\Hb\ lines whose broad components are difficult to discern. An intrinsic
anti-correlation in \Hb\ may thus be diluted by  the effects of absorption,
which suppress the broad component of the \Hb\ line, artificially narrowing the
overall profile (this assumes that the absorption only affects the broad line
component of the emission lines and not the narrow component). 

Searching for an anti-correlation between \ax\ and the Balmer line FWHM in
absorbed AGN may best be made using \Ha\ rather than \Hb, as this is the
stronger line and is modified less by dust. This is demonstrated particularly
well by F278\_010, a $z$=0.09 Seyfert 1.9 (\ax=--1.5$\pm$0.3) which is plotted
in Fig 10. The \Ha\ line is well-defined and its broad component has a FWHM of
7100~\kms, while only the narrow component of \Hb\ can be seen, presumably
because this object suffers from strong dust absorption which has removed the
broad \Hb\ emission. This may happen to a lesser degree in AGN with
intermediate values of \ax, artificially narrowing the \Hb\ line and thus
masking a true correlation. 

\beginfigure{10} 
\psfig{figure=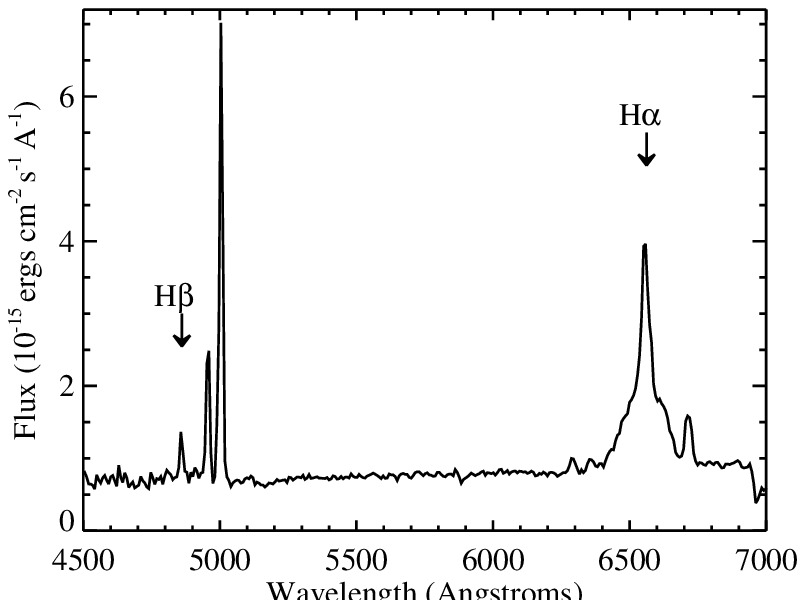,height=2.5in,width=3.3in,angle=0} 
\caption{{\bf Figure 10.} The optical spectrum of F278\_010, a $z$=0.09 Seyfert
1.9 galaxy, plotted in the rest-frame of the AGN. Note the very broad \Ha\ line
and the lack of a broad \Hb\ component.} 
\endfigure

There is some indication (\pcorr=91 per cent) of an anti-correlation between
the FWHM of \Ha\ and the X-ray spectral index, \ax\ (in the sense that  objects
with narrow lines have softer X-ray spectra and vice-versa; see Fig 8c). This
is in the same sense as the correlation between \hbfwhm\ and \ax\ seen in AGN
with soft X-ray spectra by Laor \etal (1994,1997). Removing the three outliers
with \ax$<$--0.5 and \hafwhm$<$5500~\kms, reduces \pcorr\ to  83~percent, while
remaining visually suggestive of an anti-correlation.  Recalling that the slope
of the X-ray spectrum in the RIXOS AGN is thought to be a mean indicator of the
level of the absorption, an inverse dependence  of \Ha\ FWHM on \ax\ would then
suggest that when the absorption is low, so that we have an unobscured line of
sight to the nucleus (and \ax\ is soft), the velocity of the Balmer-line
emitting region is relatively low, and rises as the absorption increases.

\beginfigure*{11} 
\psfig{figure=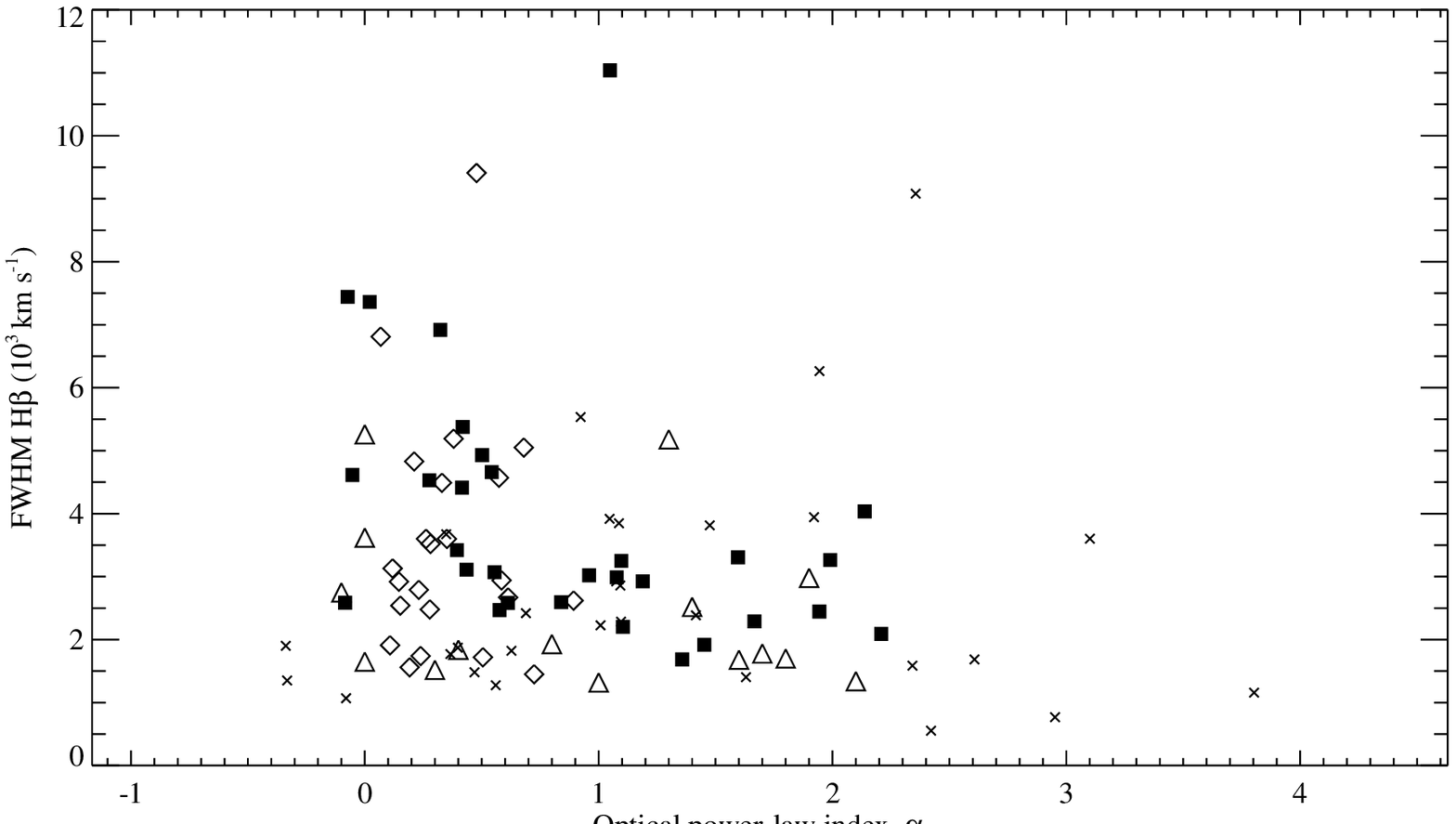,height=4in,width=7in,angle=0} 
\caption{{\bf Figure 11.} \Hb\ FWHM plotted as a function of the optical
power-law index, \aopt, for the RIXOS, Laor \etal (1997) and Puchnarewicz \etal
(1992; \ie the Ultra-Soft Survey, or `USS' AGN) samples. The RIXOS AGN are
drawn as solid squares, the Laor \etal quasars as diamonds and the USS AGN are
plotted as triangles. RIXOS AGN with poor measurements of the \Hb\ FWHM are
plotted as crosses.} 
\endfigure

Absorption in the BBF and Laor \etal (1997) samples is unlikely to be the cause
of this apparent dependence however, as these AGN generally have a much softer
\ax\ and suffer less from the effects of absorption. The relationship between
\hbfwhm\ and \ax\ seen in the BBF and Laor \etal samples is more likely to be
related to the {\sl intrinsic} strength of the soft X-ray component  (see \eg
BBF for a review); the differences between these and the RIXOS sample are
discussed in more detail in Section 5.6.1.

\subsubsection{Optical slope and Balmer line FWHM}

We find a strong correlation (\pcorr=99 per cent) between the \hbfwhm\ and the
slope of the optical continuum, in the sense that AGN with harder optical/UV
spectra have broader line widths, and vice-versa. However, when we combine the
\hbfwhm\ and \aopt\ values from other samples with the RIXOS AGN, the
correlation disappears: see Fig 11 where we compare the RIXOS data with the
Laor \etal (1997) and Puchnarewicz \etal (1992) samples. 

In the optical/UV part of the spectrum, the continuum slope softens (reddens)
as the amount of absorption increases, thus in a situation analagous to that
for \ax, the FWHM of the broad \Hb\ components will be most difficult to
measure in sources with the softest \aopt. This selection effect is reflected
in Fig. 11, which shows an overall lack of broad \Hb\ lines when \aopt\ is
soft. On the strength of the data shown in Fig 11, we conclude that there is
probably no dependence between  \hbfwhm\ and \aopt\ in AGN in general.

\subsubsection{Other line correlations with continuum slope}

There are weak (\pcorr=98 per cent) anti-correlations between the \Hb\ and
\oiii\ luminosities and the slope of the optical continuum (\aopt), although
these may be induced by the \pcorr$>$99.995 per cent  correlation between \aopt\
and \luv\ (\luv\ is also strongly correlated with \lhb\ and \loiii). 

The \mgii\ luminosity shows a correlation with \aos\ and a weaker dependence on
\aox, although these may be induced by strong correlations ($>$99.995 per cent)
between \luv\ and \aos\ and between \luv\ and \aox\ for AGN with measured
\mgii\ emission. This is also true for \lciii.  The \mgii\ EW is strongly
anti-correlated with \aox; see Section 4.1.

\section{Discussion}

Understanding the production of emission lines in AGN is a complex task and
many different factors must be taken into account, including the physical
conditions and geometry of the line-emitting gas (\eg density, temperature,
opacity, distance from the ionizing source, distribution of the gas), the
nature of the incident continuum (\eg intensity, spectral shape and degree of
anisotropy) and the effects of possible orientation-dependence in any assumed
AGN model. 

\subsection{The RIXOS sample}

When considering the properties of the RIXOS AGN in the context of BLR  models,
the exact nature of the sample must be taken into account. The RIXOS AGN are
X-ray selected and drawn from medium-deep ($>$8~ksec) observations made with
the {\sl ROSAT} PSPC, so with this sample we are probing the relatively faint
X-ray emitting population, \ie AGN with an observed flux of $>3\times10^{-14}$
erg s$^{-1}$ (between 0.4 and 2.4~\keV). The use of the {\sl ROSAT} `hard' band
(\ie above 0.4~keV) means that the sample is insensitive to the properties of
any soft X-ray excess (which dominates at lower energies, \eg Pravdo \etal
1981; Arnaud \etal 1985;  Turner \& Pounds 1989; Puchnarewicz \etal 1992). One
consequence of this is that the mean \ax\ for the RIXOS AGN (\ax=1.07$\pm$0.63;
Mittaz \etal, 1997) is significantly harder than that of samples
selected using softer X-rays (\eg for Walter \& Fink 1993, the mean \ax\ is
1.50$\pm$0.48) or by UV-excess methods (for the Laor \etal 1997 sample of PG
quasars, the mean \ax\ is 1.63$\pm$0.07). Another is that AGN whose PSPC
spectra are dominated by a very  soft X-ray excess (see \eg Puchnarewicz \etal
1995; BBF) have been selected against in the RIXOS AGN sample.

By ignoring the spectrum below 0.4~\keV, this sample can also encompass AGN 
with moderate amounts of intrinsic absorption by cold gas (\ie with column
densities, \nh, of a few times 10$^{21}$ cm$^{-2}$).  In Paper I, where we
considered the properties of the optical/UV and X-ray continua of the RIXOS AGN,
we concluded that this is a sample of objects whose intrinsic spectra are
absorbed by cold gas and dust with \nh\ in the range 0 to $\sim3\times 10^{21}$
cm$^{-2}$. This is an important result because it implies that for the sample
overall,  correlations between the optical/UV and X-ray parameters may not
necessarily be indicative of intrinsic dependences  but rather a reflection of
the level of absorption by dust in the optical/UV and by cold gas in the soft
X-rays. 

In the following discussion, we assess the results presented in this paper in
the context of present models for the physical nature of the BLR, bearing in
mind the significance of possible absorption by gas and dust and its effect on
observed correlations. 

\subsection{Models of BLR geometry}

Recently, significant progress in characterizing the geometry of the BLR has
been made  with the reverberation-mapping of the BLRs in nearby Seyferts (see
\eg Peterson 1993 and references therein).  This technique has prompted many
changes to the original `standard' model of the BLR (Davidson \& Netzer 1979;
Kwan \& Krolik 1981), where cold line-emitting clouds are pressure confined by
a hot, intercloud medium and HILs and LILs  are emitted from different parts of
the same cloud. Some of the latest models, which are based on the intensive
NGC~5548 monitoring campaign (Clavel \etal 1991; Peterson \etal 1991), suggest
that the HIL  and LIL  are emitted from spatially distinct regions; an inner,
roughly spherical high-ionization zone and an outer, flattened low-ionization
zone (Krolik \etal 1991; O'Brien, Goad \& Gondhalekar 1994). 

Other models include those of Wills \etal (1993)  who proposed two possible
geometries largely based on FWHM-EW correlations (see Section 4.2.1). The first
is the existence of an ``intermediate width emission-line region'' (ILR) and a
`very broad line region' (VBLR) in addition to the usual NLR; the ILR is
effectively a higher velocity ($\sim$ 2000~\kms), higher density extension of
the NLR. A similar `transition line region' was suggested by van Groningen \&
de Bruyn (1989) based on asymmetries and broad components in the \Hb\ and
\oiii\ lines of bright Seyfert 1 nuclei (see also Mason, Puchnarewicz \& Jones
1996). The VBLR has a high density, lies closer to the central black hole and
has velocities of $\sim$7000~\kms. It emits a relatively invariant broad
component while the ILR produces a more dynamic `core'; differences in the
relative proportions of these features cause the observed correlations.  The
second Wills \etal (1993) model, the `bipolar' model (more appropriate for
radio-loud AGN), invokes HILs from a bipolar outflow while the  LILs come from
a flattened ensemble of dense clouds; in this case the correlations are a
consequence of orientation effects. 

\beginfigure*{12} 
\psfig{figure=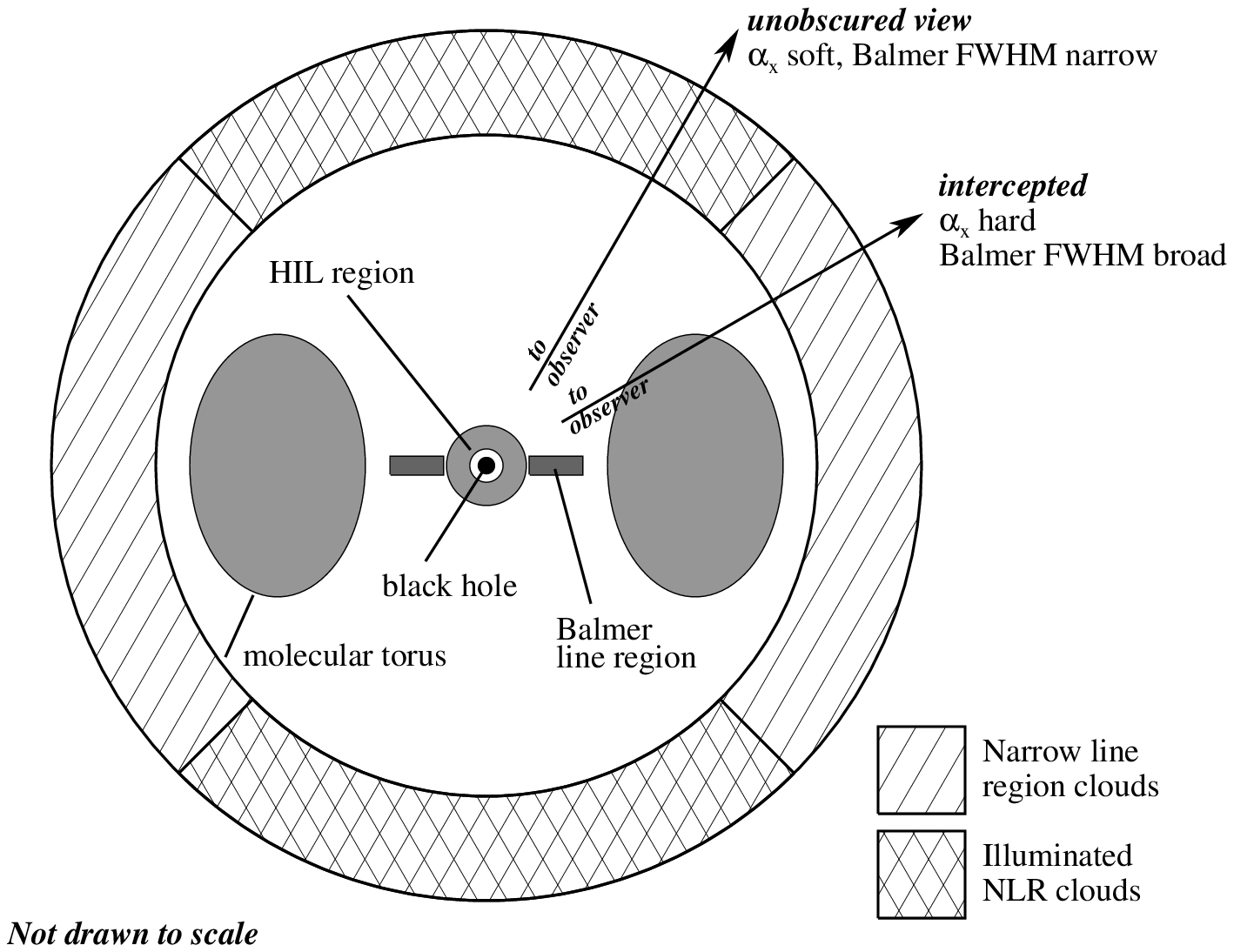,height=4.8in,width=6.0in,angle=0}
\caption{{\bf Figure 12.} A cross-section through an AGN and an illustration
of the possible orientation-dependence in the emission properties. The BLR is
made up of an inner spherical HIL region (which emits \ciii\ and \mgii), and an
outer flattened LIL zone which emits the broad Balmer lines.  While the NLR may
completely surround the inner regions, if the molecular torus is
optically-thick, then only clouds which do not lie in the shadow of the torus
will be illuminated by the nucleus. When an AGN is viewed at angles close to 
face-on so that the nucleus is unobscured, \ax\ is soft and the broad
components of the  Balmer lines are narrow. As the viewing angle increases, the
soft X-rays are absorbed, hardening \ax, while the line-of-sight velocity of
the Balmer line-emitting regions increases (\ie the FWHM increase). \ciii\ and
\mgii, which are emitted from the spherical HIL region,  show no dependence of
FWHM on \ax. }
\endfigure

\subsection{Emission line strengths}

Beginning with the strengths of the optical and UV emission lines, we now
examine the overall properties of the RIXOS AGN with respect to models of AGN
geometry, and taking into account the possible effects of absorption.

In Sections 3.1 and 3.2, we found that the strengths of the Balmer lines
(luminosities and EWs) are low relative to optically-selected AGN, whereas the
EWs of \oiii\ are comparable with optically-selected samples. The EWs  of
\ciii\ and \mgii\ are also typical of other AGN (see Section 3.3). We consider
whether these results may be explained in a geometry similar to the `unified
model' of AGN (\eg Antonucci 1993), a cross-section through which is
illustrated in Figure 12.

\subsubsection{Broad line region}

The broad line regions (including the HIL and broad Balmer line region) lie
within the molecular torus. If they are seen at viewing angles which pass
through the torus, the line and continuum  luminosities will be reduced by the
same factor, so that the EWs should remain unchanged relative to face-on
orientations (this assumes that the optical/UV continuum is also emitted from
within the torus, and that the line and continuum emission is isotropic). 
While this appears to be true for the UV lines (whose EWs are `normal' compared
to other AGN), the Balmer lines have relatively low EWs: this implies  that the
Balmer line fluxes are intrinsically weak. 

The reason for this may be systematic differences in the  shape of the ionizing
continuum incident on the BLR between X-ray and non-X-ray selected AGN.  An
important point is that the photons most directly linked to the production of
\ciii\ and \mgii\ are different from those for \Ha\ and \Hb\ (Krolik \& Kallman
1988; see also Section 4.3). \ciii\ responds to relatively low energy EUV 
photons (13.6-24.5~\eV) and \mgii\ to soft X-rays (600-800~\eV). The Balmer
lines also respond to these energies, but also to photons which reach further
into the EUV at low and high energies (\ie up to 54.4~\eV\ and also from
300-600~\eV), \ie they will be more sensitive to the relative peak flux of the
big bump component. Thus if the big bumps are relatively weak in X-ray selected
AGN, then the \Ha\ and \Hb\ lines will also be weak, but \ciii\ and \mgii\
could be less strongly affected.

\subsubsection{Narrow line region}

While NLR-type gas clouds may completely surround the torus, only those which
lie in the cone of ionizing light defined by the torus will emit the narrow
lines observed (see Fig 12). Furthermore, because the NLR is probably greatly
extended relative to the torus, our viewing angle to the NLR is expected to
have little effect on the observed narrow line flux (but see Baker 1997 for
evidence of possible anisotropies in \oiii\ emission). Thus when an AGN is
observed at viewing angles which pass through the torus, the optical continuum
will be low but the \oiii\ emission should be unaffected, and NLR EWs in
absorbed AGN should be relatively high compared to unobscured samples. Based on
absorption columns for the RIXOS AGN derived from the modelling in Paper I, we
estimate that the mean \oiii\ EW for this sample should be twice that of an
unabsorbed sample of AGN (Section  3.2). However the EWs of the RIXOS AGN are 
{\sl typical} of optically-selected objects. This may be because  the \oiii\
ionizing flux is also relatively low for the RIXOS AGN as discussed in Section
5.3.1. The observed \oiii\ EWs would then be reduced, effectively counteracting
the effects of absorption by the torus.

\subsection{Emission line response to the continuum}

The presence of Baldwin effects in \ciii\ and \mgii\ show {\sl (a)} that the UV
continuum luminosity is related to the EUV/soft X-ray luminosities (\ie the
ionizing continua of these lines); and {\sl (b)} that the luminosities in both
lines increase more slowly than that of their {\sl local} (UV) continua.
Several mechanisms have been proposed as the cause of the Baldwin effect; the
following assume dependences on luminosity, \eg as the luminosity increases
then either {\sl (1)} the ionization parameter (which describes the number of
available ionizing photons per atom or ion) decreases  (Mushotzky \& Ferland
1984); {\sl (2)} the shape of the ionizing continuum softens (Schultz 1982;
Zheng \& Malkan 1993); or {\sl (3)} the BLR cloud density decreases (Rees,
Netzer \& Ferland 1989). Also, Shields, Ferland \& Peterson (1995) proposed
that the non-linear response of the emission lines to the continuum flux is due
to a mixture of optically-thick and -thin clouds in the BLR.

Although the actual {\sl ionizing} continuum of \ciii\ lies in the EUV and thus
cannot be measured directly, the ionizing continuum of \mgii, which lies
between 0.6-0.8~\keV, has been observed for the RIXOS AGN, so that we are able
to determine the response of the \mgii\ to changes in its ionizing flux. A
linear fit to a plot of \lmgii\ versus 0.7~\keV\ luminosity indicates that the
luminosity of the line is increasing {\sl faster} than its ionizing continuum.  
This favours models which propose changes in the physical conditions of the
line-emitting gas, rather than changes in the continuum shape (and a linear
line response) for the cause of the Baldwin effect, at least in \mgii. 

We could find no evidence of a Baldwin effect in the  Balmer lines or \oiii\
(see Section 4.4), suggesting that the luminosity in the EUV/soft X-ray regions
is not strongly correlated with the optical continuum (\ie at
$\lambda\ga$5000~\AA). This may be due to the effects of galactic contamination
which are more significant at longer wavelengths. With the UV line responses,
these results are consistent with the presence of a single component which
dominates throughout the optical/UV to soft X-rays and we identify that
component with the big bump. 

\subsection{FWHM-EW correlations}

Evidence regarding the structure of the BLR may be found from relationships
between the EWs and FWHM of the emission lines. For example, in the RIXOS AGN
we find strong evidence (\pcorr$>$99 per cent) for a correlation between the
FWHM and EW of \mgii. Reports of similar correlations for \Hb\ have also been
made, while {\sl anti-}correlations have been observed in Ly$\alpha$ and \civ\
(see Section 4.2.1). Thus while relationships between FWHM and EW seem to be
typical of several different emission lines,  the actual nature of the
relationship depends on whether the lines are of a high- or low-ionization
species, \ie in general there is an anti-correlation in the HILs and a
correlation in the LILs.

\beginfigure*{13} 
\psfig{figure=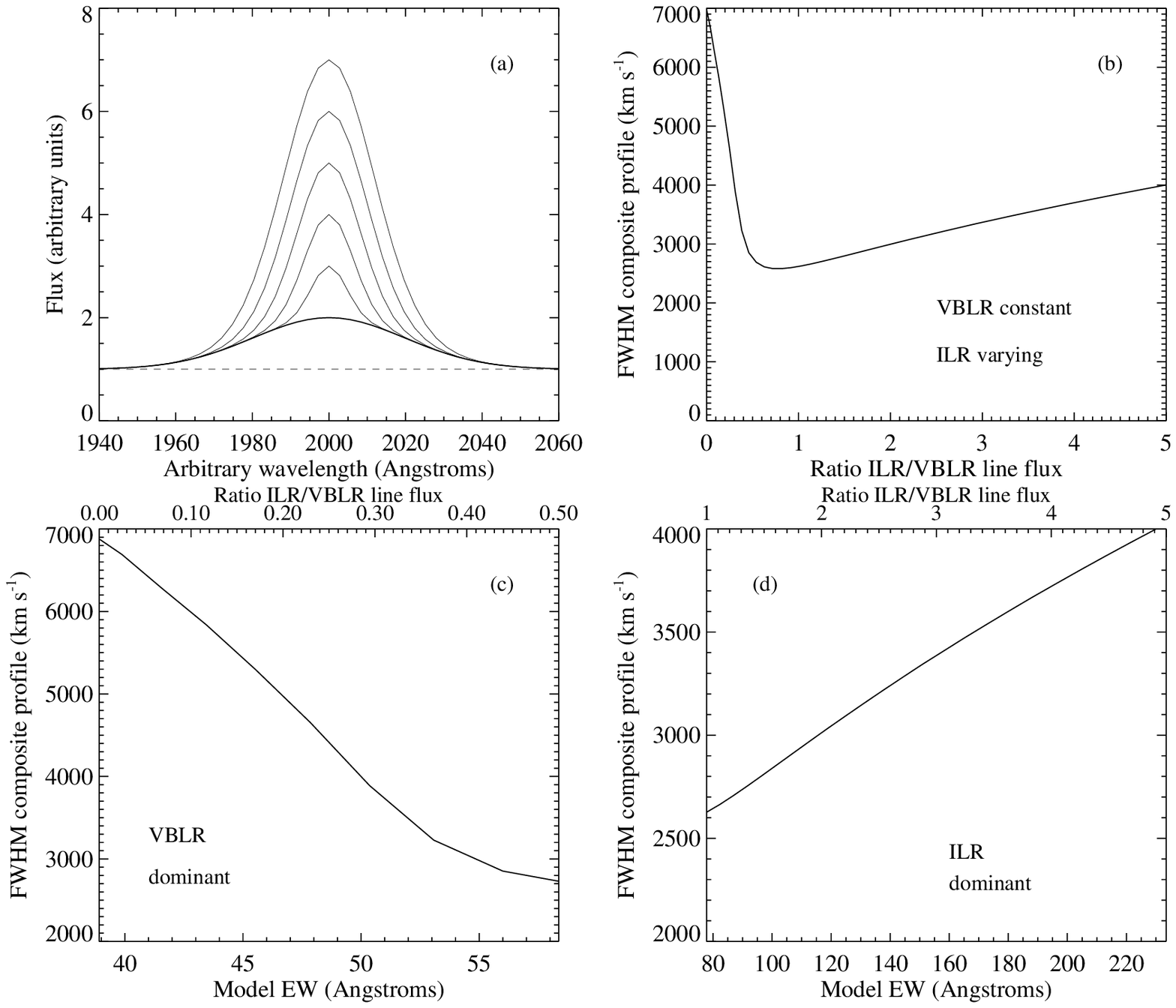,height=6in,width=7in,angle=0}
\caption{{\bf Figure 13.} A model of the emission line profiles for a
two-component VBLR plus ILR representation of a typical AGN emission line. The
model assumes a constant VBLR component and a varying contribution from the
ILR. {\sl (a)} Predicted line profiles for an arbitrary line, where the ILR
peak flux varies from 0 to five times that of the VBLR component. The continuum
is drawn as a dashed line, the profile with zero ILR flux as a thick, solid
line and profiles with increasing ILR contributions as thin solid lines. {\sl
(b)} The model FWHM over the full range of ILR/VBLR ratio (\ie where the {\sl
peak} flux in the ILR component is varied from 0 to 10 times that of the peak
VBLR flux). {\sl (c)}  A close-up view at low ILR/VBLR flux ratios, predicting
an anti-correlation between FWHM and EW in observed data when the VBLR
dominates. {\sl (d)} At high ILR/VBLR flux  ratios, the intrinsic dependence of
ILR FWHM on the ILR line strength dominates and a correlation is observed
between FWHM and EW.}
\endfigure

In the case of the HILs, Wills \etal (1993) proposed that the emission lines
have two components, one relatively constant component from the VBLR, and a
more dynamic `core' from the ILR (a similar model was proposed by Francis \etal
1992). Essentially, as the core increases relative to the broad underlying
component, the EW increases but the line becomes narrower at half maximum,
leading to the observed anti-correlation.  The LILs generally have lower
velocities than the HILs, thus they are more likely to dominate in the ILR. 
However, an opposite dependence is seen in the LILs [\Ha, \mgii\ (this paper;
see Section 4.2.1) and \Hb\ (Osterbrock 1977; Gaskell 1985; Osterbrock \& Pogge
1985; Goodrich 1989)]; this suggests an intrinsic correlation between the EW
and FWHM of lines emitted from the ILR. 

Thus the implications of the various FWHM-EW correlations in the context of the
VBLR/ILR model suggest {\sl (1)} that the HIL emission originates mostly  from
the VBLR with an additional component from the ILR; {\sl (2)} that the LIL
emission is emitted mostly from the ILR; {\sl (3)} that the VBLR component is
relatively invariant and shows no dependence between the FWHM and EW; and {\sl
(4)} that the ILR component is much more dynamic and exhibits an intrinsic
correlation between the velocity of the gas which comprises it and the strength
of the lines it emits (\ie the lines are narrow when their EW is low).

To test whether such a scheme can reproduce the observed dependences in the
HILs and LILs, we have constructed a simple model of an emission line which is
made up of two components (\ie from an ILR and a VBLR), and measured  the FWHM
and EW as the relative contribution of each changes. Each individual component
is represented by a Gaussian. The VBLR component has a FWHM of 7000~\kms\ and
its peak flux is constant. The peak flux in the ILR component varies from 0 to
10 times that of the VBLR; its FWHM increases linearly with the peak flux, from
1000~\kms\ to 4000~\kms, \ie in this model we are assuming an intrinsic
correlation between the strength and the FWHM of a line in the ILR. A composite
model profile is constructed over a range of ILR to VBLR flux ratios and the
FWHM and total flux are measured for each. 

The resulting FWHM of the composite profile as a function of the ILR to VBLR
flux ratio is shown in Fig 13b (the model FWHM is plotted as a function of EW in
Fig 13c-d; for a constant continuum flux, the EW is linearly dependent on  the
ILR/VBLR flux ratio). When the VBLR flux is strong, the changing relative
contribution from the ILR component dominates the change in profile and an
anti-correlation is observed. Then as the ILR flux increases, the  VBLR
component is less important and the underlying correlation between the line
strength and its FWHM is dominant. Thus this model can simultaneously satisfy
the FWHM-EW dependences observed in \civ, Ly$\alpha$, \mgii\ and the Balmer
lines. 

\subsubsection{Line strength and velocity in the ILR}

The correlation between the EW and FWHM of \Hb\ was suggested by Gaskell (1985)
to be due to radiative acceleration of the line-emitting clouds. For this
model, AGN with weak and narrow Balmer lines have relatively high density LIL
clouds. It decouples line FWHM from its dependence on the angle of the line of
sight, and argues against orientation as the cause of the FWHM-\ax\
correlations. However, intermediate-dispersion spectroscopy of the narrow-line
Seyfert 1 galaxy RE~J1034+396 suggests that, for this AGN at least, it is more
likely that the low-velocity line-emitting gas has a relatively  {\sl low},
rather than high, density, when compared with more typical AGN (Mason \etal
1996).

Alternatively, Osterbrock \& Pogge (1985) proposed that the dependence of \Hb\
FWHM on EW may be due to a smaller cloud covering factor (\ie leading to a
lower line flux) when the lines have low velocity. A low-velocity,
weakly-emitting LIL region could be a consequence of a more distant outer BLR
which also has a smaller covering factor. Such a model supports a relationship
between the  relative position of the outer LIL regions and the amount of cold
absorbing gas which lies beyond, \ie that the amount of cold gas is high when
the covering factor of the LIL region is high, and the LIL regions lie closer
to the centre. 

For the orientation-dependent model, a simple explanation for an increase in
line EW as the viewing angle increases, is if the line emission is anisotropic
and beamed back towards the central ionizing source. If the LIL region is
disc-shaped, then the lines will be emitted preferentially along the plane of
the disc, so that as the viewing angle increases, both the FWHM and EW
(assuming the continuum to be relatively isotropic) would also increase,
leading to the observed correlation. 

\subsubsection{Location of the \ciii\ emission}

The \ciiifull\ line is only emitted at relatively low densities (\logne$<$9.5
cm$^{-3}$), yet it is generally very broad;  indeed the \ciii\ line tends to be
broader than \civ, \eg in the B94 sample, the mean \ciii\ to \civ\ FWHM ratio
is 1.2. This suggests {\sl (1)} that any \ciii\ emission from the ILR is
relatively weak (since this has low velocity); and {\sl (2)} that it is emitted
from regions closer to the black hole than \civ.

Although Krolik \etal (1991) concluded that \ciii\ could not be emitted from
the same region as other HILs, O'Brien \etal (1994) found that it was possible
if they allowed for negative responsivity in the BLR (\ie where the emissivity
of a line {\sl decreases} with increasing ionization parameter).  They proposed
that the ionization parameter decreases with radius throughout the HIL region
and that the inner parts (where the ionization parameter is high) are probably
optically thin at the Lyman limit, permitting \ciii\ emission from the HIL
region. If \ciii\ is emitted from the inner regions of the HIL region, this
might explain why its FWHM is generally broader than that of \civ, despite its
lower state of ionization and its low critical density. 

No relationship has been found between the FWHM and EW of \ciii; neither was
any found in the \ciii\ data of B94.  This suggests that there is no intrinsic
FWHM-EW relationship in the VBLR (\ie similar to that which {\sl is} inferred 
for the ILR).

\subsection{FWHM and the X-ray spectral slope}

\subsubsection{Balmer lines}

A tendency for the \Hb\ line to be narrow when \ax\ is soft has been seen in
other samples of AGN (Puchnarewicz \etal 1992; Laor \etal 1994,1997; BBF), yet
none is seen in the RIXOS sample.  However, the RIXOS AGN are probably absorbed
in the optical/UV and soft X-rays, so that  their \Hb\ lines may be
significantly modified by dust absorption in many cases (Section 4.5.2).  This
would mean that their FWHM are no longer a reliable indicator of the intrinsic
velocity of the broad Balmer line regions. The \Ha\ line, which is stronger and
less affected by dust than \Hb, should provide a more accurate measurement of
the velocity of the Balmer line regions in dust-absorbed AGN.  Indeed we find
that the FWHM of \Ha\ does tend to be narrow when the X-ray spectrum is soft,
and vice-versa, \ie in the same sense as that seen for \Hb\ in the unabsorbed
AGN. 

Thus the \Ha\ FWHM-\ax\ dependence in the RIXOS AGN seems to extend the
relationship between Balmer line width and soft X-ray slope to much harder
X-ray slopes (see Fig. 8c), to \ax$\sim$0 and possibly as low as \ax=--2.0
(although only one data point falls below \ax=0, it does extrapolate from the
distribution at softer \ax). However the underlying physical reason for the
trends seen in the comparison samples may be different to that seen in the
RIXOS AGN;  the Puchnarewicz \etal (1992) and BBF samples are both dominated by
AGN with very soft X-ray spectra while the Laor \etal objects are bright, PG
quasars which have relatively strong UV excesses and a mean \ax\ of $\sim$1.6:
these objects are expected to suffer little intrinsic absorption and  their
values of \ax\ are thus more likely to reflect the intrinsic  strength of the
soft X-ray excess. In addition, their \Hb\ profiles should be relatively
unobscured, so that the \Hb\ FWHM versus \ax\ correlation suggests an intrinsic
link between the soft X-ray component and the `true' Balmer line velocity.

For the RIXOS AGN, we have argued that  the range in  \ax\ reflects the degree
of cold gas absorption (Paper 1) and  {\sl not} the changing strength of the
soft component. In other words, despite a change in the physical interpretation
of the slope of the soft X-ray spectrum, the correlation between the soft X-ray
slope and the velocity of the Balmer line regions appears to remain. Any
plausible model for this relationship must incorporate this change.

\subsubsection{UV lines}

There are no correlations between the UV line FWHM and the soft X-ray slope.
Since the \ciii\ and \mgii\ lines are generally dominated by the broad
component and strong enough, despite any absorption,  to permit a reliable
measurement of their FWHM, this probably reflects a `real' lack of dependence.
It suggests that the regions where \ciii\ and \mgii\ dominate are different
from those which emit the Balmer lines. 

\subsection{Orientation dependence}

A dependence of Balmer line width on the X-ray spectral slope for the RIXOS AGN
implies that the Balmer line-emitting regions `know' the amount of  cold gas
which surrounds them. This may be either because the soft X-ray slope has a
direct, physical influence on the Balmer line gas, or because the soft X-ray
slope and Balmer line velocity both depend on a third parameter. 

A `third parameter' upon which Balmer line width and  \ax\ (or equivalently for
the RIXOS AGN, the amount of absorption) may both depend, is the orientation of
the nucleus itself to our line of sight.  Previous studies have shown that the
Balmer lines may be emitted from a flattened, possibly disc-shaped region. If
it can be assumed that the Balmer line clouds are circulating in the disc and
that the FWHM are dominated by their velocity along the line of sight, then the
lines will be narrow when our line of sight is perpendicular to the disc (\ie
when the disc is seen face-on), and they will broaden as the viewing angle
increases (the viewing angle is defined to be the angle between the line of
sight and the axis of the disc). With regard to \ax, in Paper I we showed that
the cold gas and dust absorption may arise in the molecular torus which lies
beyond the BLR. When our viewing angle to the molecular torus is small (\ie
more face-on), any  nuclear soft X-ray emission will have a clear line of sight
to the observer; as the viewing angle increases, more of the torus' material
will intervene, hardening \ax. Therefore, if the Balmer line region lies in the
same plane as the molecular torus, then as the viewing angle increases, the
Balmer line FWHM will increase and \ax\ will harden, producing the correlation
observed (this is illustrated in Fig 12). A similar geometry was proposed by
Baker (1997), who suggested that absorption by dust between the BLR and NLR is
due to orientation dependence, and that this is the cause of the correlation
between radio core dominance and optical luminosity in radio-loud AGN.

A continuation of this model to much softer AGN is possible if the strength of
the unabsorbed soft X-ray component is also related to the viewing angle. This
is possible for the accretion disc model, if the inner edge of the disc is
puffed-up so that it is geometrically-thick (\eg Czerny \& Elvis 1987;  Madau
1988). In this case the soft X-rays are emitted in a cone-shaped region along
the axis of the disc so that the disc spectrum shifts to lower energies (and
thus out of the soft X-ray band) as the viewing angle increases. Therefore if
the disc is co-planar with the Balmer line regions {\sl and} the molecular
torus, then it is possible that the Balmer line FWHM versus \ax\ correlation
may continue from ultrasoft AGN to those like the absorbed RIXOS sources.

\subsubsection{Dust absorption and the [OIII] EW}

The significance of orientation effects on quasar emission line properties was
studied by BG and Boroson (1992), using a sample of 81 (mostly radio-quiet)
UV-excess objects from the Bright Quasar Survey (Schmidt \& Green 1983). It had
already been suggested that the optical continuum in {\sl radio-loud} quasars
is intrinsically anisotropic; \ie that it is brighter when the nucleus is
viewed face-on (Browne \& Wright 1985; Jackson \etal 1989; Baker 1997). Studies
of the permitted lines in radio-loud quasars have indicated that their line
FWHM may also be orientation-dependent, \ie narrower when observed face-on
(Miley \& Miller 1979; Boroson \& Oke 1984; Wills \& Browne 1986).  Boroson
(1992)  tested the hypothesis that the optical continuum and line emission in
{\sl radio-quiet}  quasars is similarly anisotropic by searching, at fixed
\loiii, for a correlation between \hbfwhm\ and \oiiiew\ (the latter will be
angular-dependent if \loiii\ is isotropic and the optical continuum is {\sl
an}isotropic).  No correlation was found, and Boroson (1992) concluded that
this type of orientation effect may {\sl not} be significant in the general
quasar population, or at least, not in objects with UV excesses. 

\beginfigure{14} 
\psfig{figure=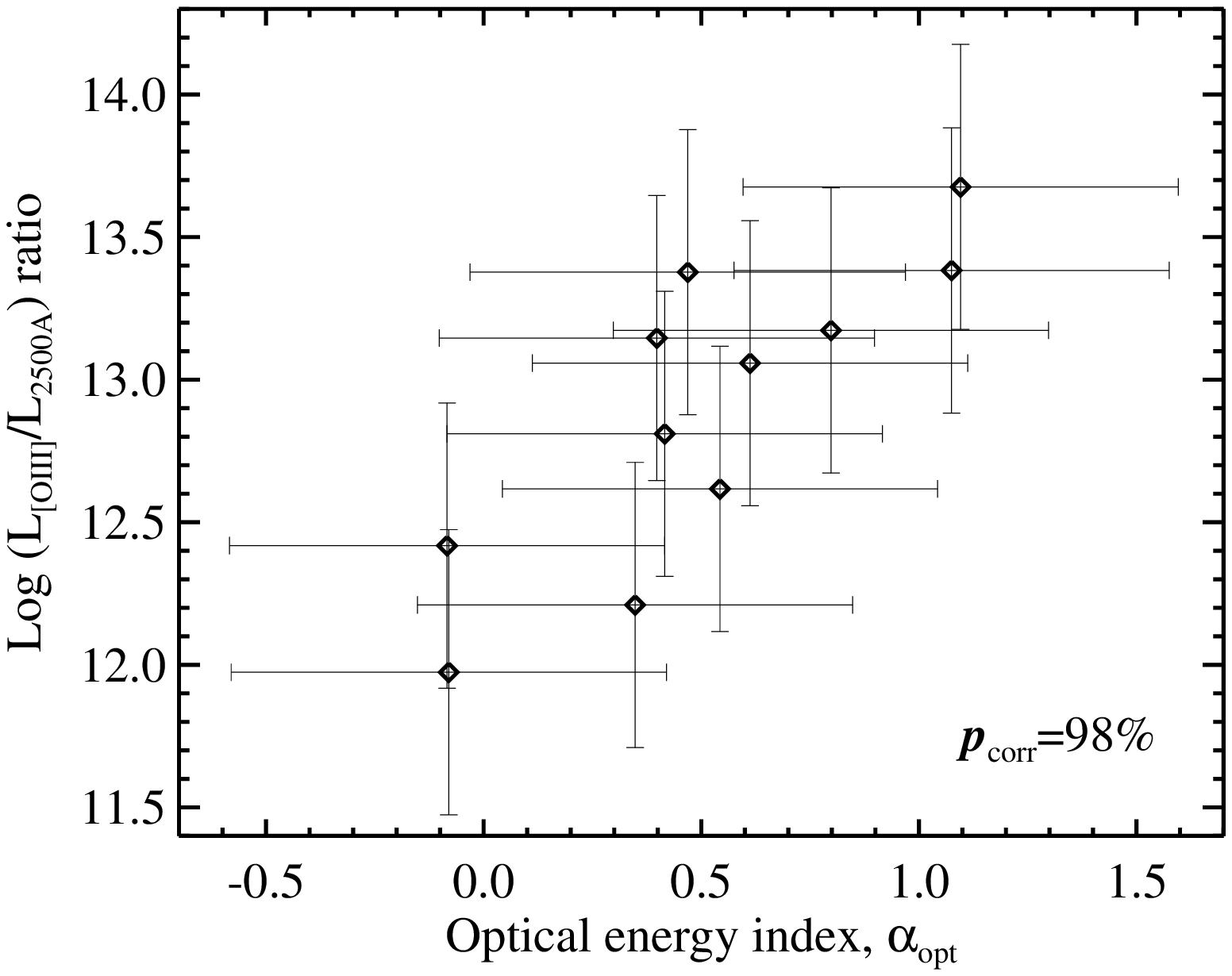,height=2.5in,width=3.2in,angle=0} 
\caption{{\bf Figure 14.} The correlation between the \loiii/\luv\ ratio and
the slope of the optical continuum, \aopt. The data shown are only for those
AGN where \loiii\ and \luv\ have been measured directly.}
\endfigure

We are unable to perform a similar test with the RIXOS AGN, because  of the
effects of dust absorption; this modifies the observed strength of the nuclear
optical continuum (which is supposed to indicate the inclination angle), and
removes the broad \Hb\ component so that the \hbfwhm\ can not be used to
`measure' the intrinsic Balmer line region velocity. Furthermore, the \oiiiew\
will itself be affected by dust absorption (sections 3.2 and 5.3), so although
we cannot reproduce the Boroson (1992) test, we are able to make a further test
for the position of the absorbing medium using \aopt\ and the \oiiiew. If the
dust does lie between the source of the optical continuum and the \oiii\
region, then the \oiiiew\ should increase as \aopt\ softens. No dependence is
seen however (see Table 3), although since the optical continuum around
5000~\AA\ may contain significant emission from the host galaxy,  the overall
effect may be diluted. If instead we look at the dependence of the \loiii/\luv\
ratio on \aopt, (at 2500~\AA, any galactic emission is weak and the effects of
dust absorption are greater), we do find a strong correlation (\pcorr=98 per
cent for objects where the rest-frame 2500~\AA\ flux has been measured
directly, see Fig 14; \pcorr=99 per cent when extrapolated values of \luv\ are
included). This is consistent with effects of dust absorption and suggests that
the dust does indeed lie between the BLR and the NLR.

\subsubsection{UV line emission}

The \ciii\ lines are often thought to be produced in the HIL region which, in
the Krolik \etal (1991) and O'Brien \etal (1994) models, are roughly spherical.
This means that their FWHM would not be expected to reflect any changes in
viewing angle; if, for the RIXOS AGN, \ax\ is an indicator of the viewing
angle, then the lack of a FWHM-\ax\ dependence in \ciii\ is consistent with a
spherical \ciii-region in the orientation-dependent model. 

The lack of a FWHM-\ax\ correlation for \mgii\ implies that this line may also
be emitted from a spherical region (in the orientation model), rather than a
flattened, disc-like geometry which has been suggested for other LILs. Also,
the  mean FWHM of \mgii\ is low relative to \ciii\ (Section 3.3), indicating 
that it is emitted from regions which lie further out from the black hole than
\ciii\ (assuming that the cloud velocities reflect the gravitational potential
of the black hole; see \eg Krolik \etal 1991). 

We note that in the Wills \etal model where the HILs are emitted primarily from
a  paraxial flow, while a second, flattened emission line region produces the
LILs, an orientation dependence in the \ciii\ and \mgii\ emission would be
seen. Since no such dependence is implied for the RIXOS AGN, this model is
probably inappropriate for this sample.

\subsection{Low intrinsic velocity}

Rather than an orientation dependence,  the observed emission line FWHM may
reflect intrinsic differences in the velocity of the line-emitting regions, for
example the lines may be narrow because {\sl (a)} the line region lies further
out from the central black hole, and/or {\sl (b)} the mass of the black hole is
relatively small so that, if the cloud velocity is Keplerian in nature, the
orbital velocities are low (see \eg Laor \etal 1994,1997; BBF). For unobscured
AGN, where \ax\ reflects the strength and/or the temperature of the soft X-ray
component, this would imply a direct relationship between \ax\ and FWHM, \eg
when \ax\ is soft, the BLR may be physically pushed-out or extended.
Alternatively, the BLR may simply be `lit-up' at greater distances due to a
`fine-tuning' of the necessary conditions (\ie ionization parameter and
density) for line emission (see \eg Baldwin \etal 1996).

These physical interpretations suggest that the Balmer line regions are
responding directly to the shape of the soft X-ray spectrum. In this case,  a
correlation would only be seen if the observer saw the same spectrum as that
`seen' by the Balmer line clouds. However for the RIXOS AGN, if the absorption
does occur {\sl beyond} the BLR (as suggested here and in Paper I) then no
correlation should be seen (since the intrinsic slope of the soft X-ray
spectrum is not known and \ax\ now reflects the amount of absorption). Yet the
\Ha\ FWHM does appear to be dependent on \ax, although the weakness of the
formal correlation (\pcorr=91 per cent) may indicate that the relationship is
beginning to break down as \ax\ hardens. If a dependence of \Ha\ FWHM on \ax\
{\sl is} confirmed to hard X-ray slopes, then it would imply (for these kinds
of models) that the velocity of the outer BLR `knows' the amount of absorbing
gas which lies beyond.

The lack of FWHM-\ax\ correlations for \ciii\ and \mgii\ would suggest that
\ax\ has {\sl no} effect on these line-emitting regions, while there {\sl is}
an observable difference to the broad Balmer line gas (whether this is direct
or indirect). Thus any mechanism which operates on the broad Balmer line gas is
probably not significant in the regions where \ciii\ and \mgii\ are produced. 

This has implications for the suitability of these kinds of models. Take, for
example, the case where the high-velocity Balmer line gas might be removed when
\ax\ is soft (which leads to relatively narrow Balmer line emission); the
mechanism which removes portions of the BLR must be able to destroy the gas
very selectively, without also disturbing the \ciii\ and \mgii-producing
regions. It is also unlikely that the BLR has been `pushed-out' en masse, since
both UV {\sl and} Balmer line FWHM would then change in response to \ax.
However, it is possible that the Balmer line  region has been extended leaving
the inner regions unchanged;  if the outer (low-velocity) parts of the BLR
dominate the Balmer line emission, (see \eg Rees \etal 1989), then the Balmer
line widths could vary independently of the UV line  FWHM. In such a situation,
it only remains then to explain why the outer BLR extends to much lower
velocities when the level of cold gas and dust absorption is low (and
vice-versa).

\subsection{BLR geometry and MgII emission}

The behaviour of the \mgii\ line appears to be quite contrary, especially in
the context of the orientation-dependent model. Like the Balmer lines, its FWHM
is {\sl correlated} with its own EW, suggesting that its `broad' component is
emitted from the ILR region (Table 1 shows that `very broad' components are
also sometimes seen, and these  are probably emitted from the VBLR, but they
were disregarded when the correlations were made). However, the \mgii\ FWHM
shows {\sl no} dependence on \ax; for the orientation-dependent model, this
suggests that its emission region must be different from that of the Balmer
lines, \eg  it may have a spherical geometry, like \ciii. These two observed
dependences are thus contradictory in the orientation model discussed here, \ie
the FWHM-EW correlation supports a flattened \mgii\ region, whereas the
lack of a FWHM-\ax\ correlation favours a spherical zone.

If instead, the observed FWHM distribution is not due to an angular dependence,
some of these restrictions are relaxed. The \mgii\ line may be produced within
an ILR which has an intrinsic EW-FWHM dependence, perhaps due to a low covering
factor at large distances/low velocities, and no anisotropic properties are
required.  The lack of a FWHM-\ax\ correlation for \mgii\ then implies that
only the very outer, low-velocity regions of the ILR gas, where the Balmer
lines are produced, would be affected by (or respond to) changes in \ax. This
would favour an extension of the outer BLR to lower velocities when \ax\ is
soft (and absorption is low) and a smaller radial extent of the BLR when \ax\
is hard (\ie the level of absorption is high).

\section{Conclusions}

Using the AGN identified by the RIXOS survey (Mason \etal, in preparation), we
have investigated the optical- and UV-emission line properties of 160 X-ray
selected Seyfert 1s and quasars. Line luminosities, EWs and FWHM of each
emission line subsample have been compared with optical/UV and X-ray continuum
parameters (slopes and luminosities) and we have looked at the relationships
between the lines themselves.

The Balmer lines of the RIXOS AGN are weak relative to optically-selected AGN
whereas the \oiii\ lines are more typical. This is consistent with  the
presence of a dust absorber lying between the BLR and the NLR (perhaps in the
molecular torus), and a weakened ionizing continuum incident on the Balmer and
\oiii-emitting regions (Sections 3.2 and 5.7.1). The EWs and FWHM of the UV
lines (\mgii\ and \ciii) are indistinguishable from those of radio-quiet AGN
(which are expected to dominate this sample), implying that their ionizing flux
may be comparable to that of optically-selected AGN. 

We find Baldwin effects in \mgii\ and \ciii, but none in the optical lines. 
Although \mgii\ increases more slowly than its local continuum, it rises {\sl
faster} than its ionizing continuum which lies in the soft X-rays
(0.6-0.8~\keV). This suggests that, for \mgii\ at least, the Baldwin effect is
caused by changes in the physical conditions of the line-emitting gas or on the
ionization parameter, as a function of continuum luminosity.

We find a correlation between the FWHM and EW of \mgii\ and discuss this
and other FWHM-EW dependences for both high- and low-ionization lines. We
extend the VBLR/ILR model (proposed previously to explain anti-correlations
between the FWHM and EWs of HILs) to the LIL regions, with the additional
premise that there is an intrinsic correlation between FWHM and EW in the ILR.
We present a simple demonstration of this model and show that it can predict
the various relationships between both HILs and LILs. A relationship between
FWHM and EW in the ILR may be due to a reduction in the covering fraction of
the LIL regions when its velocity is low, and thus perhaps also when it lies
relatively further out from the black hole. This scenario tends to support a
link between the amount of cold, dusty gas which lies beyond the
BLR and the distance to and covering fraction of the outer BLR, \ie that when
the BLR extends to greater radii from the centre, there is less cold gas
beyond.

In the orientation-dependent model, an obvious explanation for the FWHM-EW
correlations seen in the LILs would be anisotropic line emission from a
disc-shaped LIL region. However the behaviour of the \mgii\ line does not fit
easily into this geometry; although there is a correlation between the EW and
FWHM of \mgii,  there is no dependence of \mgii\ FWHM on \ax, which would be
expected in the orientation model if \mgii, like the Balmer lines, was also
emitted from a disc-like LIL zone.

Although an anti-correlation between \hbfwhm\ and \ax\ has been seen in 
unabsorbed samples of AGN, none is observed in the RIXOS sources. This is
probably due to population differences between the samples, \ie the lack  of an
\hbfwhm\ dependence on \ax\ in the RIXOS AGN may well be due to the presence of
dust which absorbs the broad component, rendering the apparent FWHM an
unreliable indicator of the `true' velocity of the Balmer line-emitting region
in dust-absorbed AGN. The evidence for a dependence of \Ha\ FWHM on \ax\ in the
RIXOS AGN, albeit weak, supports this hypothesis, since the dust absorption has
a lesser effect on the broad component of \Ha, and it is, of course, also the
stronger line. An anti-correlation between \hafwhm\ and \ax\ in the RIXOS AGN
implies that the Balmer line regions `know' how much absorbing material lies
beyond. This may be due to orientation effects or perhaps because the amount of
surrounding gas and dust affects the distance at which the Balmer line regions
actually form. The FWHM of the UV lines, \ie \ciii\ and \mgii, are  broader
than the Balmer lines and independent of \ax; this is consistent with their
emission from an inner, spherical region as has been suggested for NGC~5548. 

\section*{Acknowledgments}

We thank all in the RIXOS team for their work in obtaining and reducing the
data, and we are grateful to Paul O'Brien for his advice. We are also very
grateful to the referee, Belinda Wilkes, for her thorough and thoughtful report
on this paper, which inspired several improvements. The RIXOS project has 
received observing time under the International Time Programme offered by the
CCI of the Canarian Observatories and has received financial support by the
European Commission through the Access to Large-Scale Facilities Activity of
the Human Capital and Mobility Programme. This research has made use of data
obtained from the UK {\sl ROSAT} Data Archive Centre at the Department of
Physics and Astronomy, University of Leicester (LEDAS) and we would especially
like to thank Mike Watson and Steve Sembay for their kind assistance. We also
thank the Royal Society for a grant to purchase equipment essential to the
RIXOS project. IPF and FCG would like to thank the Consejeria de Educacion
Cultura y Deportes del Gobierno de Canarias for financial support.


\section*{References}

\beginrefs
\bibitem Alexander T., Netzer H., 1994, MNRAS, 270, 781
\bibitem Antonucci R., 1993, Ann. Rev. Astron. Astrophys., 31, 473
\bibitem Arnaud K. A., Branduardi-Raymont G., Culhane J. L., Fabian A. C.,
Hazard C., McGlynn T. A., Shafer R. A., Tennant A. F., Ward M. J., 1985, MNRAS,
217, 105
\bibitem Baker J. C., 1997, MNRAS, 286, 23
\bibitem Baldwin J. A., 1977, ApJ, 214, 679
\bibitem Baldwin J. A., \etal, 1996, ApJ, 461, 664
\bibitem Blumethal G. R., Keel W. C., Miller J. S., 1982, ApJ, 257, 499
\bibitem Boller Th., Brandt W. N., Fink H., 1996, A\&A, 305, 53 (BBF)
\bibitem Boroson T. A., 1992, ApJ, 399, L15
\bibitem Boroson T. A., Green R. F., 1992, ApJS, 80, 109
\bibitem Boroson T. A., Oke J. B., 1984, ApJ, 281, 535
\bibitem Brotherton M. S., Wills B. J., Steidel C C., Sargent W. L. W., 1994,
ApJ, 423, 131 (B94)
\bibitem Browne I. W. A., Wright A. E., 1985, MNRAS, 213, 97
\bibitem Cardelli J. A., Clayton G. C., Mathis J. S., 1989, 345, 245
\bibitem Clavel J. \etal, 1991, ApJ, 366, 64
\bibitem Collin-Souffrin S., Dyson J. E., M$^c$Dowell J. C., Perry J. J., 1988,
MNRAS, 232, 539
\bibitem Czerny, B., Elvis, M. 1987, ApJ, 321, 305
\bibitem Davidson K, Netzer H., 1979, Rev. Mod. Phys., 51, 715
\bibitem Edwards A. C., 1980, MNRAS, 190, 757
\bibitem Ferland G. J., 1993, University of Kentucky Department of Physics
              and Astronomy Internal Report
\bibitem Ferland G. J., Netzer H., 1979, ApJ, 229, 274
\bibitem Ferland G. J., Peterson B. M., Horne K., Welsh W. F., Nahar S. N.,
1992, ApJ, 387, 95
\bibitem Fiore F., Elvis M., Siemiginowska A., Wilkes B. J., M$^c$Dowell J. C.,
Mathur S., 1995, ApJ, 449, 74
\bibitem Francis P. J., Hewitt P. C., Foltz C. B., Chaffee F. H., 1992, ApJ, 
398, 476 
\bibitem Gaskell C. M., 1985, ApJ, 291, 112
\bibitem Gioia I. M., Maccacaro T., Schild R. E., Stocke J. T., Liebert J. W.,
Danziger I. J., Kunth D., Lub J., 1984, ApJ, 283, 495
\bibitem Goodrich R. W., 1989, ApJ, 342, 224
\bibitem Gorenstein P., 1975, ApJ, 198, 40
\bibitem Green P. J., 1996, ApJ, 467, 61
\bibitem Hewitt A., Burbidge G., 1989, ApJS, 69, 1
\bibitem Jackson N., Browne I. W. A., Murphy D. W., Saikia D. J., 1989, Nature,
338, 485
\bibitem Kriss G. A., Canizares C. R., 1982, ApJ, 261, 51
\bibitem Krolik J. H., Kallman T. R., 1988, ApJ, 324, 714
\bibitem Krolik J. H., Horne K., Kallman T. R., Malkan M. A., Edelson R. A.,
Kriss G. A., 1991, ApJ, 371, 541
\bibitem Kwan J. Y., Krolik J. H., 1981, ApJ, 250, 478
\bibitem Laor A., Fiore F., Elvis, M., Wilkes, B. J., M$^c$Dowell, J. C.,
1994, ApJ, 435, 611
\bibitem Laor A., Fiore F., Elvis, M., Wilkes, B. J., M$^c$Dowell, J. C.,
1997, ApJ, 477, 93
\bibitem Madau, P., 1988, ApJ, 327, 116
\bibitem Mason K. O., Puchnarewicz E. M., Jones, L. R., 1996, MNRAS, 283, L26
\bibitem Miley G. K., Miller J. S., 1979, ApJ, 228, L55
\bibitem Mittaz J. P. D., \etal, 1997, MNRAS, submitted
\bibitem Mushotzky R. F., Done C., Pounds K. A., 1993, Ann. Rev. Astron.
Astrophys., 31, 717
\bibitem Mushotzky R. F., Ferland G. J., 1984, ApJ, 278, 558
\bibitem Netzer H., 1993, ApJ, 411, 594
\bibitem Nicholson K. L., Mittaz J. P. D., Mason K. O., 1997, MNRAS, 285, 831
\bibitem O'Brien P. T., Goad M. R., Gondhalekar P. M., 1994, MNRAS, 268, 845
\bibitem Osterbrock D. E., 1977, ApJ, 215, 733
\bibitem Osterbrock D. E., Pogge R. W., 1985, ApJ, 297, 166
\bibitem Peterson B. M., \etal, 1991, ApJ, 368, 119 
\bibitem Peterson B. M., 1993, PASP, 105, 247
\bibitem Pfeffermann, E. \etal, 1986, Proc S. P. I. E., 733, 519
\bibitem Pravdo S. H., Nugent J. J., Nousek J. A., Jensen K., Wilson A. S., 
Becker R. H., 1981, ApJ, 251, 501
\bibitem Puchnarewicz E. M., Mason, K. O., C\'ordova, F. A., Kartje, J.,
              Branduardi-Raymont, G., Mittaz,~J.~P.~D., Murdin,~P.~G., 
              Allington-Smith,~J., 1992, MNRAS, 256, 589
\bibitem Puchnarewicz E. M., Mason K. O., Romero-Colmenero E., Carrera F. J., 
Hasinger G., M$^c$Mahon R., Mittaz J. P. D., Page M. J., Carballo R., 1996a, MNRAS,
281, 1243 (Paper I)
\bibitem Puchnarewicz E. M., Mason K. O., Siemiginowska A., Pounds K. A., 1995,
MNRAS, 276, 20
\bibitem Puchnarewicz E. M., Mason K. O. and Carrera F. J., 1996b, MNRAS, 283,
1311
\bibitem Rees M. J., Netzer H., Ferland G. J., 1989, ApJ, 347, 640
\bibitem Ryter C., Cesarsky C. J., Audouze J., 1975, ApJ, 198, 103
\bibitem Schmidt H., Green R. F., ApJ, 1983, 269, 352
\bibitem Schultz H., 1992, in Physics of Active Galactic Nuclei, ed W. J.
Duschl \& S. J. Wagner (Berlin:Springer), 235
\bibitem Shields J. C., Ferland G. J., Peterson B. M., 1995, ApJ, 441, 507
\bibitem Stark, A. A., Gammie, C. F., Wilson, R. F., Ball, J., Linke, R. A.,
              Heiles, C., Hurwitz, M., 1992, ApJS, 79, 77
\bibitem Steidel C. C., Sargent W. L. W., ApJ, 1991, 382, 433 (SS91)
\bibitem Stephens S. A., 1989, AJ, 97, 10
\bibitem Tananbaum, H., \etal, 1979, ApJ, 234, L9
\bibitem Turner, T. J., Pounds, K. A., 1989, MNRAS, 240, 833
\bibitem van Groningen E., de Bruyn A. G., 1989, A\&A, 1989, 211, 293
\bibitem Walter, R., Fink, H. H., 1993, A\& A, 274, 105
\bibitem Wills B. J., Brotherton M. S., Fang D., Steidel C. C., Sargent W. L.
W., 1993, ApJ, 415, 563
\bibitem Wills B. J., Browne I. W. A., 1986, ApJ, 302, 56
\bibitem Wills B. J., Netzer H., Wills D., 1985, ApJ, 288, 94
\bibitem Zheng W., Malkan M. A., 1993, 415, 517
\endrefs
\bye